\documentclass[]{pasj01}

\begin{document} 
\Received{}
\Accepted{}

\title{SILVERRUSH. VI. A simulation of Ly$\alpha$ emitters in the
reionization epoch and a comparison with Subaru Hyper Suprime-Cam 
survey early data}

\author{Akio K.\ \textsc{Inoue}\altaffilmark{1}%
}

\author{Kenji \textsc{Hasegawa}\altaffilmark{2}}
\author{Tomoaki \textsc{Ishiyama}\altaffilmark{3}}
\author{Hidenobu \textsc{Yajima}\altaffilmark{4,5}}
\author{Ikkoh \textsc{Shimizu}\altaffilmark{6}}
\author{Masayuki \textsc{Umemura}\altaffilmark{7}}
\author{Akira \textsc{Konno}\altaffilmark{8,9}}
\author{Yuichi \textsc{Harikane}\altaffilmark{8,10}}
\author{Takatoshi \textsc{Shibuya}\altaffilmark{8}}
\author{Masami \textsc{Ouchi}\altaffilmark{8,11}}
\author{Kazuhiro \textsc{Shimasaku}\altaffilmark{9,12}}
\author{Yoshiaki \textsc{Ono}\altaffilmark{8}}
\author{Haruka \textsc{Kusakabe}\altaffilmark{9}}
\author{Ryo \textsc{Higuchi}\altaffilmark{8}}
\author{Chien-Hsiu \textsc{Lee}\altaffilmark{13}}


\altaffiltext{1}{Department of Environmental Science and Technology, 
Faculty of Design Technology, Osaka Sangyo University,
3-1-1 Nakagaito, Daito, Osaka 574-8530, Japan}
\email{akinoue@est.osaka-sandai.ac.jp}

\altaffiltext{2}{Department of Physics, Graduate School of Science,
Nagoya University, Furo-cho, Chikusa-ku, Nagoya, Aichi 464-8602, Japan} 

\altaffiltext{3}{Institute of Management and Information Technologies,
Chiba University, 1-33, Yayoi-cho, Inage-ku, Chiba 263-8522, Japan}

\altaffiltext{4}{Frontier Research Institute for Interdisciplinary
Sciences, Tohoku University, Sendai 980-8578} 

\altaffiltext{5}{Astronomical Institute, Tohoku University, Sendai
980–8578, Japan}

\altaffiltext{6}{Theoretical Astrophysics, Department of Earth \& Space
Science, Osaka University, 1-1 Machikaneyama, Toyonaka, Osaka 560-0043, Japan}

\altaffiltext{7}{Center for Computational Sciences, University of
Tsukuba, 1-1-1 Tennodai, Tsukuba, Ibaraki 305-8577, Japan}

\altaffiltext{8}{Institute for Cosmic Ray Research, The University of
Tokyo, Kashiwa-no-ha, Kashiwa 277-8582}

\altaffiltext{9}{Department of Astronomy, Graduate School of Science,
The University of Tokyo, Hongo, Bunkyo-ku, Tokyo, 113-0033, Japan}

\altaffiltext{10}{Department of Physics, Graduate School of Science,
The University of Tokyo, Hongo, Bunkyo-ku, Tokyo, 113-0033, Japan}

\altaffiltext{11}{Kavli Institute for the Physics and Mathematics of the
Universe (Kavli IPMU), WPI, The University of Tokyo, Kashiwa, Chiba
277-8583, Japan}

\altaffiltext{12}{Research Center for the Early Universe, Graduate
School of Science, The University of Tokyo, 7-3-1 Hongo, Bunkyo, Tokyo
113-0033, Japan}

\altaffiltext{13}{Subaru Telescope, National Astronomical Observatory of
Japan, 650 North A'ohoku Place, Hilo, HI 96720, USA}



\maketitle

\begin{abstract}
The survey of Lyman $\alpha$ emitters (LAEs) with Subaru Hyper 
 Suprime-Cam, called SILVERRUSH (Ouchi et al.), is producing massive 
 data of LAEs at $z\gtrsim6$.
Here we present LAE simulations to compare the SILVERRUSH data. 
In 162$^3$ comoving Mpc$^3$ boxes, where numerical radiative transfer 
 calculations of reionization were performed, LAEs have been modeled 
 with physically motivated analytic recipes as a function of halo mass.
We have examined $2^3$ models depending on the presence or absence of
 dispersion of halo Ly$\alpha$ emissivity, dispersion of the halo 
 Ly$\alpha$ optical depth, $\tau_\alpha$, and halo mass dependence
 of $\tau_\alpha$.
The unique free parameter in our model, a pivot value of $\tau_\alpha$,
 is calibrated so as to reproduce the $z=5.7$ Ly$\alpha$ luminosity
 function (LF) of SILVERRUSH. 
We compare our model predictions with Ly$\alpha$ LFs at $z=6.6$ and
 $7.3$, LAE angular auto-correlation functions (ACFs) at $z=5.7$ and
 $6.6$, and LAE fractions in Lyman break galaxies at $5<z<7$.
The Ly$\alpha$ LFs and ACFs are reproduced by multiple models,
 but the LAE fraction turns out to be the most critical test. 
The dispersion of $\tau_\alpha$ and the halo mass dependence of 
 $\tau_\alpha$ are essential to explain all observations reasonably.
Therefore, a simple model of one-to-one correspondence between halo 
 mass and Ly$\alpha$ luminosity with a constant Ly$\alpha$ escape 
 fraction has been ruled out.
Based on our best model, we present a formula to estimate the 
 intergalactic neutral hydrogen fraction, $x_{\rm HI}$, from 
 the observed Ly$\alpha$ luminosity density at $z\gtrsim6$.
We finally obtain $x_{\rm HI}=0.5_{-0.3}^{+0.1}$ as a volume-average at
 $z=7.3$.
\end{abstract}

\section{Introduction}

Cosmic reionization is one of the central issues in modern astronomy.
Understanding this major phase transition in the early Universe is 
closely related to the formation of the first generation of galaxies.
There are three key questions to understand reionization: 
the history, sources and topology. The epoch of reionization is
constrained to the redshift range of $6<z<9$ by the Gunn-Peterson
trough found in distant QSO spectra (e.g., \cite{Fan2006}) and the
Thomson scattering optical depth measured from polarization in the
Cosmic Microwave Background (CMB) (e.g., \cite{Planck2014}). However,
constraints on the evolution of the hydrogen neutral fraction, 
$x_{\rm HI}$, in the intergalactic medium (IGM) during the epoch are
still weak \citep{GreigMesinger17}. Leading candidates of
reionization sources are star-forming galaxies if they inject $\sim20\%$
of the produced Lyman continuum into the IGM (e.g., \cite{Inoue2006}). 
However, this escape fraction, $f_{\rm esc}$, of the Lyman continuum is
uncertain, while efforts to measure or constrain $f_{\rm esc}$ directly
(e.g., \cite{Micheva2017}) and indirectly (e.g., \cite{Khaire2016}) are
on-going. There are two major types of the reionization topology:
inside-out \citep{Iliev2006} and outside-in \citep{Miralda2000}. The
topology may depend on the types of the dominant ionizing sources,
i.e. galaxies or AGNs, owing to the difference of the mean-free-path (or
mean energy) of ionizing photons. In theoretical models, it also depends
on the treatment of recombination \citep{Choudhury2009}. Currently, it
is unknown yet which topology was realized in the real Universe.

Lyman $\alpha$ emitters (LAEs) are considered to be a useful probe of
reionization because Ly$\alpha$ is the resonant line of neutral hydrogen
and sensitive to $x_{\rm HI}$ (e.g., \cite{Dijkstra2014}). A decrease
in Ly$\alpha$ luminosity functions (LFs) at $z>6$ following the no
evolution at $3<z<6$ \citep{Ouchi2008} is interpreted as an increase in
$x_{\rm HI}$ at $z>6$ due to reionization 
\citep{Kashikawa2006,Kashikawa2011,Ouchi2010,Ota2010,Shibuya2012,Konno2014}.
A decrease of the LAE number fraction in Lyman break galaxies (LBGs) at
$z>6$ is also interpreted as a reionization signature 
\citep{Stark2010,Stark2011,Ono2012,Schenker2013,Furusawa2016}.
In the inside-out reionization scenario, the clustering of LAEs is
expected to be enhanced as $x_{\rm HI}$ increases \citep{mcquinn07}.
\citet{Ouchi2010} compared the observed LAE angular correlation
functions (ACFs) with the model prediction by \citet{mcquinn07} and put
a constraint on the neutral hydrogen fraction of the IGM as
 $x_{\rm HI}\lesssim0.5$ at $z=6.6$.

There is further progress in LAE surveys at $z\gtrsim7$.
\citet{Ota2017} has performed the deepest survey of LAEs at $z\sim7$
with Subaru/Suprime-Cam and confirmed a decrease of the Ly$\alpha$ LF at 
$z=7.0$ compared to those at $z=5.7$ and $6.6$. On the other hand,
\citet{Zheng2017} has reported no difference in the bright-end of the
Ly$\alpha$ LF ($L_{\rm Ly\alpha}>10^{43.5}~{\rm erg~s^{-1}}$) 
between $z=5.7$ and $z=6.9$ based on the widest survey of LAEs at
$z\sim7$ so far with DECam on the NOAO/CTIO 4~m Blanco Telescope. 
Clearly, wider and deeper LAE surveys are required to settle the issue.
Hyper Suprime-Cam (HSC;
\cite{Miyazaki2012,Miyazaki2017,Komiyama2017,Kawanomoto2017,Furusawa2017})
mounted on the 8.2~m Subaru Telescope is the most ideal
instrument for such surveys. Indeed, we are conducting surveys of LAEs
at $z=5.7$, $6.6$ and $7.3$, called Systematic Identification of LAEs
for Visible Exploration and Reionization Research Using Subaru HSC 
(SILVERRUSH; \cite{Ouchi2017}), under the Subaru Strategic Program with
HSC \citep{Aihara2017}. Early data results are already available in a 
series of papers
\citep{Ouchi2017,Shibuya2017,Shibuya2017b,Konno2017,Harikane2018,Higuchi2018}.
We are also conducting additional HSC narrowband observations including
an LAE survey at $z=7.0$ \citep{Itoh2018}, 
called CHORUS (Cosmic HydrOgen Reionization
Unveiled with Subaru; Inoue et al.~in prep.).

To interpret SILVERRUSH and CHORUS data and deduce the information of
reionization, we need to make use of a model of LAEs in that epoch. In
literature, there are many such models adopting a range of simplification 
(e.g., \cite{mcquinn07,Iliev2008,Kobayashi2010,Zheng2010,Dayal2011,Jensen2013,kakiichi16}).
For example, works with a large-scale ($>100$ comoving Mpc) and full
numerical radiative transfer simulation of reionization tend to adopt a
simplified LAE model assuming one-to-one correspondence between halo
mass and Ly$\alpha$ luminosity with a constant Ly$\alpha$ escape fraction 
(e.g., \cite{mcquinn07,Iliev2008,Jensen2013,kakiichi16}).
On the other hand, works with a detailed modeling of LAEs tend to adopt
a simplified or no radiative transfer simulation of reionization 
(e.g., \cite{Kobayashi2010,Dayal2011}).
\citet{Zheng2010} simulated Ly$\alpha$ photon transfer in galaxy halos
and the circum-galactic medium (CGM) in a large-scale reionization simulation
although the Ly$\alpha$ transfer in the interstellar medium (ISM) of
galaxies was not resolved. This is still difficult even today.

This work presents a new LAE simulation adopting a more physically
realistic LAE model in a large-scale full numerical radiative transfer 
simulation of reionization to be compared with SILVERRUSH data. Such a
model is essential when using LAEs as a probe of reionization but
overlooked so far. We will examine validity of the assumption of
one-to-one correspondence between halo mass and Ly$\alpha$ luminosity
often adopted in literature and rule out it after comparisons with early
data of SILVERRUSH. 

The structure of this paper is as follows; 
In \S2, we describe the reionization simulations consisting of 
radiation hydrodynamics simulations to produce models of halos'
emissivity and the IGM clumping factor, N-body simulations to produce
the density structure in the IGM as well as the halo distribution, and
radiative transfer simulations to compute the neutral hydrogen
distribution in the IGM.
In \S3, we make $2^3$ LAE models depending on the presence
or absence of fluctuation in the Ly$\alpha$ production, fluctuation in
the ISM/CGM opacity for Ly$\alpha$, and halo mass dependence of the 
opacity. In \S4, we show the comparisons of the simulations with early
data of SILVERRUSH and identify the best model among the $2^3$
models. In \S5, we discuss the nature of LAEs in the reionization era
expected from the best model and how to constrain the reionization
history with LAEs. The last section is devoted to our conclusions.

The cosmological parameters adopted in this paper are
$\Omega_0=0.31$, $\Omega_b=0.048$, $\lambda_0=0.69$, $h=0.68$,
$n_s=0.96$, and $\sigma_8=0.83$. These values are matching with a result
of the CMB observations conducted by the Planck satellite
\citep{Planck2014,Planck2016}.
All magnitudes in this paper are expressed in the AB system 
\citep{Oke1990}.

\section{Reionization simulation}

To model LAEs, we need to know the neutral hydrogen (H~{\sc i})
distribution in the IGM which affects the observability of LAEs. 
In this paper, we adopt a large-scale reionization simulation by
Hasegawa et al.~(in prep.) for the H~{\sc i} distribution. 
Since the star formation rates and ionizing photon escape
fractions of galaxies are regulated by radiative
feedback \citep{Umemura2012, Hasegawa2013}, high resolution radiation
hydrodynamics (RHD) simulations are desirable for simulating 
reionization. However, the volume of a cosmological RHD simulation is
generally limited due to expensive numerical costs. Therefore, 
Hasegawa et al.\ adopted a two-step approach. 
First, they performed a cosmological RHD simulation in
a computationally feasible box size to model galaxies and the IGM
clumping factor under the influence of the radiative feedback
\citep{Hasegawa2016}. Using the RHD models of galaxies/IGM, then, 
they solved radiative transfer of ionizing photons in a large-scale
$N$-body simulation box to obtain the cosmological distribution of 
$x_{\rm HI}$. 
In the following, we briefly describe the simulations, while a full 
description is presented in Hasegawa et al.~(in prep.).
Some information can be also found in \citet{Hasegawa2016}, 
\citet{Kubota2017} and \citet{Yoshiura2017}.

\subsection{Radiative hydrodynamics simulations for making recipes}

The RHD simulation was performed with $2\times512^3$ particles in a 20
comoving Mpc$^3$ box. 
The implemented physical processes in the RHD simulation are similar to
those in \citet{Hasegawa2013}, except for the
following two points: (i) stellar age dependent spectra of  
P{\'E}GASE2 \citep{pegase}
\footnote{http://www2.iap.fr/users/fioc/PEGASE.html} 
and (ii) attenuation by dust grains \citep{DraineLee84}
\footnote{https://www.astro.princeton.edu/~draine/dust/dust.html}.

The halo ionizing emissivity recipe is described as a look-up table of
average escaping Lyman continuum spectra 
($91.2\leq\lambda_{\rm rest}/{\rm \AA}\leq912$) as a function of halo
mass and its local ionization degree obtained from the RHD simulation. 
The look-up table spectra already includes the escape fraction depending
on the wavelength. It is worth mentioning that 
the contribution from massive galaxies is less dominant compared to previous 
reionization simulations (e.g., \cite{Iliev2006}), 
because the RHD simulation shows that the escape fraction decreases with
increasing halo mass \citep{Hasegawa2013,Hasegawa2016,Xu2016}: 
$\sim$10\%, 2\% and 0.2\% at $M_{\rm h}=10^{9}$, $10^{10}$ and
$10^{11}$ M$_\odot$, respectively. The values are very consistent with
those in a higher resolution RHD simulation by \citet{Xu2016} for
halos less than $\sim10^9$ M$_\odot$ which they examined.

The clumping factor, defined as 
${\cal C}_{\rm HII}=\langle n_{\rm HII}^2 \rangle/\langle n_{\rm HII} \rangle^2$ 
with $\langle ~ \rangle$ being the volume average, 
is an important parameter regulating the recombination time-scale 
in reionization simulations.
Previous studies on the clumping factor have shown that it 
becomes $\sim3$ on average by photo-heating during reionization 
\citep{Pawlik09, Finlator12}. 
\citet{Hasegawa2016} revisited the influence of the radiative feedback 
on the clumping factor and found that it depends on the density and 
ionization degree locally. Therefore, a spatially constant clumping factor 
is not correct. Hence, based on the RHD simulation, 
Hasegawa et al.~(in prep.) made a look-up table 
for the local clumping factor as a function of the local density and 
the local ionization degree within $0.6$ comoving Mpc scale, 
corresponding to the grid size of the post-processing radiative transfer 
described later in \S2.2.2. The clumping factor is larger 
as the local density increases or the local ionization degree decreases;  
${\cal C}_{\rm HII}=$2--4 in ionized regions ($x_{\rm HII}>0.5$) and 
${\cal C}_{\rm HII}>10$ in neutral overdensity regions 
($x_{\rm HII}<0.1$ and $\rho_{\rm DM}/\langle \rho_{\rm DM} \rangle>1$
with $\rho_{\rm DM}$ being the dark matter density), 
resulting in a slow (fast) reionization process in high (low)
density regions relative to a constant clumping factor model 
(Hasegawa et al.~in prep.).

\subsection{Radiative transfer in the IGM}

\subsubsection{N-body simulation}

The $N$-body simulation was run on the K computer at the RIKEN Advanced
Institute for Computational Science, using a massively parallel TreePM 
code, GreeM \citep{Ishiyama2009b,Ishiyama2012}
\footnote{http://hpc.imit.chiba-u.jp/\~ishiymtm/greem/}
with the Phantom-GRAPE software library 
\citep{Nitadori2006,Tanikawa2012,Tanikawa2013}
\footnote{http://code.google.com/p/phantom-grape/}, 
with $4096^3$ dark matter particles in a comoving box of $162$
Mpc. A particle mass is $2.46 \times 10^6 ~ \rm M_{\odot}$.   
The gravitational softening length is $1.24~ \rm kpc$.
The initial particle distribution was generated using a publicly
available code, 2LPTic (e.g., \cite{Crocce2006}) 
\footnote{http://cosmo.nyu.edu/roman/2LPT/}. 
The matter transfer function was obtained through the online
version of CAMB \citep{Lewis2000}
\footnote{http://lambda.gsfc.nasa.gov/toolbox/tb\_camb\_form.cfm}.  
The initial and final redshifts of the simulation are 127 and 5.5,
respectively.

For reionization simulations described in \S2.2.2, halos were
identified by the Friends-of-Friends (FoF) algorithm \citep{defw} with a
linking parameter of $b=0.2$.  The minimum halo mass is 
$9.80 \times 10^7 ~ \rm M_{\odot}$, which corresponds to 40 particles.
The halo catalogs were frequently stored with a constant time interval
of 9.6 Myr.  The total number of outputs are 100 from redshifts 30 to
5.5. At the same redshifts, we stored dark matter mass density on a
uniform grid calculated by the TSC (triangular shaped cloud) scheme.
The number of grid points is $256^3$, which gives $632~ \rm kpc$ spatial
resolution (comoving).

\subsubsection{Radiative transfer simulation}

Post-processing radiative transfer calculations in the $N$-body
simulation box (\S2.2.1) were performed with the recipes of halos'
ionizing emissivity and IGM H~{\sc ii} clumpiness described in \S2.1.
In the following, we present radiative transfer and chemical reaction 
equations only for H~{\sc i} but we also solved those for He~{\sc i} 
and He~{\sc ii} simultaneously. The thermal equations were also 
solved at every time-step to evaluate the photoheating rates  
(see \cite{Kubota2017} \S3.1 for a complete set of equations).

The time evolution of the neutral fraction of hydrogen, 
$x_{\rm HI}=n_{\rm HI}/n_{\rm H}$, defined as the number density ratio 
of neutral hydrogen and all hydrogen at a certain grid point 
is given by 
\begin{equation}
 \frac{dx_{\rm HI}}{dt} = -k_\gamma^{\rm HI} x_{\rm HI}
  -k_{\rm C}^{\rm HI} x_{\rm HI} n_e 
  + {\cal C}_{\rm HII}(x_{\rm HII},\rho_{\rm DM}) 
  \alpha_{\rm B}^{\rm HII} x_{\rm HII} n_e \,,
\end{equation}
where $k_\gamma^{\rm HI}$ is the H~{\sc i} photoionization rate, 
$k_{\rm C}^{\rm HI}$ is the H~{\sc i} collisional ionization coefficient, 
${\cal C}_{\rm HII}$ is the H~{\sc ii} clumping factor, 
$\alpha_{\rm B}^{\rm HII}$ is the Case B H~{\sc ii} recombination 
coefficient \citep{ost89}, and $n_e$ is the electron number density.
The clumping factor ${\cal C}_{\rm HII}$ is described as the look-up table 
depending on the ionized fraction, $x_{\rm HII}=n_{\rm HII}/n_{\rm H}$, 
and the dark matter density, $\rho_{\rm DM}$ (\S2.1). 
The photoionization rate is given by 
\begin{equation}
 k_\gamma^{\rm HI} = \sum_j \frac{1}{4\pi R_j^2} 
  \int_{\nu_{\rm HI}}^\infty \frac{L_{\nu,j}(x_{{\rm HI},j})}{h\nu} 
  \sigma_\nu^{\rm HI} e^{-\tau_{\nu,j}} d\nu\,,
\end{equation}
where $j$ is the index of the grid points, $R_j$ is the proper distance 
between the $j$-th grid point and the point interested, $\tau_{\nu,j}$ 
is the optical depth in the distance $R_j$ at the radiation frequency, 
$\nu$, $L_{\nu,j}$ is the luminosity density at the frequency $\nu$ 
at the $j$-th grid point, $\sigma_\nu^{\rm HI}$ is the H~{\sc i} 
photoionization cross section for the radiation at the frequency $\nu$, 
$\nu_{\rm HI}$ is the H~{\sc i} Lyman limit frequency, and $h$ is the 
Planck constant. The computational cost for ray-tracing to estimate 
$\tau_{\nu,j}$ is significantly reduced by the algorithm of 
\citet{Susa2006} and \citet{Hasegawa2009}, keeping the accuracy of 
the long-characteristic nature in equation~(2). The luminosity density 
$L_{\nu,j}$ of each grid point is given by 
\begin{equation}
 L_{\nu,j}(x_{{\rm HI},j}) = \int l_{\nu}(M_{\rm h},x_{{\rm HI},j}) 
  \phi_j(M_{\rm h}) dM_{\rm h}\,,
\end{equation}
where $\phi_j$ is the halo mass function in the volume allotted to 
the $j$-th grid point. The luminosity density at the radiation 
frequency $\nu$, $l_\nu$, is described as the look-up table as 
a function of the halo mass, $M_{\rm h}$, and the neutral fraction, 
$x_{{\rm HI},j}$, at the grid point (\S2.1).

\begin{figure}
 \begin{center}
  \includegraphics[width=5cm]{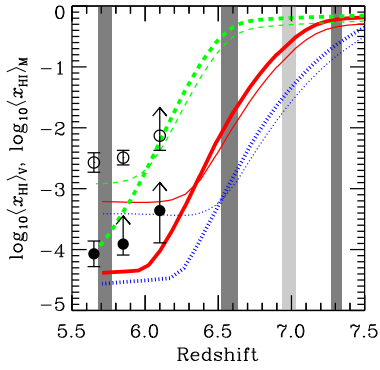} 
 \end{center}
 \caption{The evolution of the neutral hydrogen fraction, 
 $x_{\rm HI}=n_{\rm HI}/n_{\rm H}$ with $n$ being the number density, in
 our simulations. The thick and thin lines correspond to the volume
 average, $\langle x_{\rm HI} \rangle_{\rm V}=\langle n_{\rm HI} /
 n_{\rm H} \rangle$, and mass-density-weighted average, $\langle x_{\rm
 HI}\rangle_{\rm M}=\langle n_{\rm HI} \rangle/\langle n_{\rm
 H}\rangle$, respectively. The solid (red), dashed (green) and dotted
 (blue) lines are the {\it mid}, {\it late} and {\it early} models,
 respectively. The filled and open circles with error-bars are the
 observational estimates of the volume average and mass average of
 $x_{\rm HI}$ taken from Fan et al.~(2006). The vertical shades indicate
 the redshift ranges explored with HSC NB816, NB921, and NB101 filters
 from SILVERRUSH (Ouchi et al.~2017) and NB973 filter from CHORUS (Inoue
 et al.~in prep.) surveys.}
 \label{fig:xHIhistory}
\end{figure}

\begin{figure*}
 \begin{center}
  \includegraphics[width=6cm]{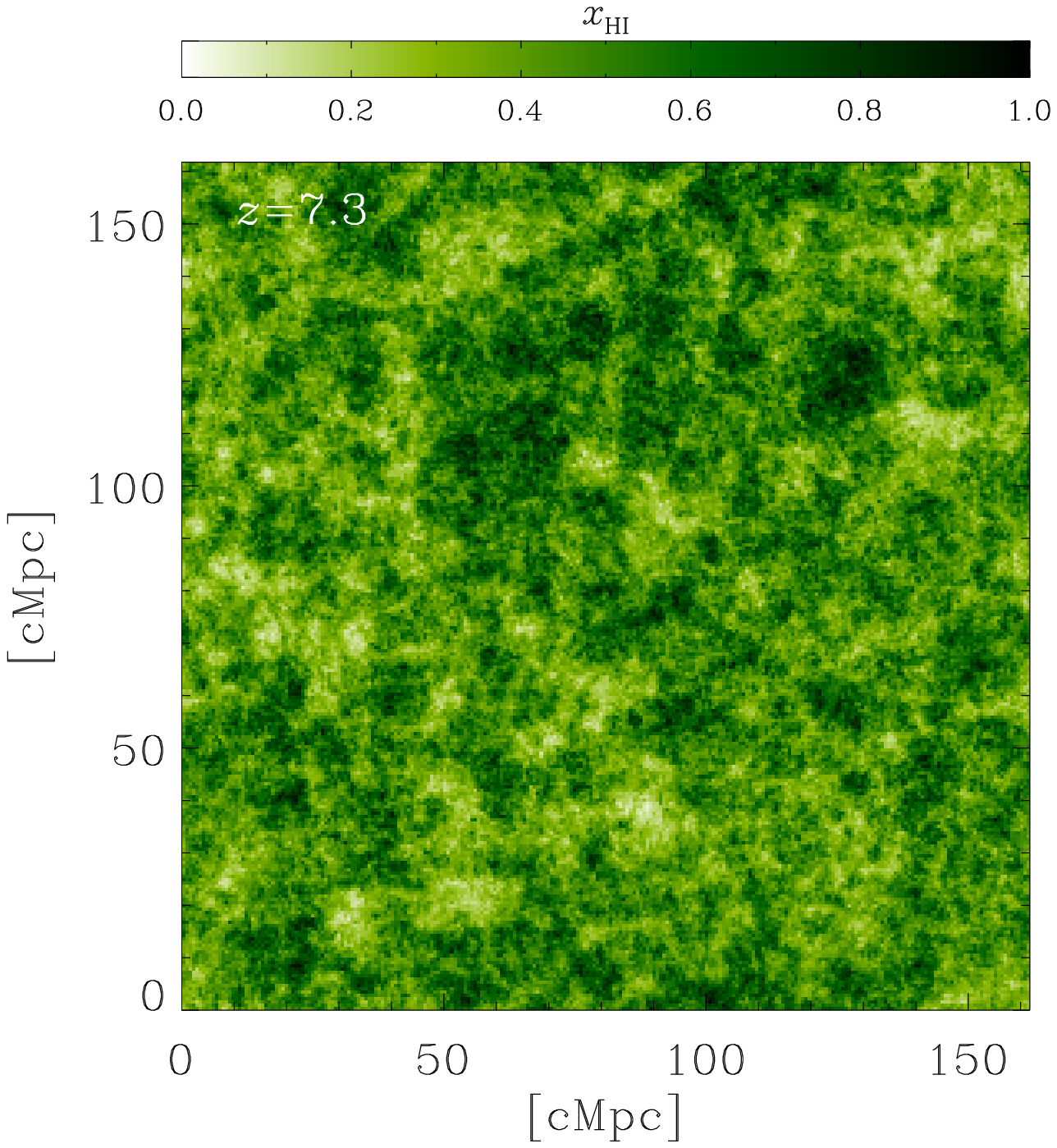}
  \includegraphics[width=6cm]{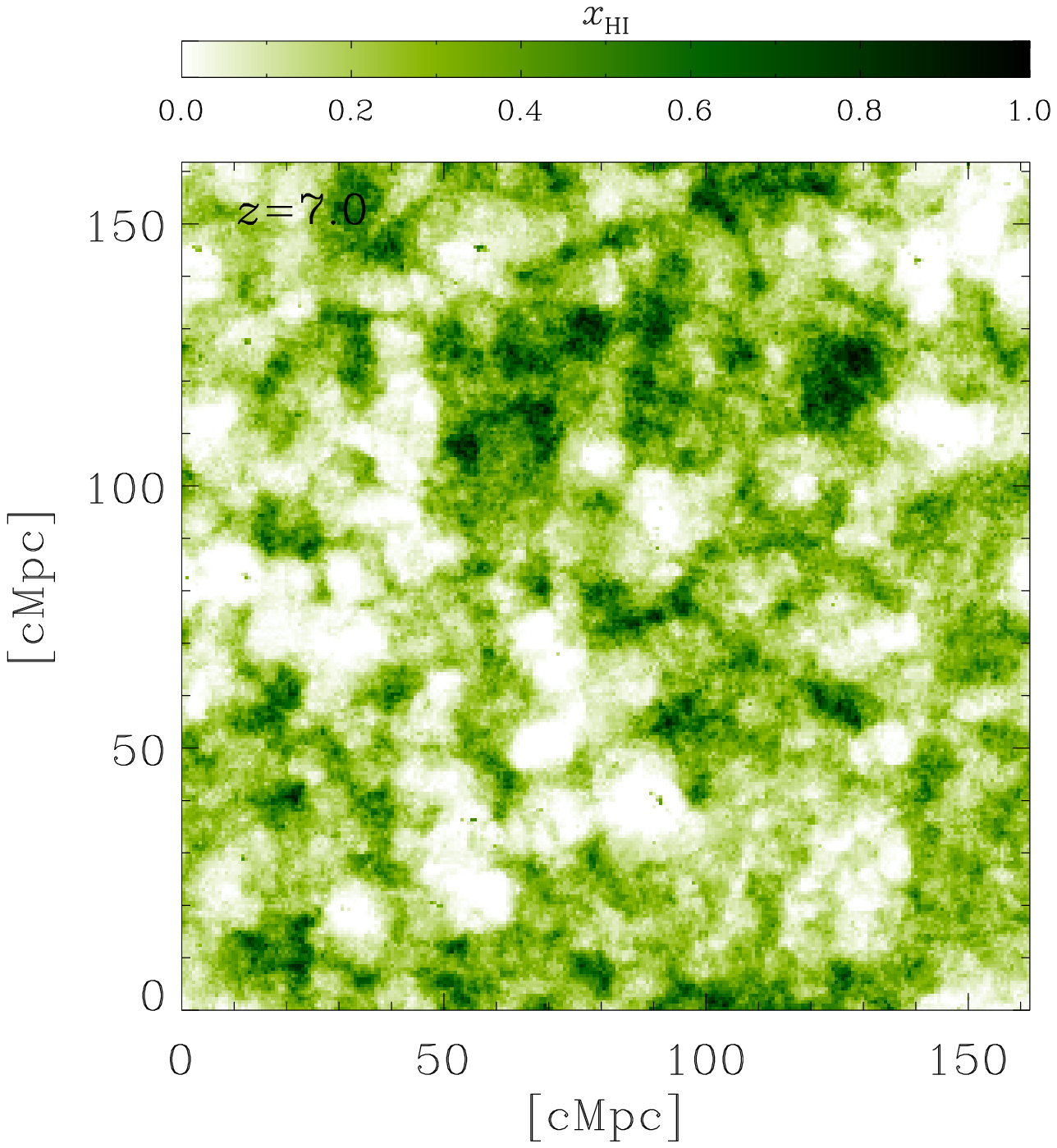}
  \includegraphics[width=6cm]{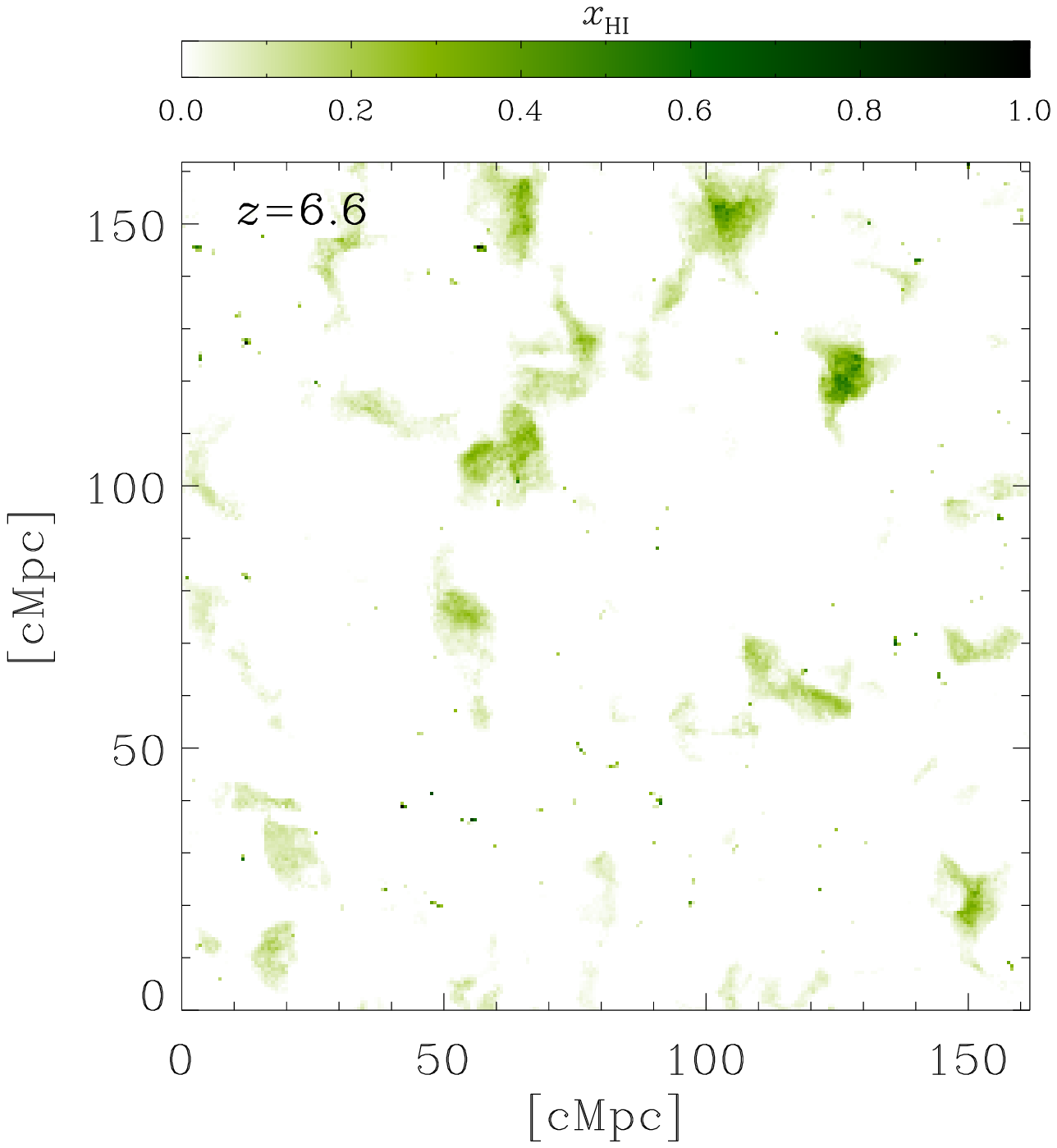}
  \includegraphics[width=6cm]{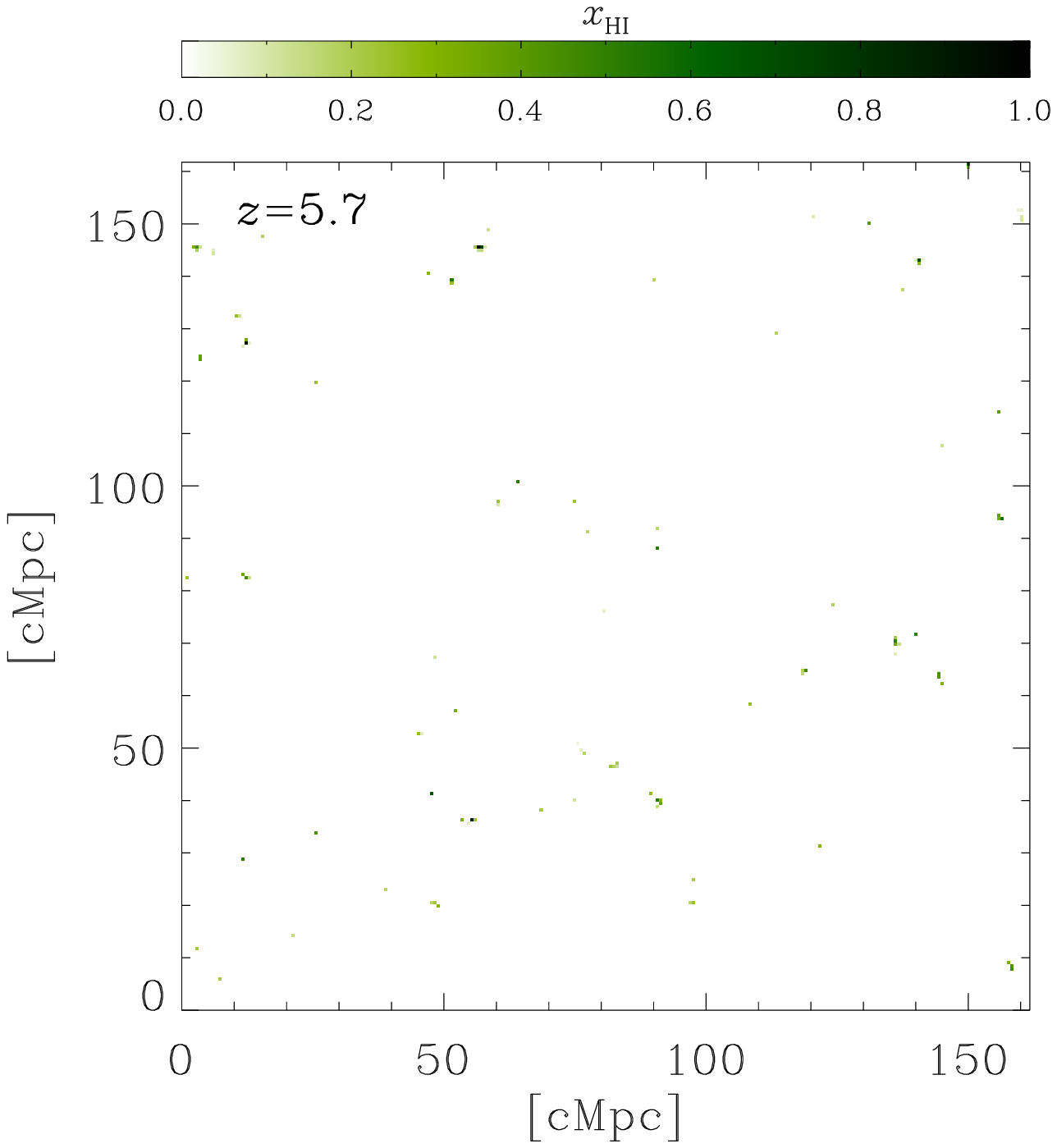}
 \end{center}
 \caption{The distribution of the neutral hydrogen fraction, $x_{\rm HI}$, 
 in the {\it mid} model. Each panel shows $x_{\rm HI}$ of mass-weighted 
 average over the line-of-sight sampled by narrowband observations. 
 (Top left) $z=7.3$ (NB101), (Top right) $z=7.0$ (NB973), 
 (Bottom left) $z=6.6$ (NB921), and 
 (Bottom right) $z=5.7$ (NB816).}\label{fig:xHImap}
\end{figure*}

The ionizing source model obtained from the RHD simulation 
(\S2.1 and $l_\nu$ in eq.~[3]) may 
have uncertainty especially in the escape fraction of ionizing photons,
because the escape fraction is known to depend on the spatial resolution
of numerical simulations (e.g., \cite{Wise14}; \cite{Kimm17};
\cite{Sumida2018}).  
Therefore, we also carried out two additional reionization
simulations, where the Lyman continuum spectra were changed to be a 1.5
times higher or lower photon production rates. Hereafter, we refer to
the high emissivity, fiducial, and low emissivity models as $early$,
$mid$, and $late$ reionization models, respectively. The simulated
reionization histories are shown in Fig.~\ref{fig:xHIhistory}. 
These three models fully agree with the latest Thomson scattering
optical depth measurement by the Planck satellite \citep{Planck2016b}: 
$\tau=0.0552$, $0.0591$, and $0.0648$ for the $mid$, $early$ and $late$
models, respectively, against the observation $0.058\pm0.012$. 
For the $mid$ model, H~{\sc i} distributions in the widths of some HSC
narrowband filters are found in Fig.~\ref{fig:xHImap}.

\section{LAE model}

The cosmological Ly$\alpha$ radiative transfer can be divided into three 
components: the production in a galaxy, the transfer in the galaxy
(including its ISM and CGM [or halo]) and the transfer in the IGM (e.g.,
\cite{mcquinn07,meiksin09,laursen11,Jensen2013,kakiichi16}). 
The observable flux of the Ly$\alpha$ line can then be expressed as 
\begin{equation}
 F_{\rm Ly\alpha} = 
  \frac{L_{\rm Ly\alpha}^{\rm int}f_{\rm esc,\alpha}^{\rm ISM}T_\alpha^{\rm IGM}}
  {4\pi d_{\rm L}^2}\,,
\end{equation}
where $L_{\rm Ly\alpha}^{\rm int}$ is the Ly$\alpha$ luminosity produced
in a galaxy, $f_{\rm esc,\alpha}^{\rm ISM}$ is the Ly$\alpha$ escape
fraction in the ISM (and halo) of the galaxy, $T_\alpha^{\rm IGM}$ is
the IGM transmission for Ly$\alpha$ photons, and $d_{\rm L}$ is the
luminosity distance toward the galaxy. In the following subsections 
(\S3.1, 3.3 and 3.4), we will describe how we model and calculate 
the three quantities related to Ly$\alpha$. In \S3.2, we will present an
Ly$\alpha$ line profile modeling.

Another important quantity, the source rest-frame equivalent width of 
the Ly$\alpha$ line, is defined as 
\begin{equation}
 EW_0 = \frac{L_{\rm Ly\alpha}^{\rm int}f_{\rm esc,\alpha}^{\rm ISM}T_\alpha^{\rm IGM}}
  {L_{\lambda_\alpha}^{\rm con}}
  = \frac{L_{\rm Ly\alpha}^{\rm int}f_{\rm esc,\alpha}^{\rm ISM}T_\alpha^{\rm IGM}}
  {L_{\lambda_{\rm UV}}^{\rm con}(\lambda_\alpha/\lambda_{\rm UV})^\beta}\,,
\end{equation}
where $L_{\lambda_\alpha}^{\rm con}$ is the continuum flux density at 
the Ly$\alpha$ wavelength $\lambda_\alpha$, $L_{\lambda_{\rm UV}}^{\rm con}$
is that at the UV wavelength $\lambda_{\rm UV}(\approx1500~{\rm \AA})$, and 
$\beta$ is the UV spectral slope ($L_\lambda\propto\lambda^\beta$). 
We will also describe modeling of the UV continuum in a later subsection (\S3.5).

\subsection{Ly$\alpha$ production in a galaxy}

\begin{figure}
 \begin{center}
  \includegraphics[width=5cm]{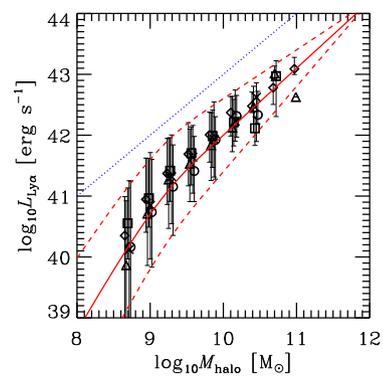} 
 \end{center}
 \caption{The total (recombination$+$collisional excitation)
 Ly$\alpha$ luminosity as a function of halo mass. The symbols and
 error-bars indicate the median and upper and lower 25-percentiles from 
 galaxies in the radiation hydrodynamics simulation by Hasegawa et al: 
 diamonds, triangles, squares, crosses, and circles for $z=5.7$, 6.0,
 6.6, 7.0, and 7.3, respectively. The solid and dashed lines show our
 simple recipe described by equations (6) and (7). The dotted line is 
 a maximum luminosity adopted here.}\label{fig:Lamodel}
\end{figure}

We make a simple recipe of the Ly$\alpha$ photon production rate as a
function of halo mass from the RHD galaxy formation simulation 
in \S2.1 by Hasegawa et al.~(in prep.) 
Fig.~\ref{fig:Lamodel} shows the total
(recombination$+$collisional excitation) luminosity of Ly$\alpha$ photons
of galaxies at $z\sim6$--7 in the simulation. The solid line on the
figure is our recipe described as 
\begin{equation}
  L_{\rm Ly\alpha,42}^{\rm int}={M_{\rm h,10}}^{1.1}
   (1-e^{-10M_{\rm h,10}})\times10^{\delta_{\rm Ly\alpha}}\,,
\end{equation}
where $L_{\rm Ly\alpha,42}^{\rm int}$ is the total intrinsic Ly$\alpha$
luminosity normalized by $10^{42}$ erg s$^{-1}$, $M_{\rm h,10}$ is the
halo mass  normalized by $10^{10}$ M$_\odot$, and 
$\delta_{\rm Ly\alpha}$ accounts for the fluctuation of the Ly$\alpha$
production. We draw a Gaussian random number with the mean of zero and
the standard deviation of 
$\sigma_{\rm Ly\alpha}$ for $\delta_{\rm Ly\alpha}$, where  
\begin{equation}
 \sigma_{\rm Ly\alpha} = 0.6 - 0.3 \log_{10} M_{\rm h,10}
\end{equation}
for $\log_{10}M_{\rm h,10}\leq2$, otherwise $\sigma_{\rm Ly\alpha}=0$.
The dashed lines in Fig.~\ref{fig:Lamodel} show the $\pm1\sigma$ range
around the fiducial equation~(6). The fluctuation in Ly$\alpha$
production may account for a different star formation history of each
halo. Jensen et al.~(2013, 2014) adopted a similar log-normal
fluctuation but their dispersion was 0.4-dex as a constant.


Since the dispersion described in equation~(7) is somewhat large, an
abnormally large Ly$\alpha$ luminosity may happen in case without any
limit. Therefore, we set an upper limit of the Ly$\alpha$ luminosity
indicated by the dotted line in Fig.~\ref{fig:Lamodel} as 
$L_{\rm Ly\alpha,42}^{\rm int}<10M_{\rm h,10}$. This choice is arbitrary but 
the RHD simulation results may suggest an upper limit around there.\footnote{
The distribution of the Ly$\alpha$ production rate at a certain halo
mass in the RHD simulation does not follow a Gaussian function but a
function with an upper truncation. In reality, a finite time-scale of
star-formation sets an upper limit on the star-formation rate in a halo,
and then, the Ly$\alpha$ production rate.}
In practice, when we obtained a Ly$\alpha$ luminosity over the limit 
in a Monte Carlo trial computing the fluctuation $\delta_{\rm Ly\alpha}$ 
and the luminosity $L_{\rm Ly\alpha}^{\rm int}$, we discard it and draw
another random number.

The simulated Ly$\alpha$ luminosity shown in Fig.~\ref{fig:Lamodel}, 
i.e. our recipe described in equation~(6), already includes the 
effect of the ionizing photon escape which reduces the Ly$\alpha$ 
photon production rate. The exponential part in equation~(6) accounts 
for this effect in lower halo mass where the ionizing photon escape 
becomes more significant (typically $>10$\% for $M_{\rm h}<10^9$
M$_\odot$; Hasegawa et al.~in prep.). 
However, the halo mass of the observed LAEs is larger than, say,
$10^{10.5}$ M$_\odot$ as found in \S5.1 and the ionizing photon
escape is not very large for these halos (typically $\sim1$\% or less).

\subsection{Ly$\alpha$ line profile through the ISM}
 
\begin{figure}
 \begin{center}
  \includegraphics[width=5cm]{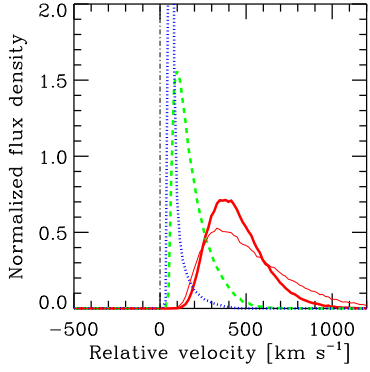} 
 \end{center}
 \caption{Examples of Ly$\alpha$ line profiles through spherical
 uniform outflowing gas. The solid (red), dashed (green), and
 dotted (blue) lines show the
 cases with $\log_{10}(N_{\rm HI}/{\rm cm^{-2}})=20$, 19, and 18,
 respectively. The thick lines are the cases with an outflowing velocity
 of $V_{\rm out}=150~\rm km~s^{-1}$ but the thin line is the case of 
 $V_{\rm out}=300~\rm km~s^{-1}$. The vertical axis is normalized so as
 that the integrated flux becomes unity.}\label{fig:lineprofile}
\end{figure}

The Ly$\alpha$ line profile is modulated by multiple resonant
scatterings during the transfer in the ISM (e.g., 
\cite{Verhamme2008,Yajima2012,Yajima2015}). Depending on the line
profile, the IGM transmission is also changed (e.g.,
\cite{Yajima2017}). When the gas in the ISM is outflowing, the line
profile shows an asymmetric shape with a red wing (e.g.,
\cite{Dijkstra2006}), resulting in a higher transmission through the IGM
in the Hubble flow.  
Recent observations indicated that a large fraction of high-redshift
LAEs have outflowing gas (e.g., \cite{Ouchi2010,Yamada2012,Shibuya2014}).
In this work, for the simplicity, we consider line profiles through
spherically uniform outflowing gas with the velocity structure of 
\begin{equation}
 V(r) = V_{\rm out} \left( \frac{r}{R_{\rm edge}}\right)\,,
\end{equation}
where $R_{\rm edge}$ is the radius of the gas distribution and 
$V_{\rm out}$ is the outflow velocity at $R_{\rm edge}$. We set 
$R_{\rm egde}=10$ kpc, while the line profile does not depend on 
$R_{\rm edge}$ but only on the H~{\sc i} column density of the gas 
distribution and $V_{\rm out}$. 
We calculated the transfer of $10^{5}$ photon packets in 100 expanding
spherical shells by using a Monte Carlo method 
\citep{Yajima2012,yajima14,Yajima2017}. Fig.~\ref{fig:lineprofile}
shows examples of line profiles. We find a significant H~{\sc i}
column density dependence but the $V_{\rm out}$ dependence is small.
In the following sections, we consider the cases with 
$\log_{10}(N_{\rm HI}/{\rm cm^{-2}})=18$, 19, or 20 and a fixed 
$V_{\rm out}=150~\rm km~s^{-1}$ which is typically observed in LAEs 
(e.g., \cite{Hashimoto2013})

\subsection{Ly$\alpha$ escape fraction from a halo}

The transfer of Ly$\alpha$ photons in a galaxy is a complex process
because of resonant scattering by neutral hydrogen. Numerical
simulations show a large dispersion of the Ly$\alpha$ escape fraction
for fixed galaxy properties (e.g., \cite{yajima14}). We try to model
this stochasticity, assuming a simple Gaussian function where both of
the mean and the dispersion are equal to $\langle\tau_\alpha\rangle$: 
\begin{equation}
 P(\tau_\alpha) = \frac{\exp\{-(\tau_\alpha-\langle\tau_\alpha\rangle)^2
  /2\langle\tau_\alpha\rangle\}}
  {\sqrt{2\pi\langle\tau_\alpha\rangle}}\,.
\end{equation}
A possible explanation for this treatment is found in Appendix~1.

The mean Ly$\alpha$ optical depth, $\langle\tau_\alpha\rangle$, 
for a halo with the mass of $M_{\rm h}$ can be described as 
\begin{equation}
 \langle\tau_\alpha\rangle = \tau_{\alpha,10} {M_{\rm h,10}}^p\,,
\end{equation}
where the pivot value, $\tau_{\alpha,10}$, at the halo mass of 
$10^{10}$ M$_\odot$ is a model parameter and the index $p$ describes 
the halo mass dependence of $\langle\tau_\alpha\rangle$. 
Here we consider two cases of $p=0$ (no halo mass dependence) and 
$p=1/3$ corresponding to a case of 
$\langle\tau_\alpha\rangle\propto M_{\rm h}/R_{\rm vir}^2$ 
where the right-hand-side means a column density and $R_{\rm vir}$ is
the virial radius of a halo ($\propto M_{\rm h}^{1/3}$). 
A similar halo mass dependence of the $p=1/3$ case is observed in
numerical simulations \citep{yajima14}.
Finally, the escape fraction of Ly$\alpha$ photons is given by 
\begin{equation}
 f_{\rm esc,\alpha}^{\rm ISM}=e^{-\tau_\alpha}\,,
\end{equation}
with $\tau_\alpha$ drawn from the probability distribution of
equation~(9).

As found in the subsequent sections (see table~1), 
$\langle\tau_\alpha\rangle\sim1$--5. For such a small 
$\langle\tau_\alpha\rangle$, we easily obtain an unphysical negative
$\tau_\alpha$ under the Gaussian probability distribution of
equation~(9). We just discard it and redraw another $\tau_\alpha$. This
treatment gives modulation to the Gaussian function of equation~(9).


\subsection{Ly$\alpha$ transmission through the IGM}

The IGM transmission of the Ly$\alpha$ line is defined as 
\begin{equation}
 T_\alpha^{\rm IGM} = 
  \frac{\int F_{\alpha,\lambda_e}~e^{-\tau_{\lambda_e}^{\rm IGM}} d\lambda_e}
  {\int F_{\alpha,\lambda_e} d\lambda_e}\,,
\end{equation}
where $F_{\alpha,\lambda_e}$ is the Ly$\alpha$ line profile as a
function of the wavelength, $\lambda_e$, in the source rest-frame and 
$\tau_{\lambda_e}^{\rm IGM}$ is the IGM optical depth at $\lambda_e$. 
The line profile is one observable at the virial radius of the halo of
interest described in \S3.2.

The IGM optical depth against Ly$\alpha$ photons along a physical length
coordinate, $l_{\rm p}$, defined from an object to an observer is 
\begin{equation}
 \tau_{\lambda_e}^{\rm IGM} = \int_{R_{\rm vir}}^{l_{\rm p,max}} 
  s_\alpha(\lambda,T) n_{\rm HI} dl_{\rm p}\,,
\end{equation}
where $s_\alpha(\lambda,T)$ is the neutral hydrogen cross section for 
Ly$\alpha$ photons, $\lambda$ is the wavelength in the gas rest-frame,
$T$ is the temperature of gas, and $n_{\rm HI}$ is the number density of
neutral hydrogen. The gas rest-frame wavelength $\lambda$ is 
$\lambda=\lambda_e/(1-v_{\rm gas}/c)$ and $v_{\rm gas}=H(z)l_{\rm p}$,
where $H(z)$ is the Hubble parameter at redshift $z$. We have neglected  
any peculiar motion of gas. We fix $H(z)$ in a simulation box with its 
redshift $z$. We define the IGM to be gas out of halos in this paper. 
Then, the integral starts at the virial radius, $R_{\rm vir}$, of the
halo of interest. The upper limit of the integral, $l_{\rm p,max}$, is
arbitrary but several tens of comoving Mpc is sufficient for
convergence.

The Ly$\alpha$ cross section of neutral hydrogen is given by 
\begin{equation}
 s_\alpha(\lambda,T) = 1.041\times10^{-13}~[{\rm cm^2}] 
  \left(\frac{T}{10^4~{\rm K}}\right)^{-1/2}
  \frac{H(a,x)}{\sqrt{\pi}}\,,
\end{equation}
where $H(a,x)$ is the Voigt function with being $\nu=c/\lambda$, 
$x=(\nu_\alpha-\nu)/\nu_{\rm D}$ and $a=\nu_{\rm L}/2\nu_{\rm D}$. 
The light speed in vacuum is $c=2.9979\times10^{10}$ cm s$^{-1}$, the 
Ly$\alpha$ frequency $\nu_\alpha=2.466\times10^{15}$ Hz and the natural 
broadening width in frequency $\nu_{\rm L}=9.936\times10^7$ Hz 
corresponding to the Einstein coefficient of $A_{21}=6.265\times10^8$ Hz. 
The thermal Doppler width in frequency $\nu_{\rm D}=\nu_\alpha v_{\rm th}/c$, 
where $v_{\rm th}=\sqrt{2k_{\rm B}T/m_{\rm p}}$ with the Boltzmann constant 
$k_{\rm B}=1.381\times10^{-16}$ erg K$^{-1}$ and the proton mass 
$m_{\rm p}=1.673\times10^{-24}$ g.
The Voigt function is calculated by using the analytic approximation of
\citet{Teppergarcia2006}.

For each simulation box, we set 6 directions ($\pm x$, $\pm y$, $\pm z$) 
for lines-of-sight and calculate $T_\alpha^{\rm IGM}$ of each halo. 
We have confirmed as a sanity check that the 6 directions' transmissions
averaged over a number of halos agree with each other and the difference
is at most $\sim0.1$\%.

\subsection{UV continuum}

We assume a simple linear relation between UV luminosity and halo mass, 
which is consistent with the prediction of a cosmological hydrodynamical
simulation by \citet{shimizu14}\footnote{In \citet{shimizu14}, they have
derived a power-law relation of $L_{\rm UV}\propto M_{\rm h}^{0.9}$ in
their eq.~(9) from a fit for all the simulated galaxies, whereas the
linear relation and dispersion in this paper are obtained from a fit for
their data binned along $M_{\rm UV}$ shown in their Fig.~5.}:  
\begin{equation}
 M_{\rm UV} = -17.2 - 2.5\log_{10}(M_{\rm h,10}) + \delta_{\rm UV}\,,
\end{equation}
where $\delta_{\rm UV}$ brings fluctuation in the UV magnitude. We draw
a Gaussian random number with the mean of zero and the standard
deviation of $\sigma_{\rm UV}$ for $\delta_{\rm UV}$, where  
\begin{equation}
 \sigma_{\rm UV} = 0.4 - 0.2 \log_{10} M_{\rm h,10}
\end{equation}
for $\log_{10}M_{\rm h,10}\leq2$, otherwise $\sigma_{\rm UV}=0$.
This is also obtained from the \citet{shimizu14} simulation results.
We here note that the fluctuation of the UV magnitude and that of the 
Ly$\alpha$ production rate are not correlated in our modeling, 
while they are probably correlated in reality. Fortunately, we will 
find that the UV magnitude fluctuation is small and has a relatively 
small impact and that the fluctuation of the Ly$\alpha$ production is 
less important than that of the Ly$\alpha$ optical depth.
Therefore, this point does not affect the conclusions of this paper.

\begin{figure}
 \begin{center}
  \includegraphics[width=7cm]{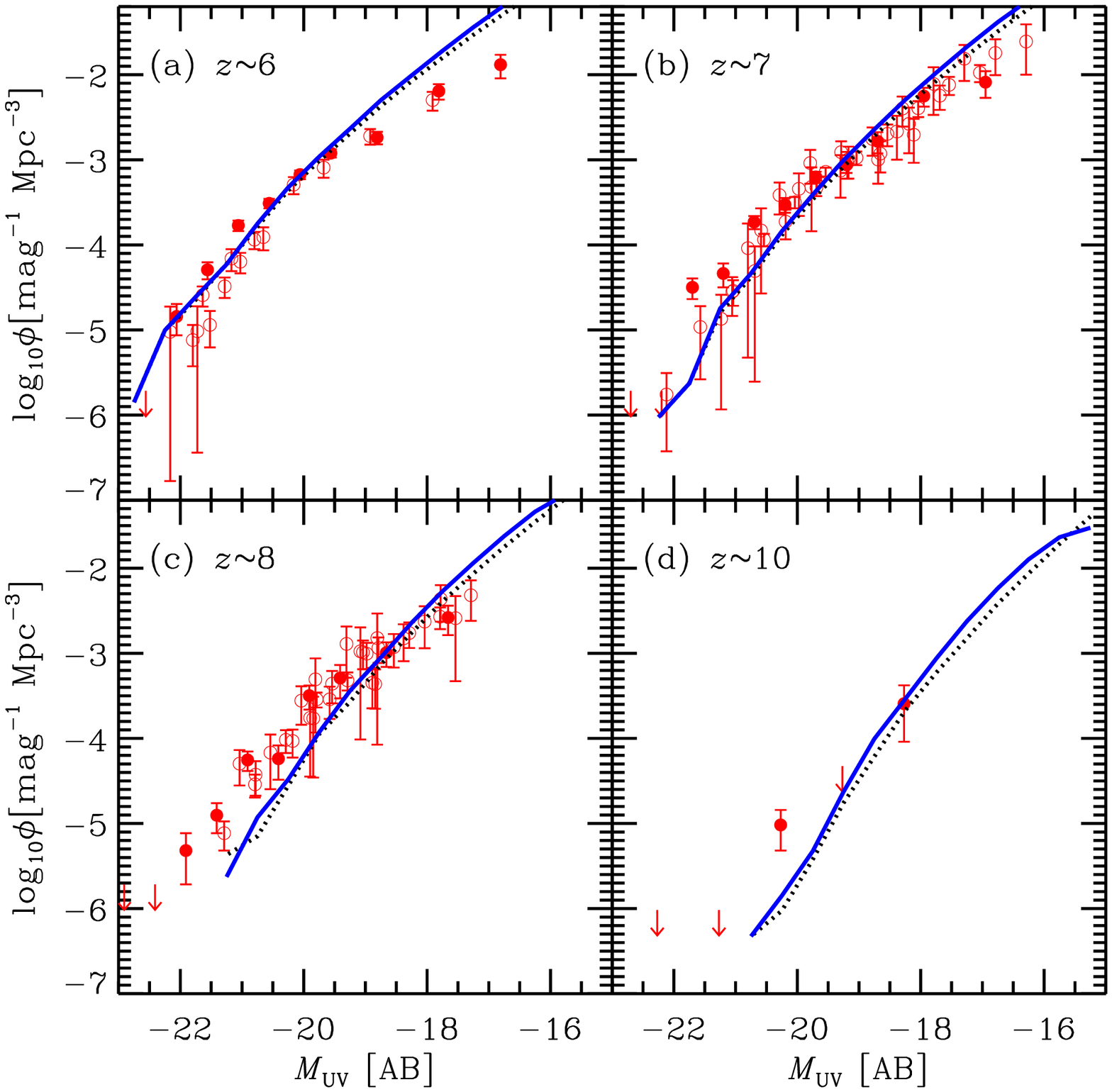} 
 \end{center}
 \caption{UV luminosity functions at (a) $z\sim6$, (b) $z\sim7$, (c)
 $z\sim8$, and (d) $z\sim10$. The filled data points are taken from
 \citet{bouwens15} and open data points are a compilation of 
 \citet{shimasaku05}, \citet{mclure09}, \citet{bradley12}, 
 \citet{oesch12}, \citet{mclure13}, \citet{oesch13}, \citet{Schenker2013},
 \citet{atek15}, and references therein. 
 The solid/dotted lines are the model predictions with/without
 dispersion described in eq.~(16).}\label{fig:UVLF}
\end{figure}

Fig.~\ref{fig:UVLF} shows a comparison with observations in UV LFs at
$z\sim6$, 7, 8, and 10. We find a reasonable agreement although there
are overpredictions at the faint-end, especially at $z\sim6$, and
underpredictions at the bright-end at $z>8$. This is probably caused by
our simple linear assumption in the halo mass and UV magnitude relation
(eq.~15). Indeed, a non-linear relation between halo mass and UV
magnitude is observed in LBGs at $4<z<6$
\citep{Harikane2016,Harikane2017}. Given the reasonable agreement of our
simple model and UV LFs at $z\sim6$ and 7 in the range of 
$-22<M_{\rm UV}<-18$, the non-linearity effect would be small in our
study.

We also assume an empirical relation between $M_{\rm UV}$ and the UV
spectral slope $\beta$, where the UV flux density per unit wavelength
interval $F_\lambda\propto\lambda^\beta$, reported by \citet{bouwens14}:
\begin{equation}
 \beta = -2.05 - 0.20(M_{\rm UV}+19.5) + \delta_\beta\,,
\end{equation}
where $\delta_\beta$ again accounts for fluctuation in $\beta$. This may
be caused by differences of the star formation history and the dust
content of each halo. We draw a Gaussian random number with the mean of
zero and the standard deviation of $\sigma_\beta=0.1$ which is obtained
from the simulation of \citet{shimizu14} and consistent with the random
errors estimated in \citet{bouwens14}. We do not account for the
systematic errors reported in \citet{bouwens14} to avoid an abnormally 
large dispersion in $\beta$ at low luminosity and it should be
reasonable because the $\beta$ value in equation~(17) agrees with the
prediction by \citet{shimizu14}.

\subsection{Virtual observations}

To make the models as realistic as possible, we have made virtual
observations of the simulation box. First we randomly select the
observation direction among six directions along the axes (i.e. $\pm x$,
$\pm y$ and $\pm z$) and select a random point in the box as the point
(0,0,$d_{\rm C}$), where $d_{\rm C}$ is the radial comoving distance
toward the redshift of the simulation output. Then we obtain redshifts
of all halos from their coordinates along the selected observation
direction and map the halos into the celestial sphere by using the other
two coordinates and the redshift of them. We then calculate the model
observed magnitudes of the halos through Subaru HSC broadband and
narrowband filters. Photometric errors are also taken into account 
according to the limiting magnitudes of observations by 
\citet{Konno2017} for $z=5.7$ and $6.6$ and by \citet{Konno2014} for 
$z=7.3$. Using these predicted apparent magnitudes, we can select mock 
LAEs by the same magnitude and color criteria as the real observations 
(e.g., \cite{Shibuya2017}).
Note that unlike the real observations, there is 
no contamination of low-$z$ objects in the mock LAEs because we do not 
make a light-cone from a series of the simulation outputs but treat 
only a single simulation output for each redshift of interest in the 
virtual observations (but we have considered the redshift difference 
within the simulation box). Our simulation box has a $1$~deg$^2$ area 
for $z\gtrsim6$, while the early SILVERRUSH data have 14~deg$^2$ for 
$z=5.7$ LAEs and 21~deg$^2$ for $z=6.6$ LAEs \citep{Shibuya2017}. 
Since the NB widths correspond to about 1/4 of the simulation box size 
in the comoving scale, we may obtain at least 4 different slices of the 
NB survey from a simulation output. The $x,y,z$-axes can produce 
different slices, resulting in 12 sets. On the other hand, the 
stochastic processes in the LAE modeling described in the previous 
subsections produce mock LAEs from different halos in different trials. 
This may allow us to have a much larger number of different sets of 
mock LAEs from a single simulation output (i.e. a single distribution of 
halos). We then repeat to produce mock LAE catalogs 100 times or 
up to 500 times, respectively, for LFs and LAE fractions or for ACFs. 
This may be equivalent to a set of observations of 100--500 patches of 
1~deg$^2$ area in the sky.

\subsection{Models and parameter calibration}

\begin{figure}
 \begin{center}
  \includegraphics[width=5cm]{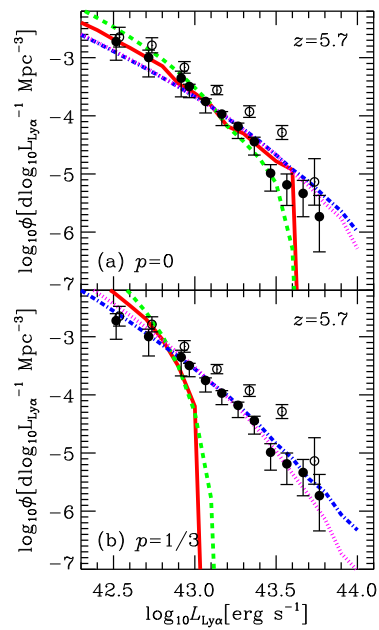}
 \end{center}
 \caption{Results of the parameter calibration with the Ly$\alpha$
 luminosity function at $z=5.7$. The filled and open circles with
 error-bars are Konno et al.~(2018) with HSC and Santos et al.~(2016),
 respectively. (a) The models without halo mass dependence on the
 Ly$\alpha$ optical depth in the ISM ($p=0$ in eq.10).
 The solid (red), dashed (green), dot-dashed (blue), and dotted
 (magenta) lines are, respectively, the models of A, B, C,
 and D. (b) The models with the halo mass dependence ($p=1/3$ in eq.10). 
 The solid (red), dashed (green), dot-dashed (blue), and dotted
 (magenta) lines are, respectively, the models of E, F, G, and H. The
 {\it mid} reionization model is adopted but the IGM neutral hydrogen
 fraction is sufficiently low at $z=5.7$ for any reionization models in
 this paper.}\label{fig:LaLFz5p7}
\end{figure}

In this paper, we consider $2^3$ models depending on the presence of
dispersion of the Ly$\alpha$ production in a halo (eq.~7), stochasticity
of the Ly$\alpha$ optical depth in the halo (eq.~9), and halo mass
dependence of the Ly$\alpha$ optical depth (eq.~10). These 8 models are 
named as A to H and listed in Table~\ref{tab:models}. There is only a
single free parameter in these models: $\tau_{\alpha,10}$, a pivot
value of the ISM Ly$\alpha$ optical depth at a halo of
$10^{10}$~M$_\odot$ (see eq.~10). We determine the value for each
model so as to reproduce the observed Ly$\alpha$ LF at $z=5.7$. Since
the IGM is highly ionized and has only a minimum effect on the LFs at
this redshift, the reionization history is not matter and we adopt 
the $mid$ model for the calibration. We use the early data of SILVERRUSH
\citep{Konno2017} as the reference observations. As we discuss in 
Appendix~2, their LFs are corrected for incompleteness and the filter 
transmission effects and then considered to be the best estimates of the
true LFs. Therefore, we compare theirs with the true LFs in our
simulation box. Fig.~\ref{fig:LaLFz5p7} shows the results of the
parameter calibration from this comparison. In fact, models E and F can
not reproduce the observed LF at all. The obtained pivot values are
listed in Table~\ref{tab:models}.

\citet{Santos2016} (and also \cite{Matthee2015}) presented significantly
higher LFs than those of \citet{Konno2017} and other previous ones
obtained with Subaru/Suprime-Cam \citep{Shimasaku2006,Ouchi2008}. Since
the reason of this discrepancy is unclear, we also calibrated
$\tau_{\alpha,10}$ with the LF at $z=5.7$ by \citet{Santos2016}. However, 
with this calibration, no model could reproduce the LAE fraction (\S4.3), 
may suggesting a possible overestimation in their LFs.

\begin{table*}
 \tbl{List of the LAE models.}{%
 \begin{tabular}{llllllll}
  \hline
  Model & $p$ & $\sigma_{\rm Ly\alpha}$ & $\sigma_{\tau_\alpha}$ 
	& $\tau_{\alpha,10}$ & Ly$\alpha$ LF & ACF & LAE fraction \\ 
  \hline
  A & 0 & --- & --- & 1.6 & $\bigtriangleup$ & $\bigtriangleup$ & $\times$ \\
  B & 0 & eq.(7) &--- & 1.9 & $\bigtriangleup$ & $\bigtriangleup$ & $\bigtriangleup$ \\
  C & 0 & --- & eq.(9) & 3.1 & $\bigcirc$ & $\bigcirc$ & $\times$ \\
  D & 0 & eq.(7) & eq.(9) & 4.6 & $\bigcirc$ & $\bigcirc$ & $\times$ \\
  E & 1/3 & --- & --- & 0.6 & $\times$ & $\bigtriangleup$ & $\times$ \\
  F & 1/3 & eq.(7) & --- & 1.0 & $\times$ & $\times$ & $\times$ \\
  G & 1/3 & --- & eq.(9) & 1.1 & $\bigcirc$ & $\bigcirc$ & $\bigcirc$ \\
  H & 1/3 & eq.(7) & eq.(9) & 2.4 & $\bigcirc$ & $\bigcirc$ & $\times$ \\
  \hline
 \end{tabular}}\label{tab:models}
 \begin{tabnote}
  Notation is as follows: $\bigcirc$ good, $\bigtriangleup$ marginal, 
  and $\times$ bad.
 \end{tabnote}
\end{table*}

\subsubsection{Models without halo mass dependence of the Ly$\alpha$
   optical depth in the halo ($p=0$)}

In the following, we present brief comments on each models.
The first four models (A--D) have no halo mass dependence of the
Ly$\alpha$ optical depth, namely $p=0$ in equation~(10).

\paragraph{Model A}

This case has neither dispersion of the Ly$\alpha$ production nor that
of the halo Ly$\alpha$ optical depth. Thus, this model is the simplest
case with a constant Ly$\alpha$ luminosity per halo mass and 
a constant Ly$\alpha$ escape fraction for all halos, which
is often found in the literature (e.g., 
\cite{mcquinn07,Iliev2008,Jensen2013,Jensen2014,kakiichi16,Kubota2017,Yoshiura2017}).
This model is shown as the solid line in Fig.~\ref{fig:LaLFz5p7} (a). 
The adopted constant escape fraction of Ly$\alpha$ photons is 0.20. 
The observed LFs are reproduced reasonably well although a sudden dearth
of the model LAEs at $>10^{43.6}$ erg s$^{-1}$ apparently seems
inconsistent with the data.

\paragraph{Model B}

This model has dispersion of the Ly$\alpha$ production but not that of
the halo Ly$\alpha$ optical depth. This case also assumes a constant
Ly$\alpha$ escape fraction for all halos.
This model is shown as the dashed line in Fig.~\ref{fig:LaLFz5p7}~(a).
Relative to Model~A, Model~B predicts a steeper LF. This is due to 
the dispersion of the Ly$\alpha$ production and the upper limit on the 
Ly$\alpha$ emissivity (see Fig.~\ref{fig:Lamodel}). In fact, the 
dispersion itself makes the LF shallower than Model~A (see also
\cite{Jensen2014}), especially at the faint-end $<10^{42.5}$ erg
s$^{-1}$ in our case. On the other hand, the luminosity upper limit in
the current model setup makes the distribution of the Ly$\alpha$
luminosity asymmetric with a cut-off in the luminous side. This causes
the steeper slope in the luminous-end than Model~A. The constant
Ly$\alpha$ escape fraction is 0.15 smaller than that of Model~A
in order to fit the observed LF.

\paragraph{Model C}

This case has no dispersion of the Ly$\alpha$ production but does have
dispersion of the halo Ly$\alpha$ optical depth. Therefore, a stochastic
Ly$\alpha$ escape from a halo is realized. This model is shown as
the dot-dashed line in Fig.~\ref{fig:LaLFz5p7}~(a). The LF becomes
shallower than Model~A because of the dispersion of the halo Ly$\alpha$
optical depth. There is a more chance to have a large escape
fraction than a small one because of asymmetry of an exponential
function of the escape fraction in eq.~(11) even under a symmetric
Gaussian probability distribution of the optical depth described in
\S3.3, leading to more LAEs in luminous bins. To fit the observed LF, an
average Ly$\alpha$ escape fraction is as small as 0.045.

\paragraph{Model D}

This model has both dispersions of the Ly$\alpha$ production and the
halo Ly$\alpha$ optical depth. This case is shown as the dotted line in
Fig.~\ref{fig:LaLFz5p7}~(a) and is very similar to Model~C with 
slightly less LAEs in luminous bins caused by the upper limit of the
Ly$\alpha$ production. The smallest average Ly$\alpha$ escape fraction
of 0.010 is required to fit the observed LF.

\subsubsection{Models with halo mass dependence of the Ly$\alpha$
   optical depth in the halo ($p=1/3$)}

The rest four models (E--H) have a halo mass dependence of the halo
Ly$\alpha$ optical depth, that is $p=1/3$ in equation~(10). This dependence comes
from the assumption that the optical depth is proportional to the
matter column density in a halo (see a discussion below eq.~10). 
In these cases, the values of the pivot optical depth at a halo of 
$10^{10}$ M$_\odot$ tend to be smaller than those in the models with 
$p=0$. This is natural because a more massive halo has a larger optical 
depth. An average optical depth over all halos would be equivalent.

\paragraph{Model E}

This model has neither of the dispersions like Model~A. The LF is shown
as the solid line in Fig.~\ref{fig:LaLFz5p7}~(b). Due to higher
Ly$\alpha$ optical depths in more massive halos, the LF becomes too 
steep and totally inconsistent with the data.

\paragraph{Model F}

This model has only the dispersion in the Ly$\alpha$ production like
Model~B. The LF is shown as the dashed line in
Fig.~\ref{fig:LaLFz5p7}~(b). Due to the dispersion of the Ly$\alpha$
production with an upper limit, the LF becomes further steeper than
Model~E and again totally inconsistent with the data.

\paragraph{Model G}

This case has only the dispersion in the halo Ly$\alpha$ optical depth
like Model~C. The LF is shown as the dot-dashed line in 
Fig.~\ref{fig:LaLFz5p7}~(b) and becomes much shallower than Models~E
and F due to the dispersion of the Ly$\alpha$ optical depth. As a
result, the LF is reasonably consistent with the data.

\paragraph{Model H}

This case has both of the dispersions like Model~D. The LF is shown as
the dotted line in Fig.~\ref{fig:LaLFz5p7}~(b) and is similar to
Model~G but slightly steeper due to the dispersion of the Ly$\alpha$
production with an upper limit.

\section{Results}

We are ready to compare our $2^3$ LAE models with the observations, 
in particular, the early results of SILVERRUSH 
(\cite{Ouchi2017,Konno2017}). Comparing the models to the LAE luminosity
functions (LFs; \cite{Konno2017}), the angular correlation functions
(ACFs; \cite{Ouchi2017}), and the LAE fraction as a function of the UV
magnitude (\cite{Ono2012}), we will select the best model in this paper.

\subsection{Ly$\alpha$ luminosity function}

\begin{figure*}
 \begin{center}
  \includegraphics[width=4cm]{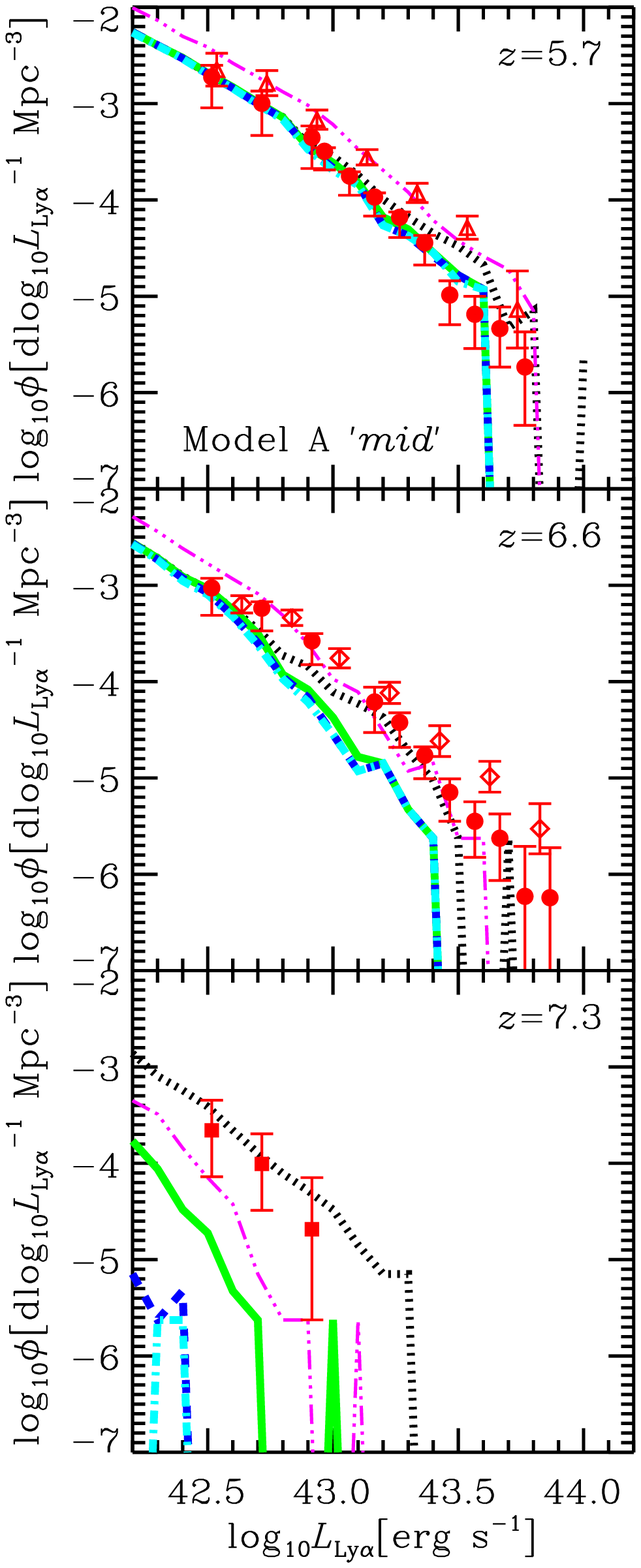}
  \includegraphics[width=4cm]{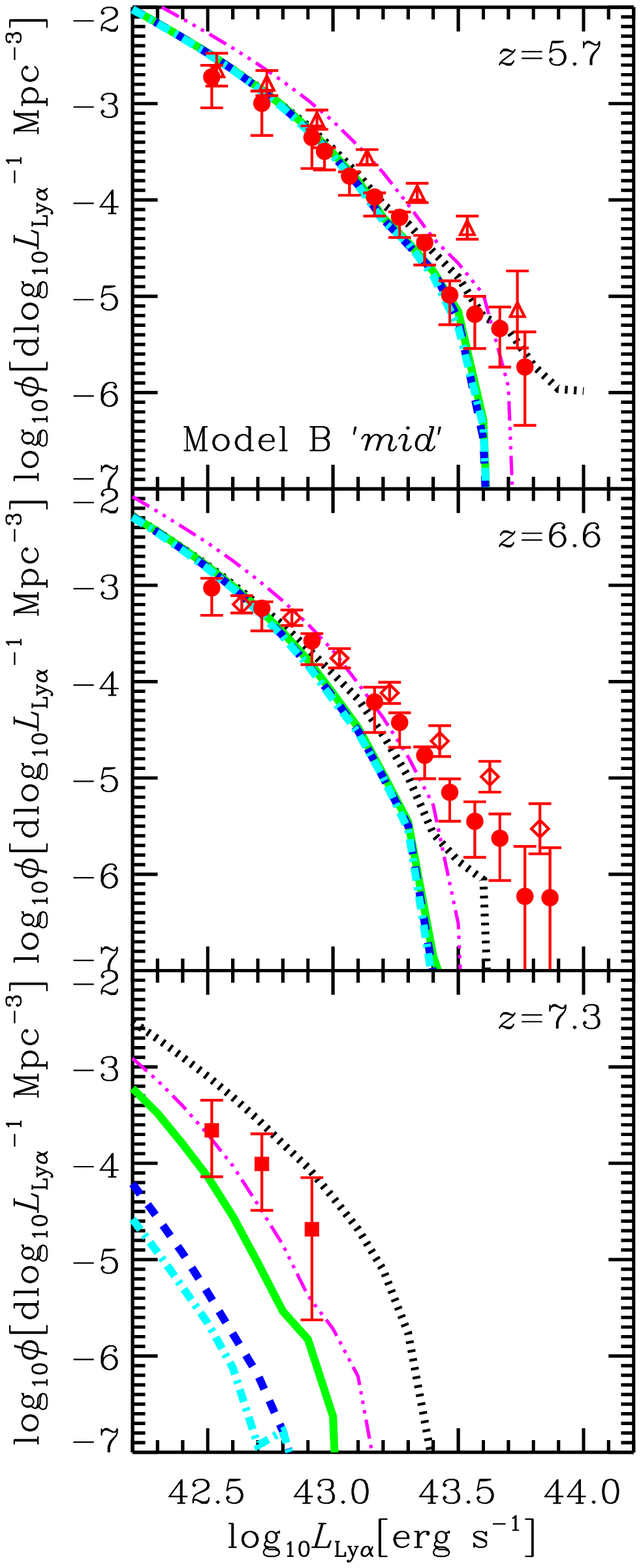}
  \includegraphics[width=4cm]{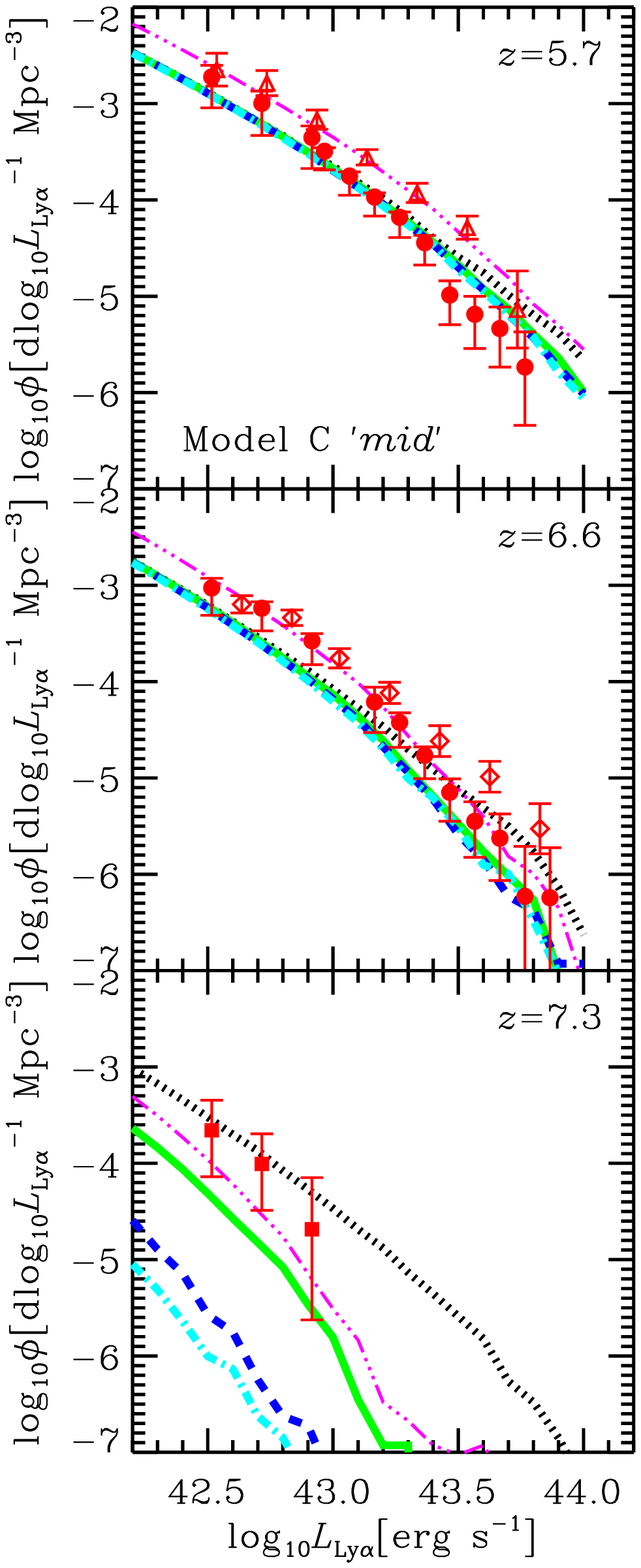}
  \includegraphics[width=4cm]{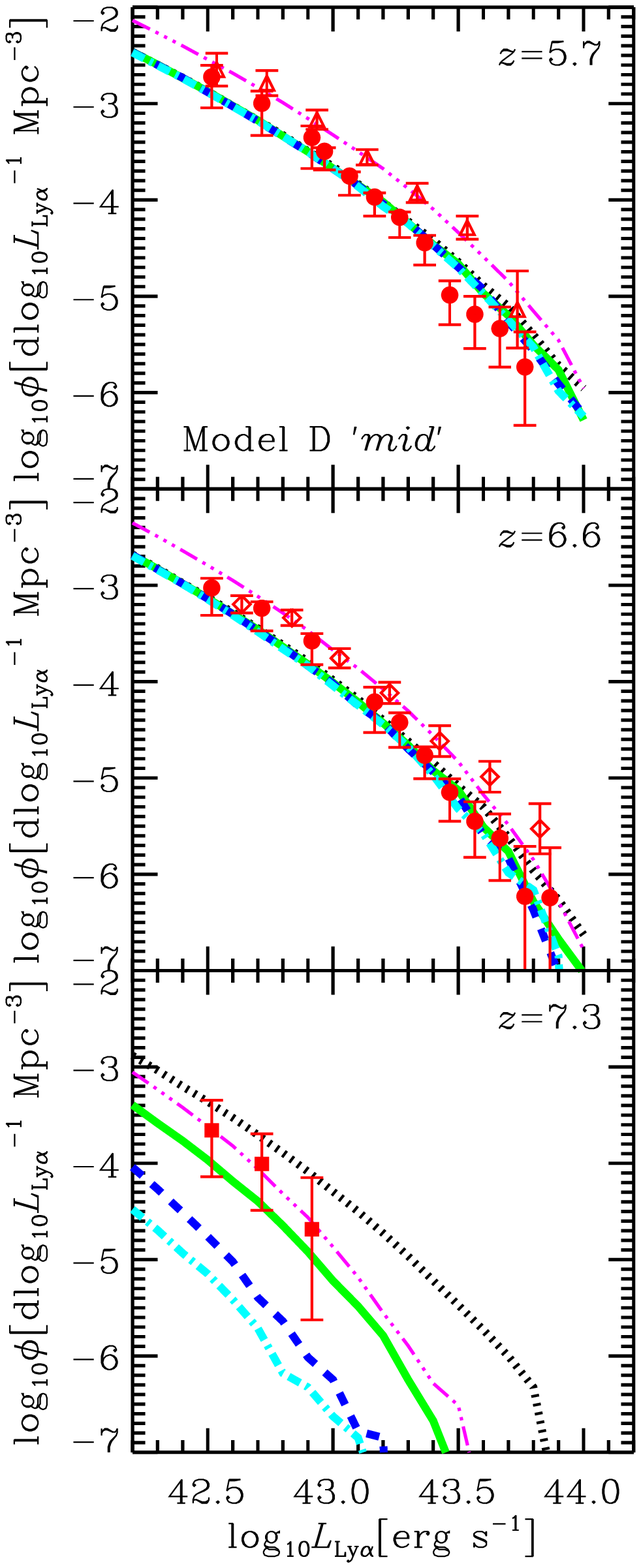}
  \includegraphics[width=4cm]{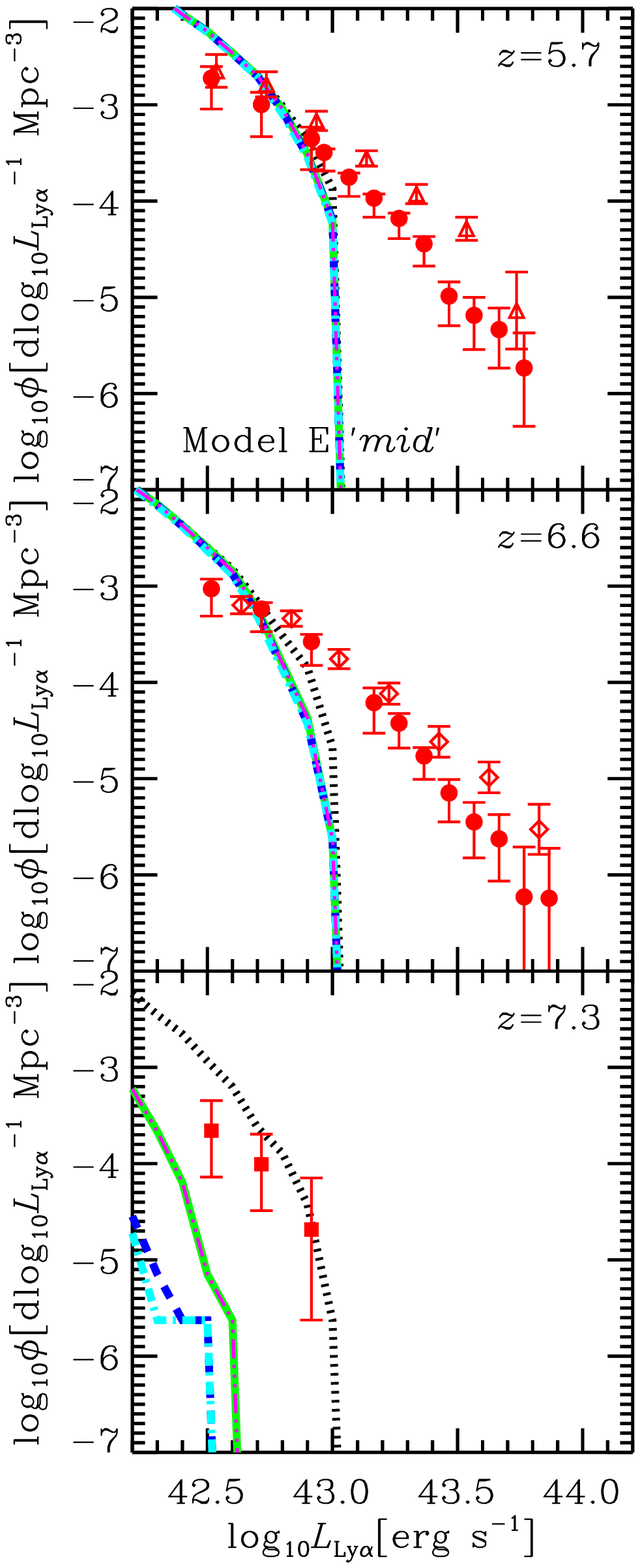}
  \includegraphics[width=4cm]{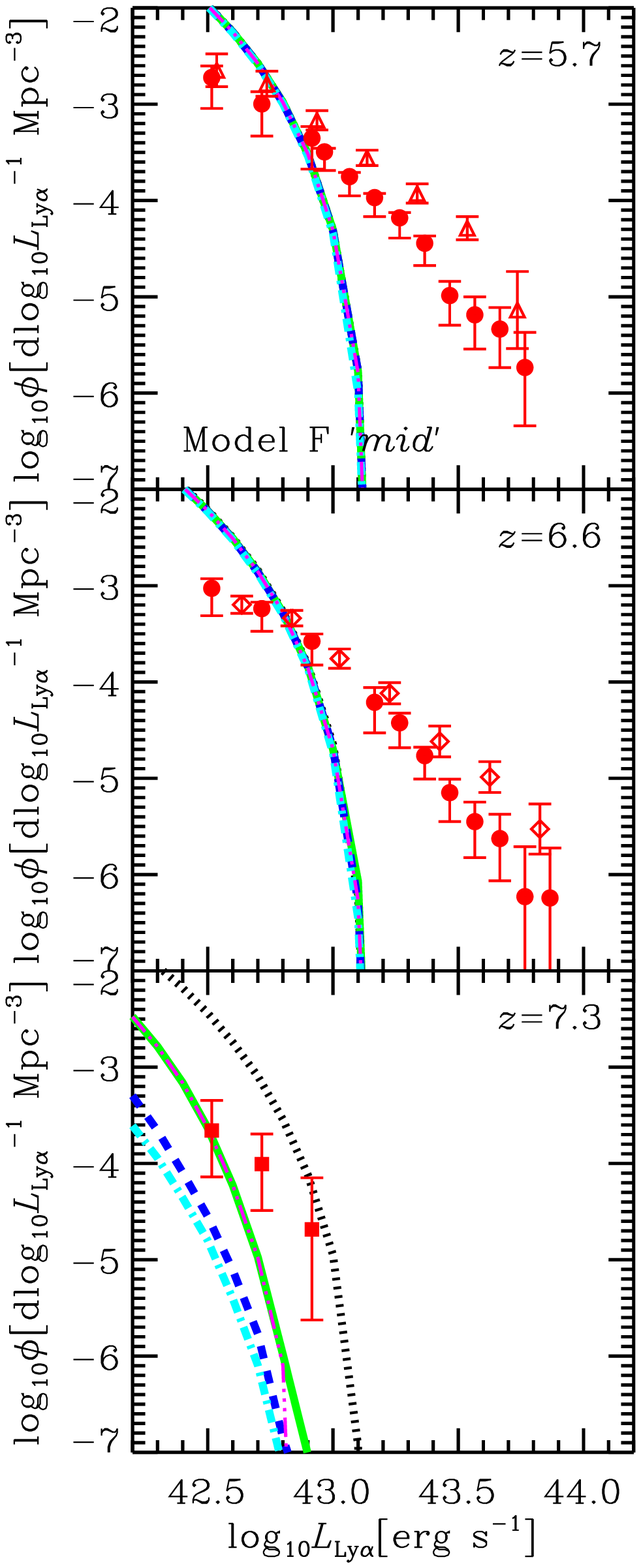}
  \includegraphics[width=4cm]{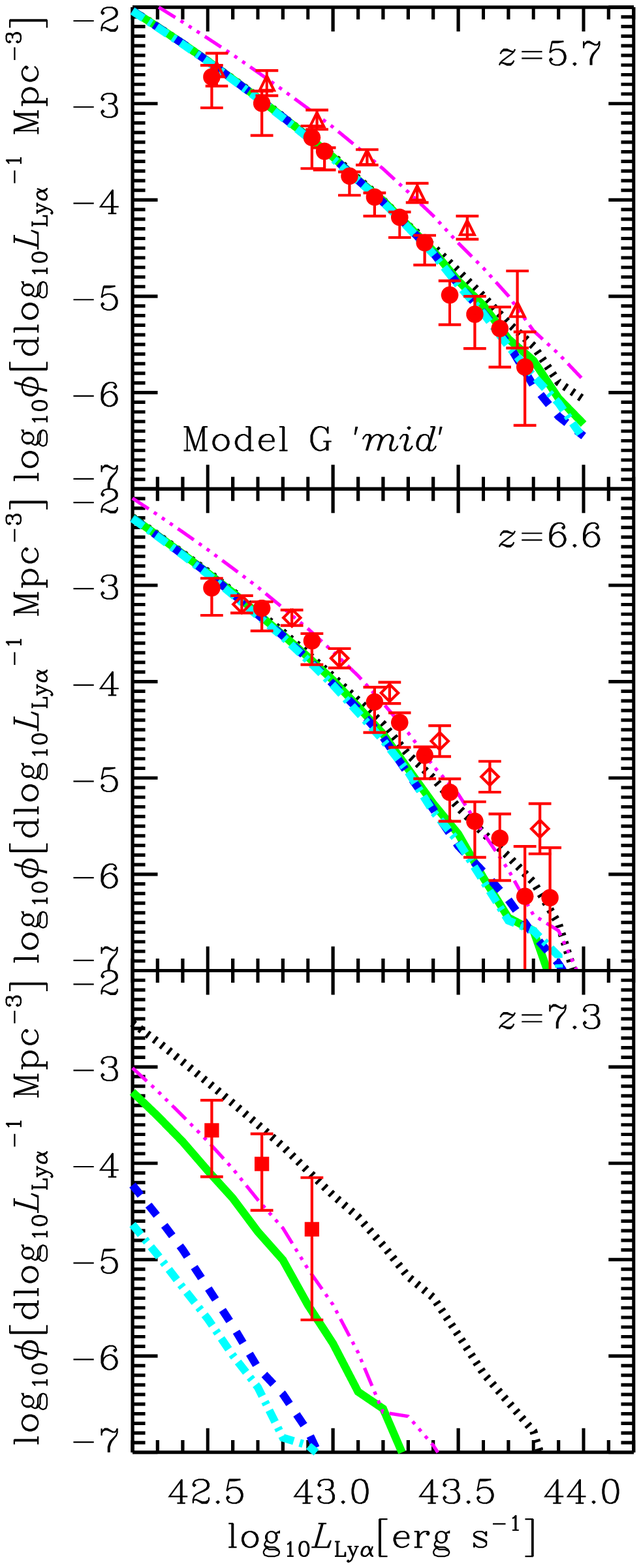}
  \includegraphics[width=4cm]{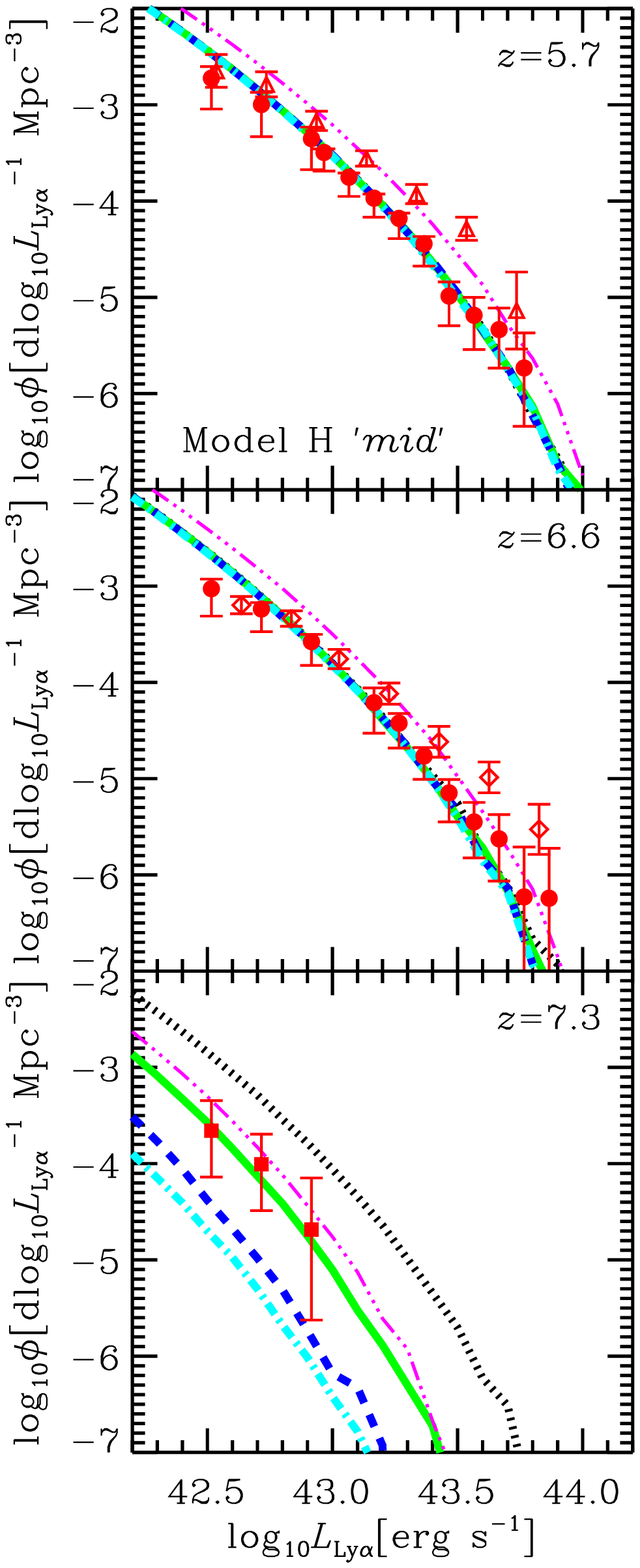}
 \end{center}
 \caption{Ly$\alpha$ luminosity functions for the $mid$ history. The
 dotted black lines are the completely transparent IGM cases. 
 The solid (green), dashed (blue), and dot-dashed (cyan) lines  
 show the cases with different Ly$\alpha$
 line profiles in a uniform outflowing gas of a velocity of 150 km
 s$^{-1}$ and an H~{\sc i} column density of 
 $\log_{10}(N_{\rm HI}/{\rm cm^{-2}})=20$, 19, and 18,
 respectively. The thin triple-dot-dashed (magenta) lines are the cases fit to
 Santos et al.~(2016) at $z=5.7$ instead of the HSC data of Konno et
 al.~(2018) and $\log_{10}(N_{\rm HI}/{\rm cm^{-2}})=20$. The filled
 symbols are the best estimates with HSC at $z=5.7$ and 6.6 (Konno et
 al.~2017) and with S-Cam at $z=7.3$ (Konno et al.~2014). The open
 symbols are the data taken from Santos et al.~(2016).} 
 \label{LaLF_mid}
\end{figure*}

\begin{figure*}
 \begin{center}
  \includegraphics[width=4cm]{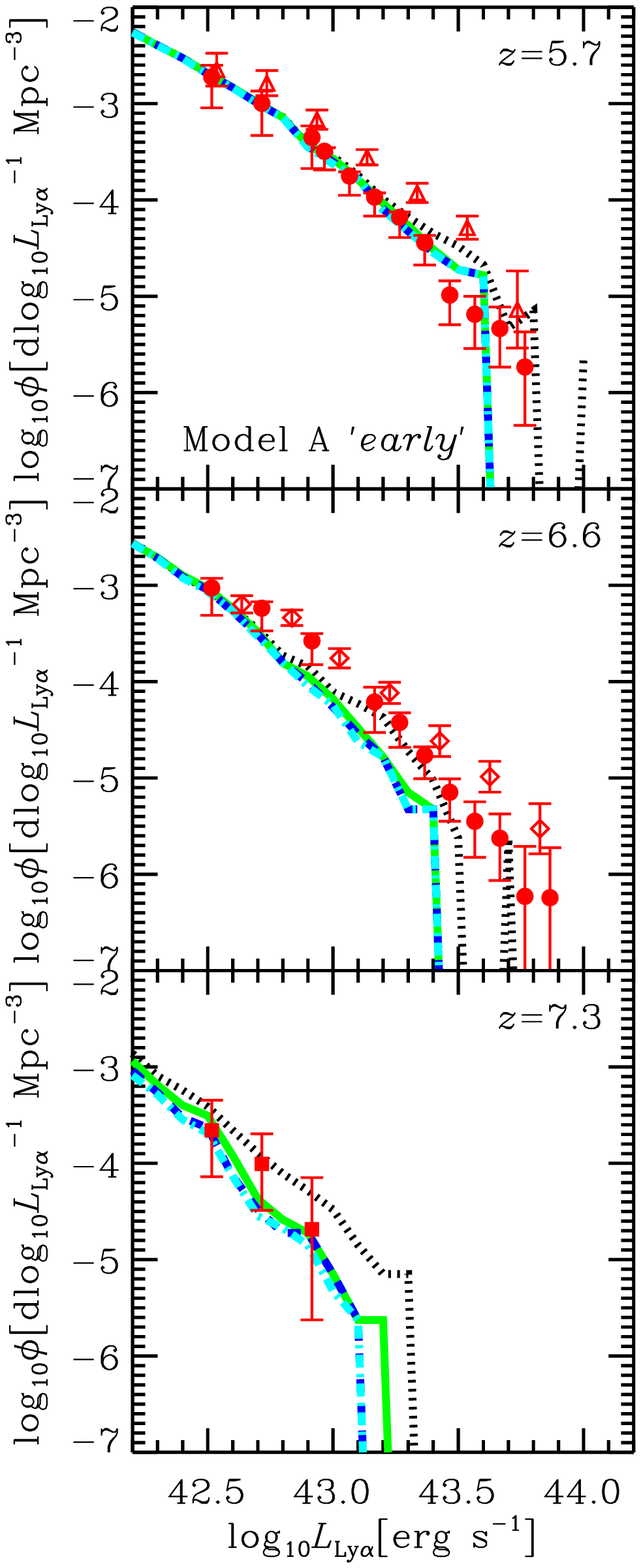}
  \includegraphics[width=4cm]{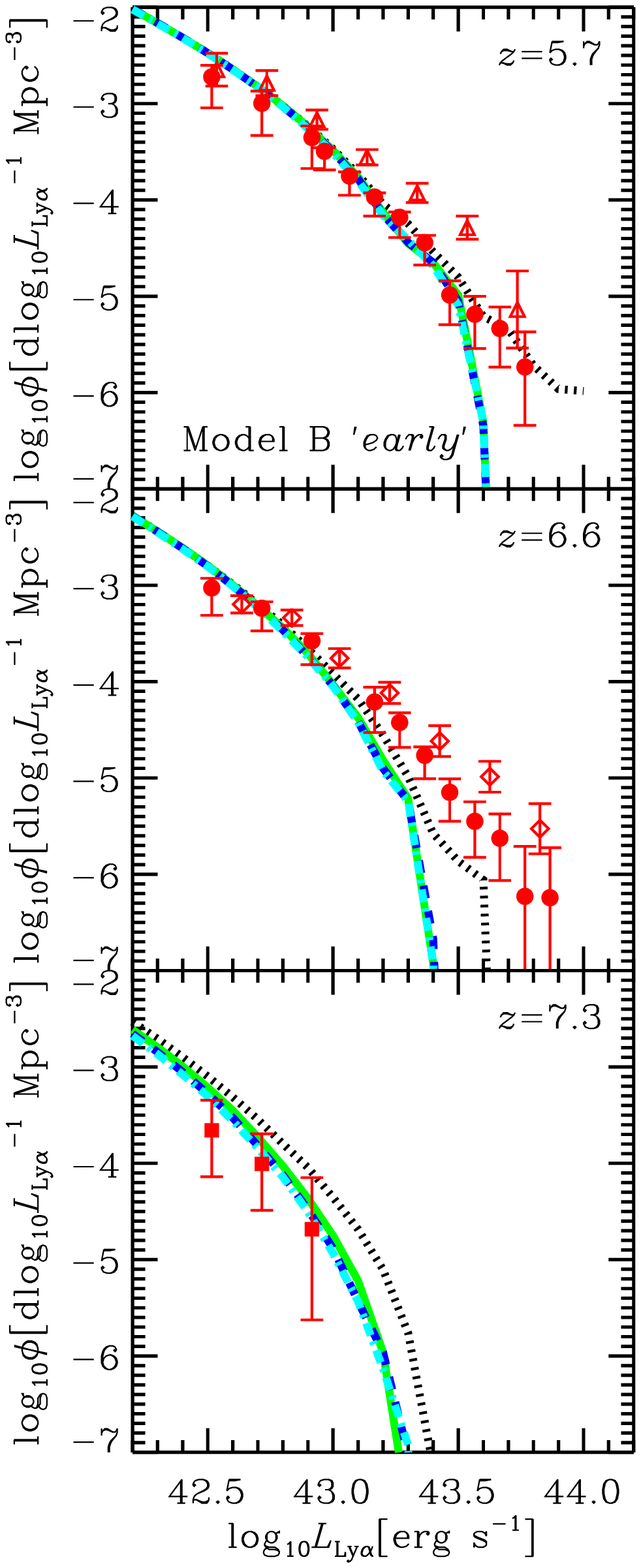}
  \includegraphics[width=4cm]{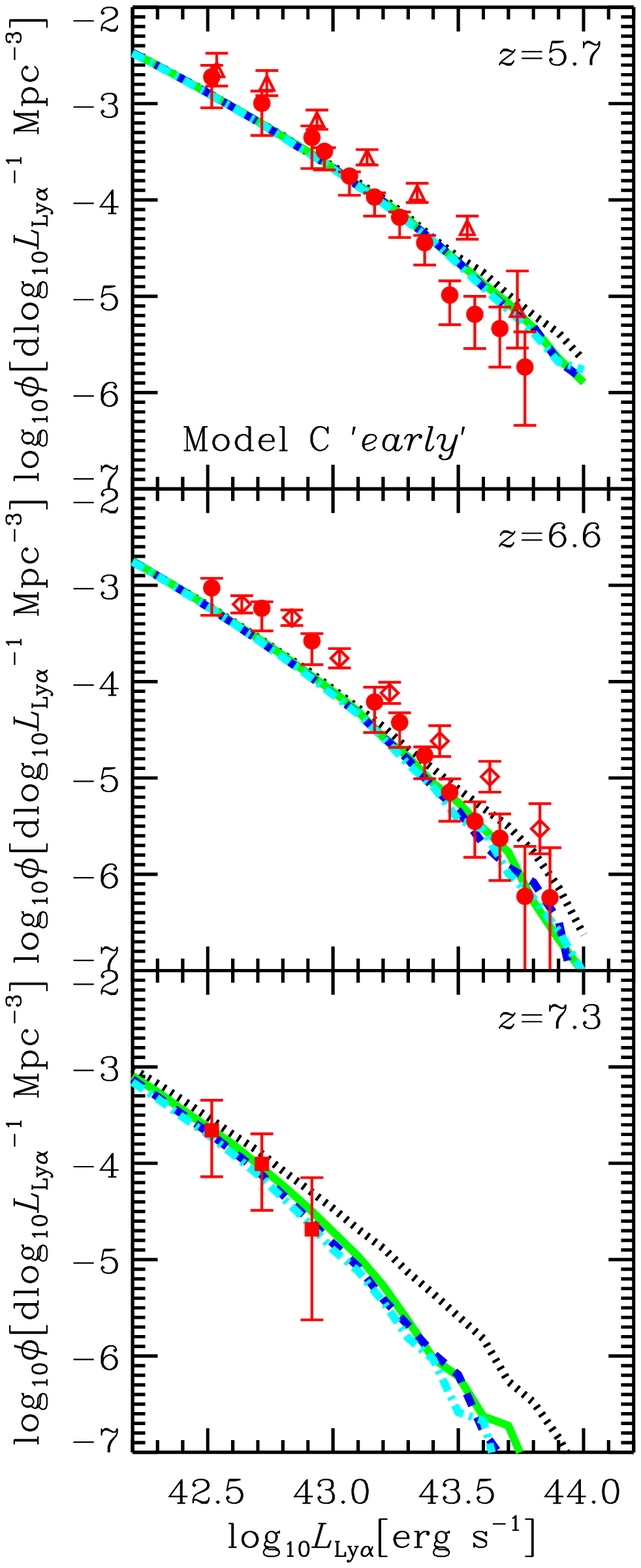}
  \includegraphics[width=4cm]{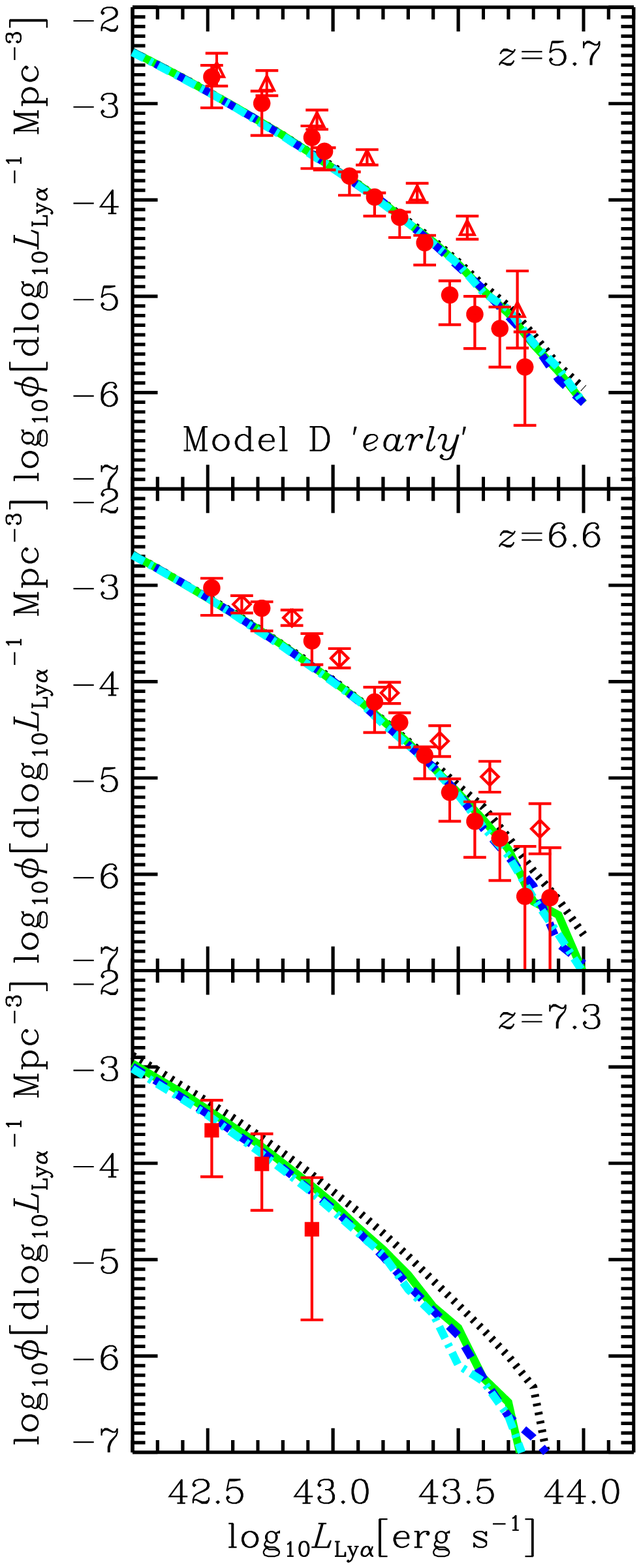}
  \includegraphics[width=4cm]{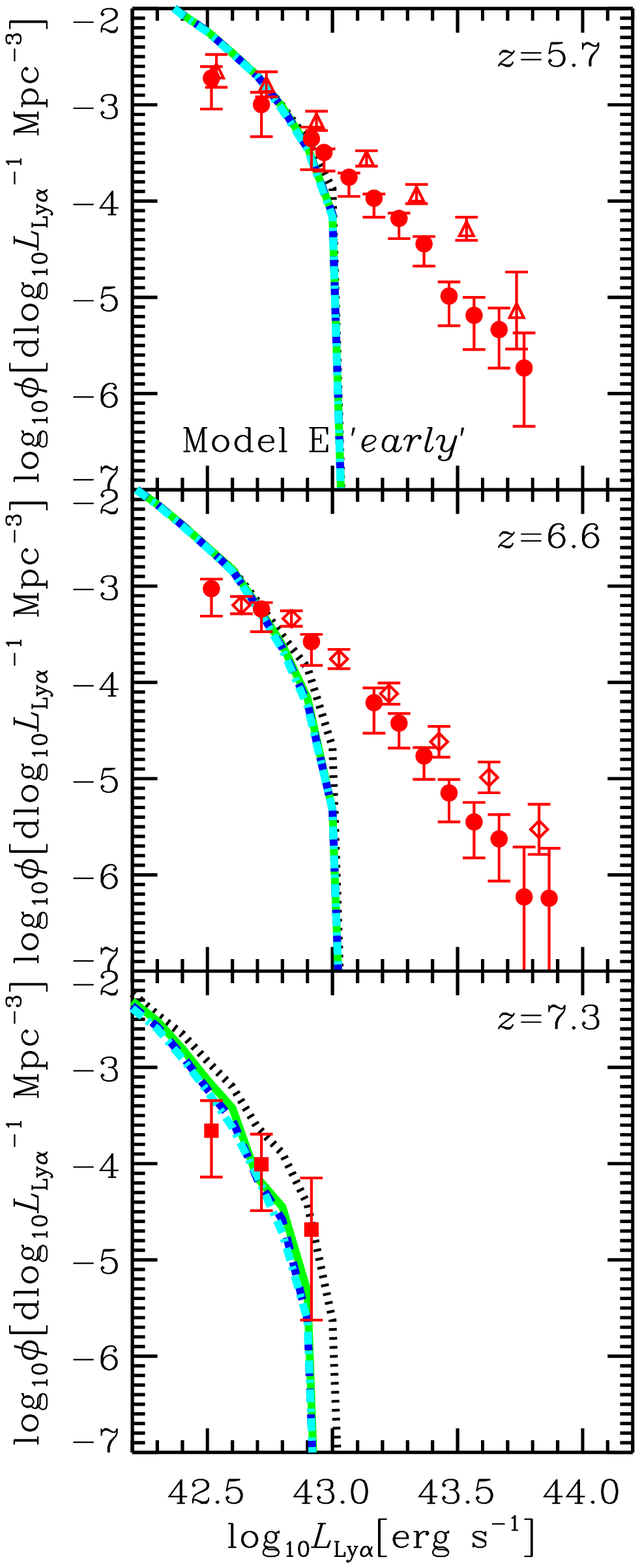}
  \includegraphics[width=4cm]{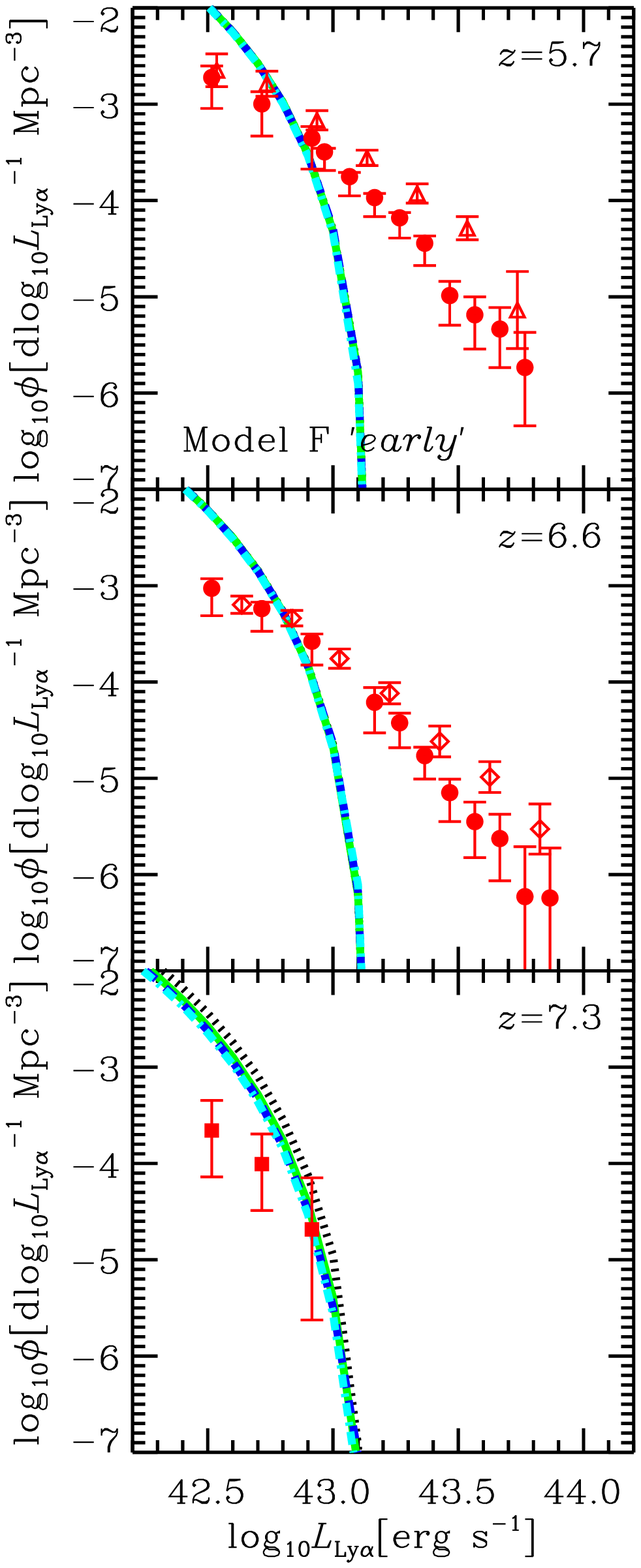}
  \includegraphics[width=4cm]{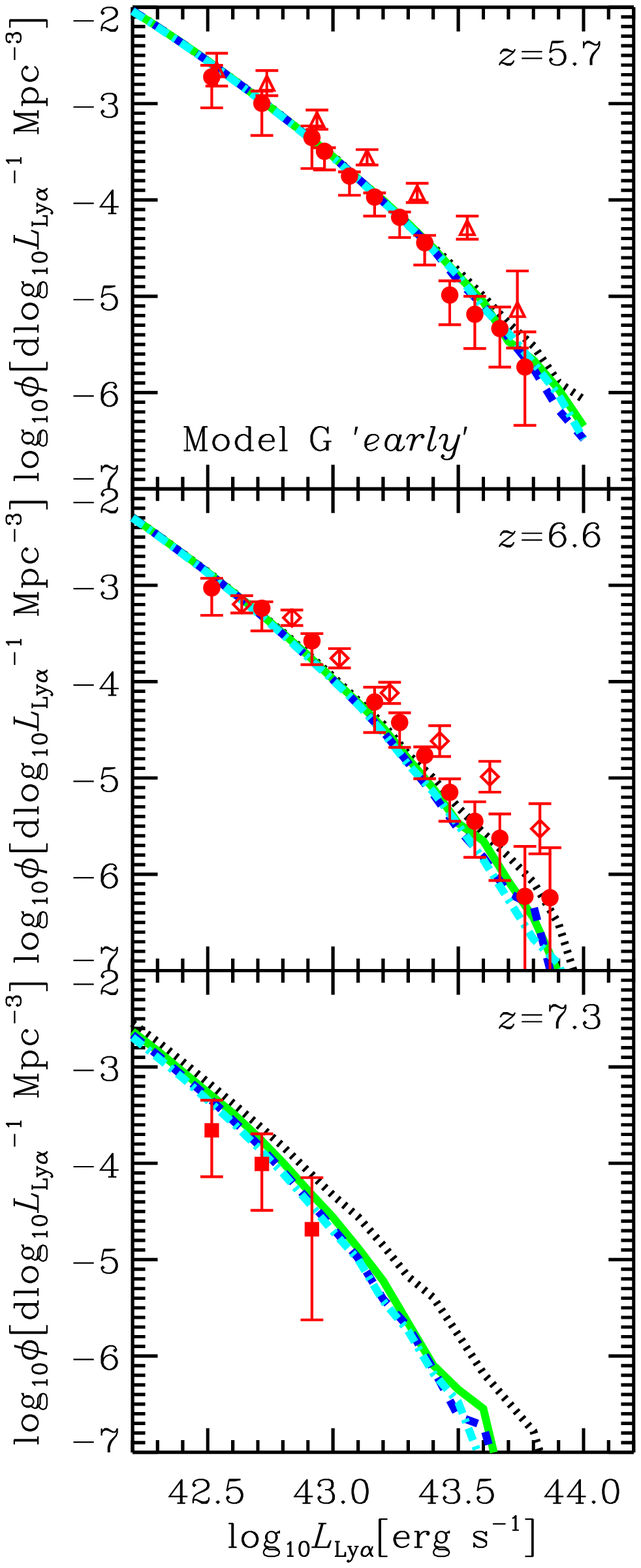}
  \includegraphics[width=4cm]{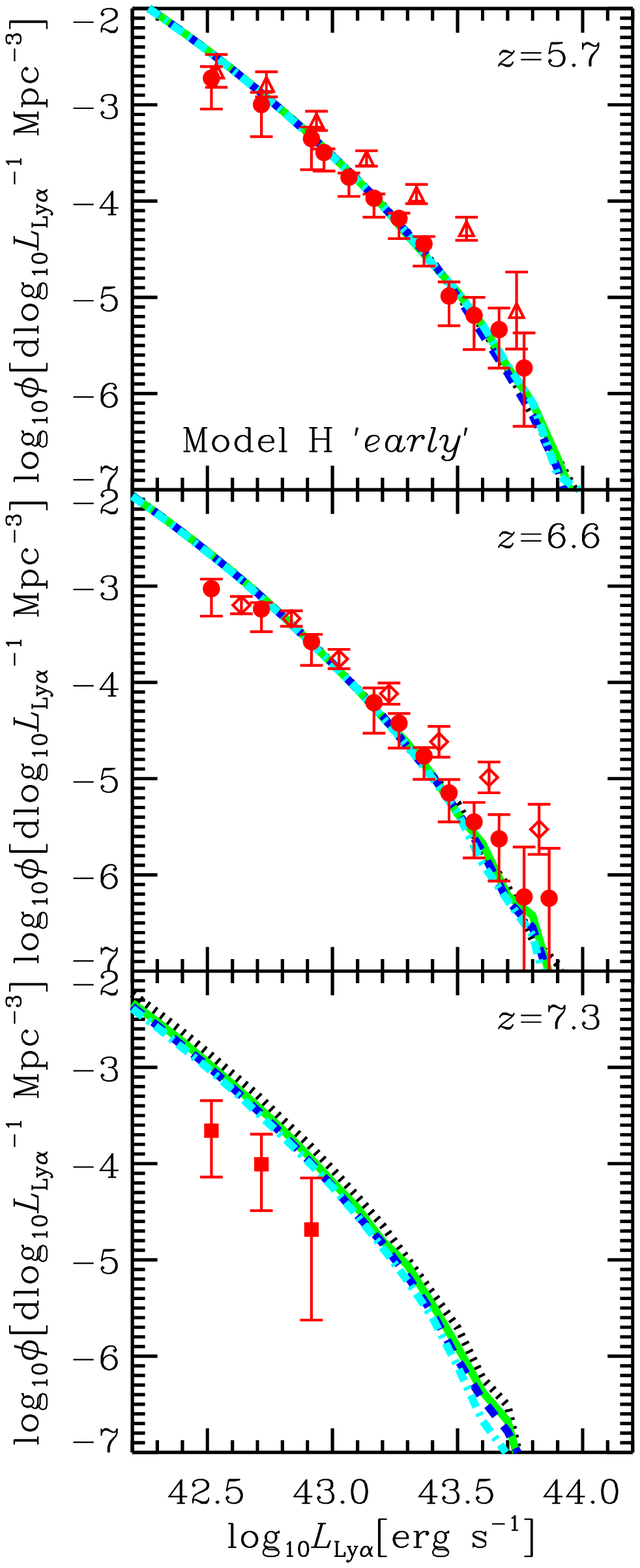}
 \end{center}
 \caption{Same as Fig.\ref{LaLF_mid} but for the $early$ history.}
 \label{LaLF_early}
\end{figure*}

\begin{figure*}
 \begin{center}
  \includegraphics[width=4cm]{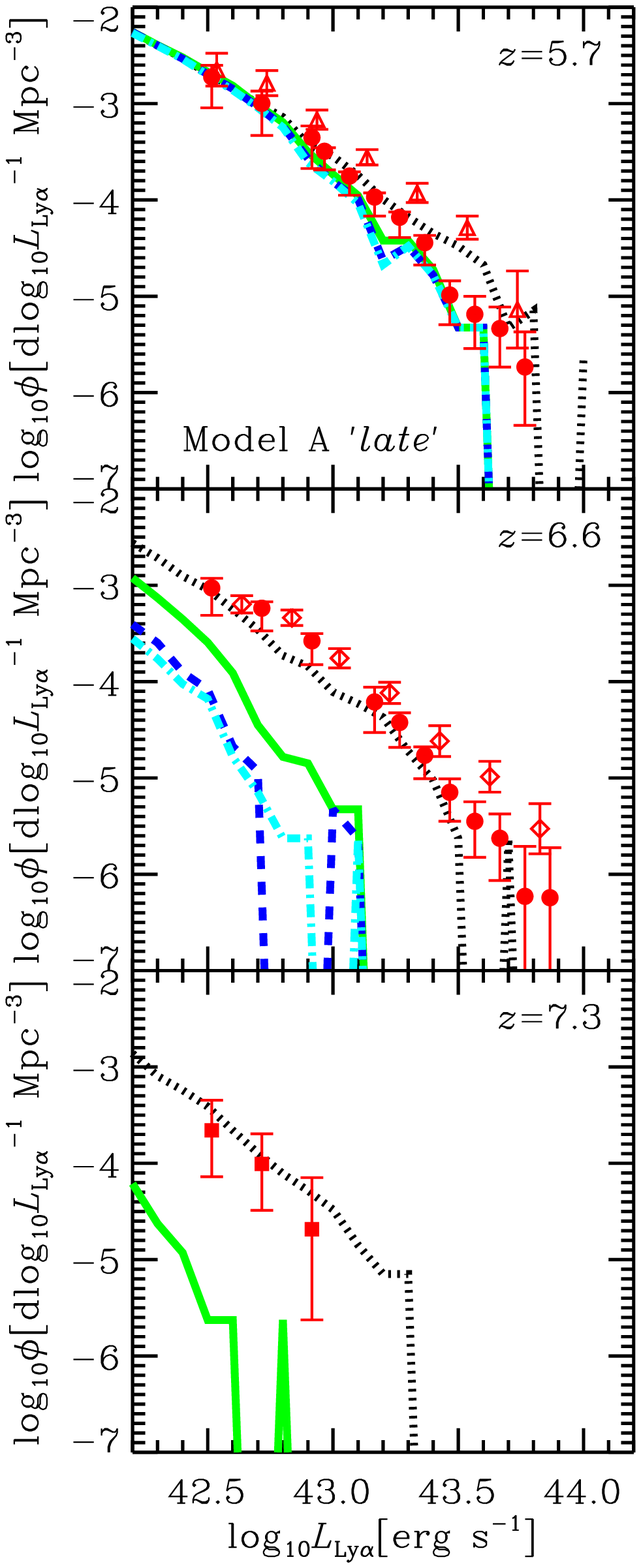}
  \includegraphics[width=4cm]{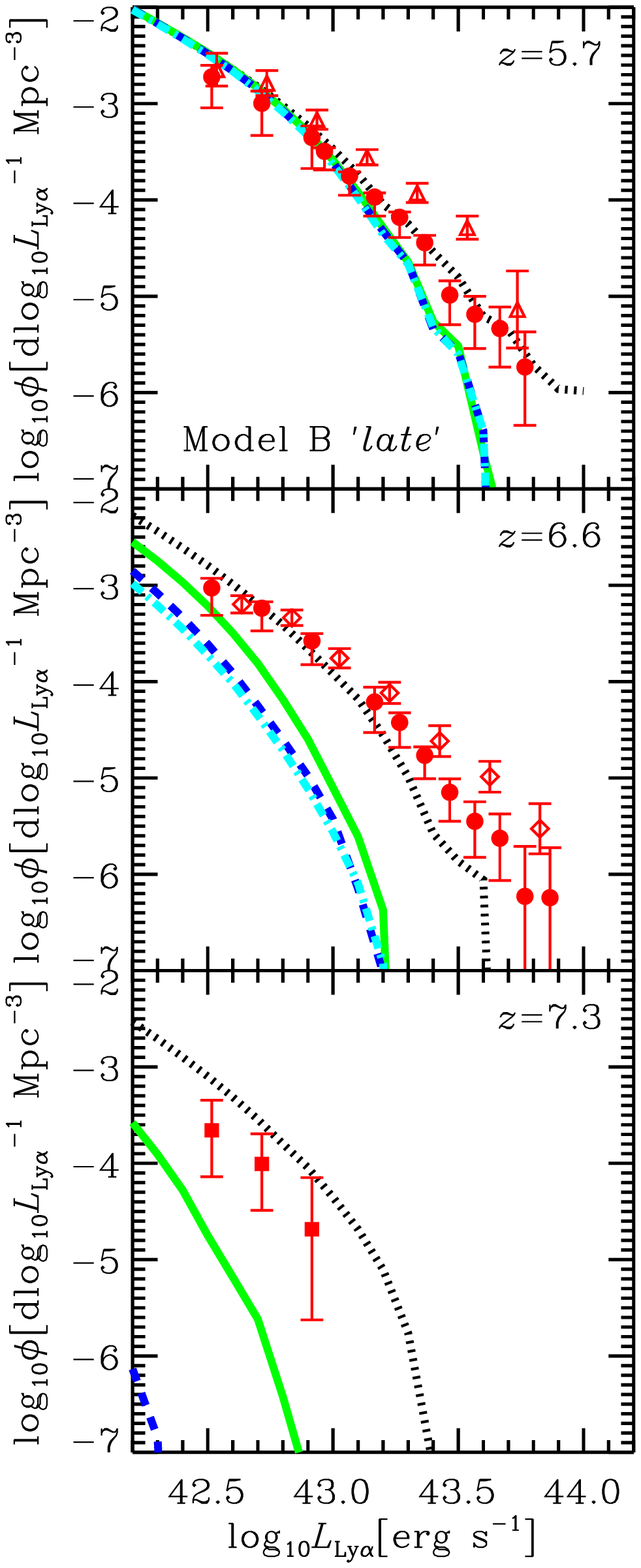}
  \includegraphics[width=4cm]{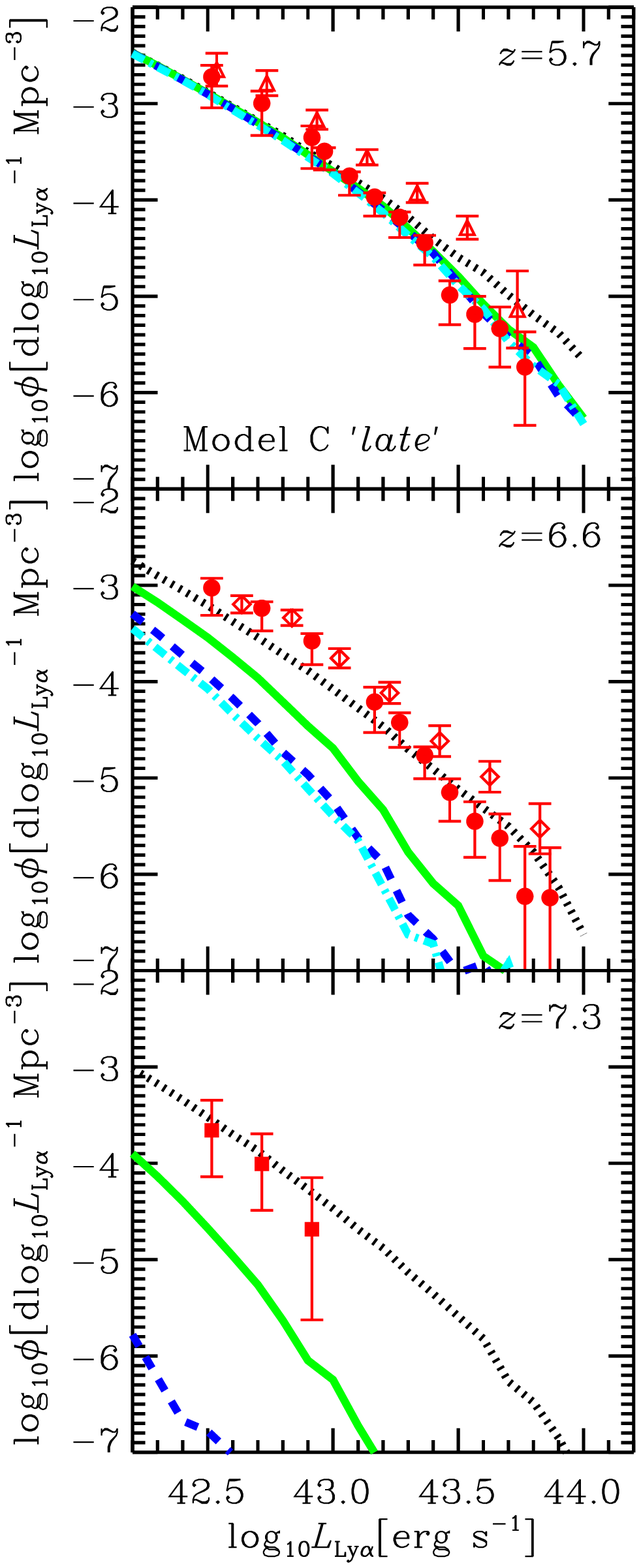}
  \includegraphics[width=4cm]{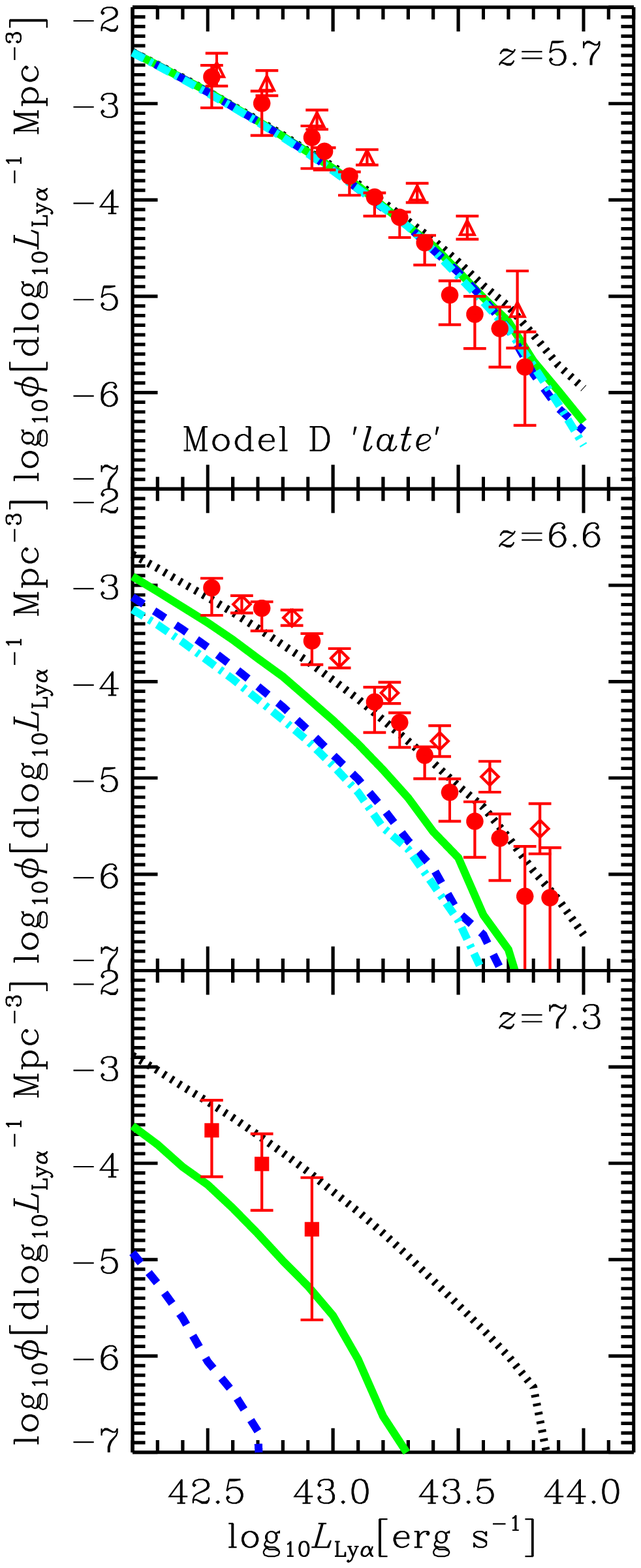}
  \includegraphics[width=4cm]{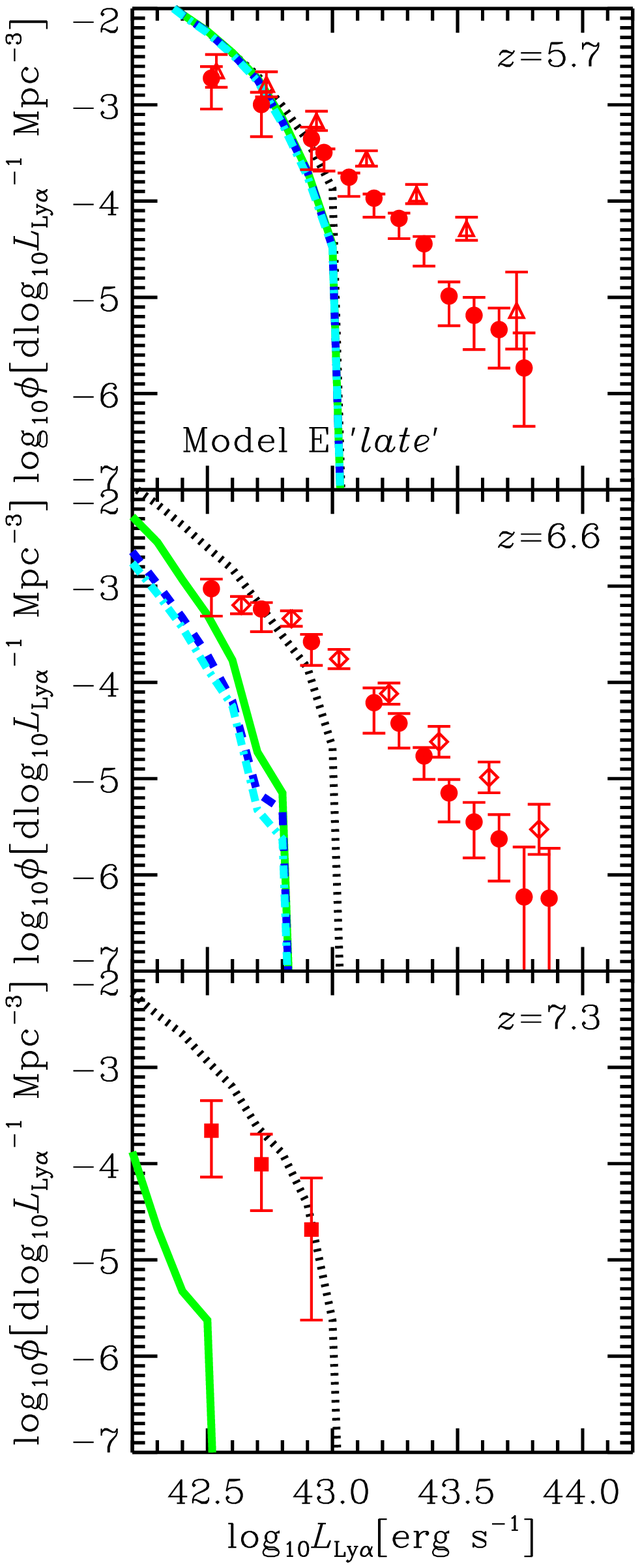}
  \includegraphics[width=4cm]{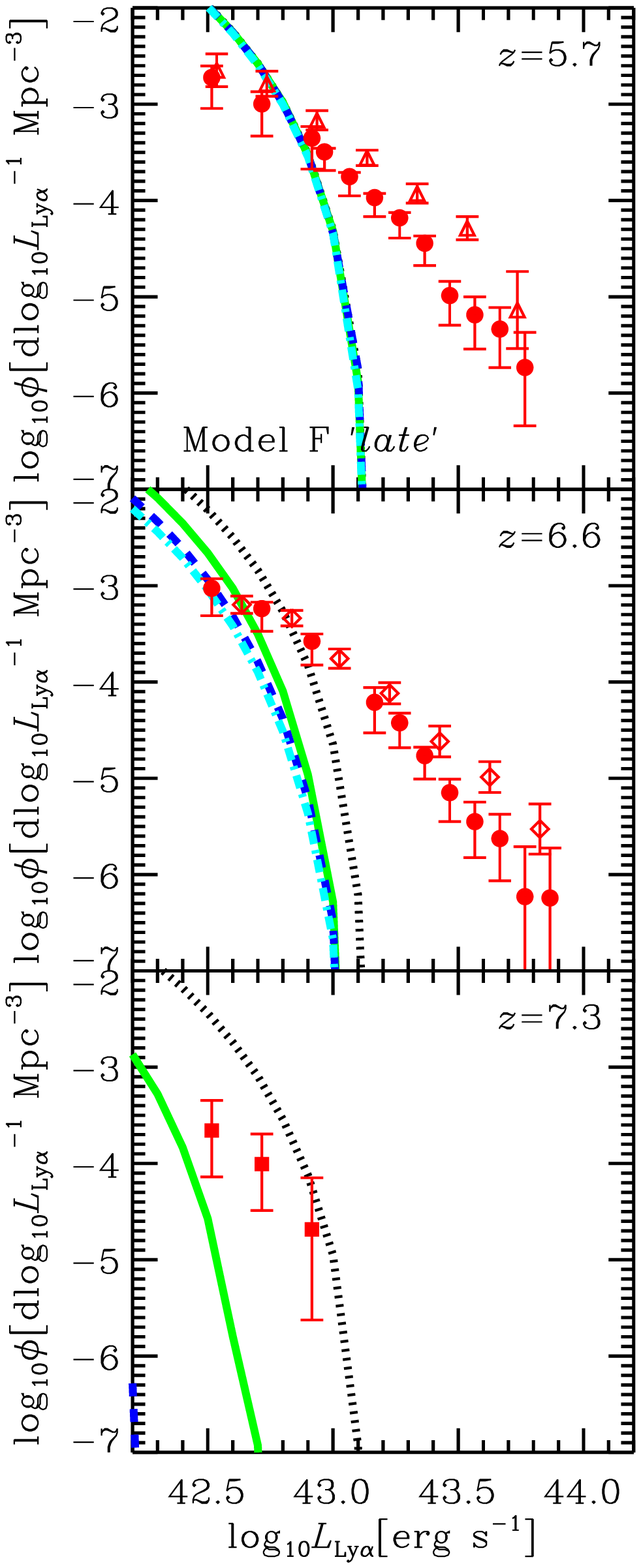}
  \includegraphics[width=4cm]{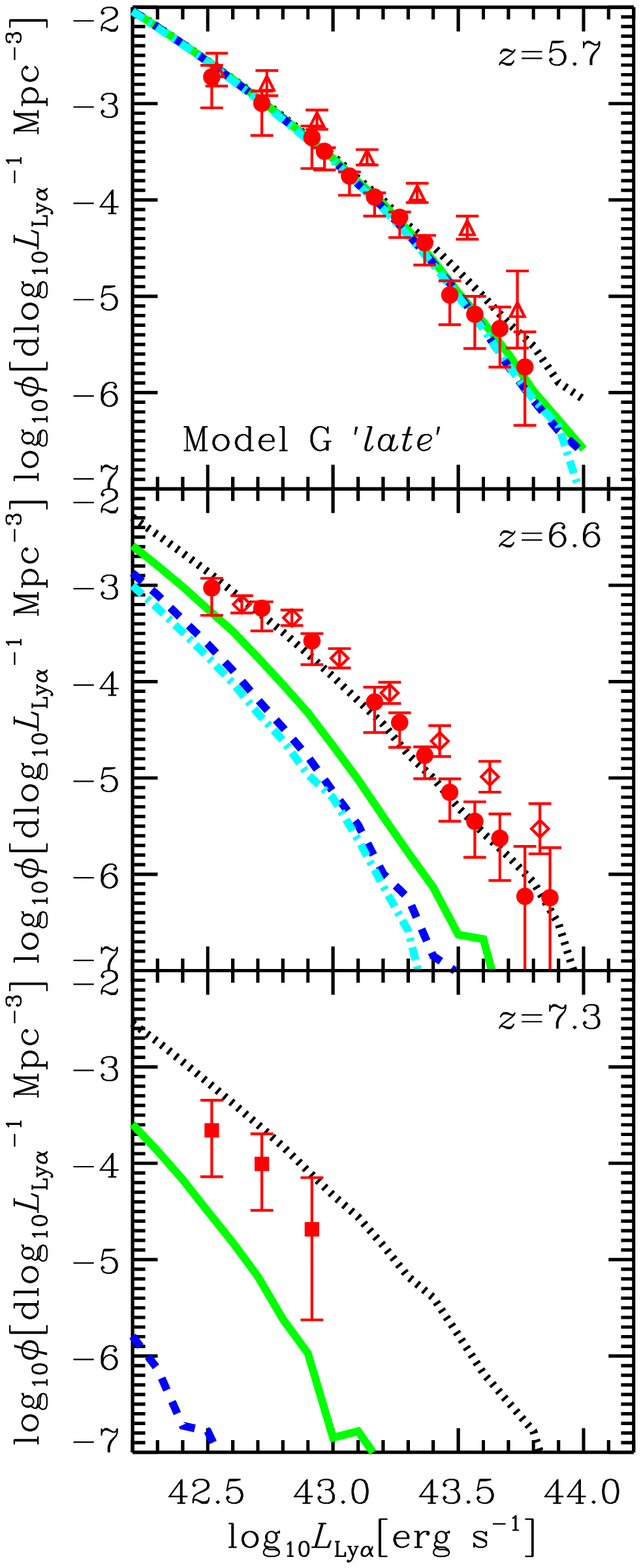}
  \includegraphics[width=4cm]{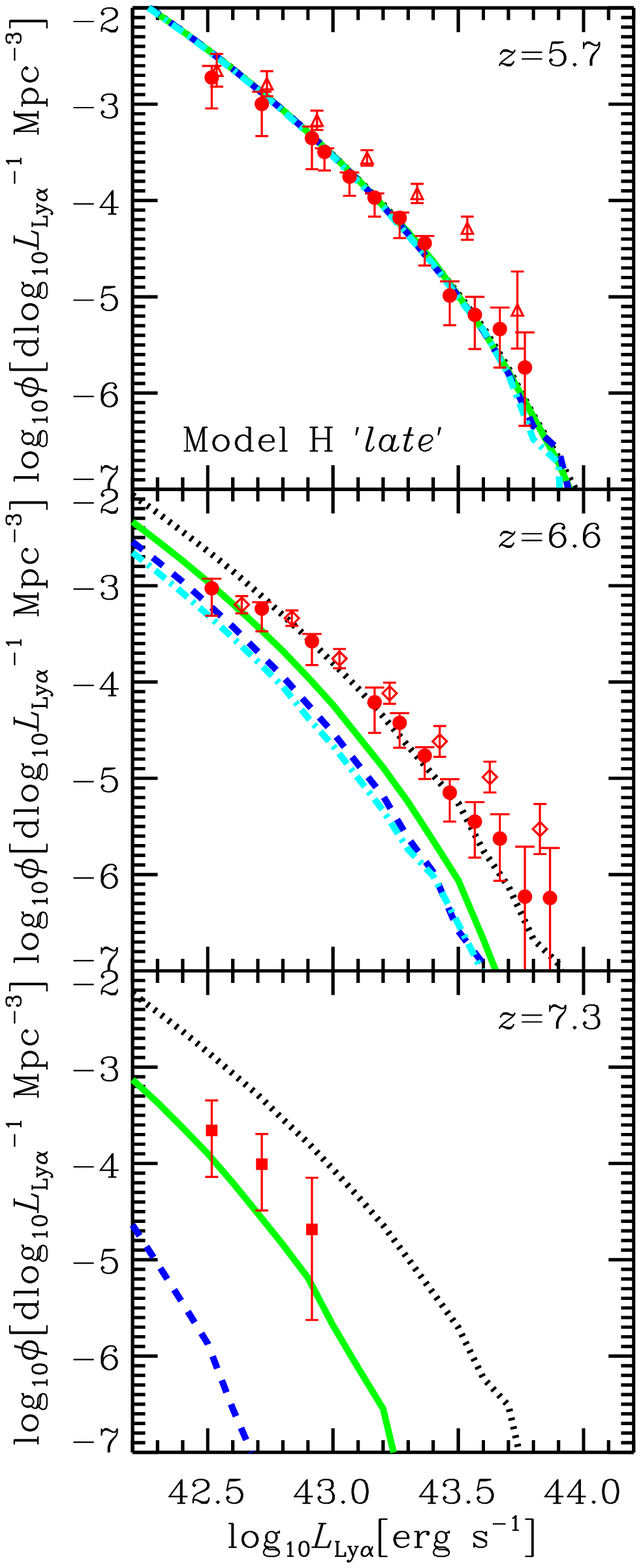}
 \end{center}
 \caption{Same as Fig.\ref{LaLF_mid} but for the $late$ history.}
 \label{LaLF_late}
\end{figure*}

We compare the true LFs in our simulation box with the SILVERRUSH early
LFs of \citet{Konno2017} directly because the observed LFs are already
corrected for incompleteness and the filter transmission effects (see
Appendix~2). 
Figs.~\ref{LaLF_mid}, \ref{LaLF_early} and \ref{LaLF_late} show the
comparisons of the observed LFs with the models for the $mid$, $early$
and $late$ reionization histories, respectively. The dotted (black) lines are 
the model LFs without the IGM transmission, namely those through a 
completely ionized IGM. Thus, these put the upper limit of the LAE 
number densities in each model. If the dotted line in a panel becomes 
far below the observed data, the model should be rejected. Such cases 
are the models E and F which predict too few LAEs brighter than 
$\simeq10^{43}$ erg s$^{-1}$. The models A and B are marginal due to 
a small number densities of LAEs at $>10^{43.5}$ erg s$^{-1}$ and at 
$z=6.6$. Other models are qualified.

The solid (green), dashed (blue) and dot-dashed (cyan) lines, 
which are overlapped each other 
in many cases, are the model LFs through the IGM transmission with 
different Ly$\alpha$ line profiles depending on the H~{\sc i} column 
density in the outflowing gas. As found in Fig.~\ref{fig:lineprofile}, 
the velocity shift of the line peak becomes larger for a larger column 
density, resulting in higher IGM transmission. This effect becomes 
more pronounced in the IGM with a higher $x_{\rm HI}$. For example, 
in the $mid$ reionization model (Fig.~\ref{LaLF_mid}), the difference is 
visible only at $z=7.3$. On the other hand, in the $early$ model 
(Fig.~\ref{LaLF_early}), the difference is small even at $z=7.3$, whereas 
it can be found also at $z=6.6$ in the $late$ model 
(Fig.~\ref{LaLF_late}). At $z=5.7$, the difference is negligible 
no matter which reionization model.

Regarding the IGM transmitted LFs (colored lines), 
the qualified models (C, D, G and H) including marginal ones (A and B) 
show an excellent agreement between the predictions and the observed 
data at $z=5.7$ where we have calibrated these models with the $mid$
reionization history. At $z=6.6$, Models C and G (and also A and B) in
the $mid$ and $late$ histories predict less LAEs than observed, while
Models C and G in the $early$ history still agree to the data. The same
thing is found at $z=7.3$. On the other hand, Models D and H in the
$mid$ and $early$ histories are consistent with the data at $z=6.6$
and 7.3 (Model~H in the $early$ history seems overprediction at
$z=7.3$). However, these models in the $late$ history also underpredict
the LAE number densities at $z=6.6$. At $z=7.3$, these models are still
touching the lower bound of the measurements. Overall, the observed
Ly$\alpha$ LFs at $z=6.6$ and $7.3$ seem to favor more ionized IGM like
the $early$ history. This point will be discussed more in \S5.2. 

In Fig.~\ref{LaLF_mid}, we also show the LFs calibrated with
\citet{Santos2016} instead of \citet{Konno2017}.\footnote{We just use
the same pivot values, $\tau_{\alpha,10}$, for the models~E and F in the
both calibrations because the LF shape is inconsistent with the
observations.} The results are qualitatively similar to those with 
\citet{Konno2017}. The predictions at $z=6.6$ are smaller than the
observed LFs (open symbols) of \citet{Matthee2015} updated by
\citet{Santos2016} in the 
bright-end. This again indicates somewhat earlier reionization than the
$mid$ history.

\subsection{Angular correlation function}

\begin{figure*}
 \begin{center}
  \includegraphics[width=4.2cm,angle=90]{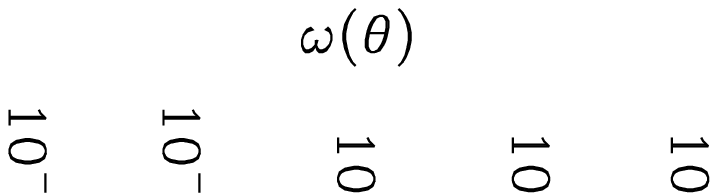}
  \includegraphics[width=4.2cm,angle=90]{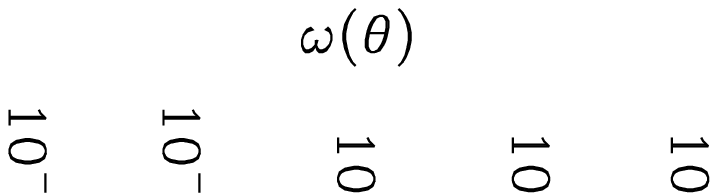}
  \includegraphics[width=4.2cm,angle=90]{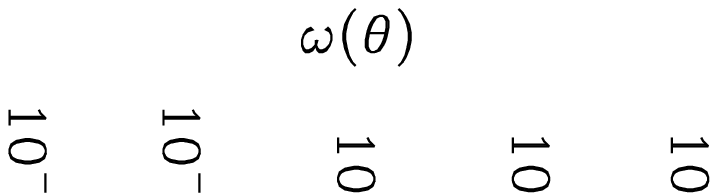}
  \includegraphics[width=4.2cm,angle=90]{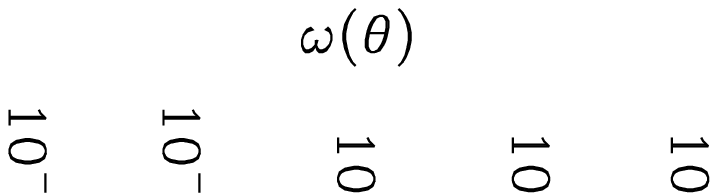}
  \includegraphics[width=4.2cm,angle=90]{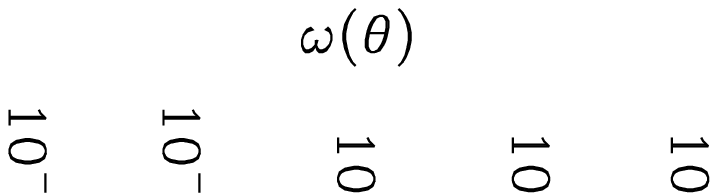}
  \includegraphics[width=4.2cm,angle=90]{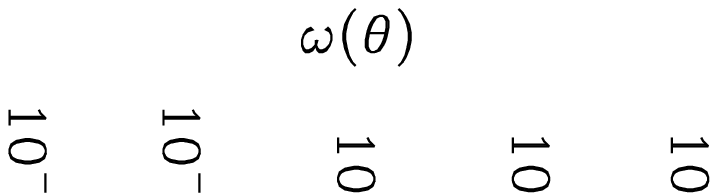}
  \includegraphics[width=4.2cm,angle=90]{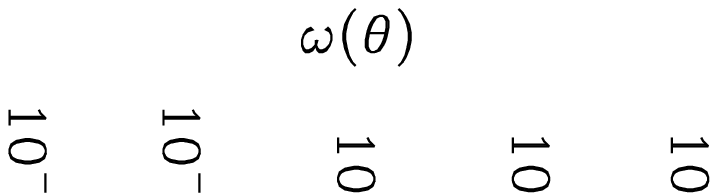}
  \includegraphics[width=4.2cm,angle=90]{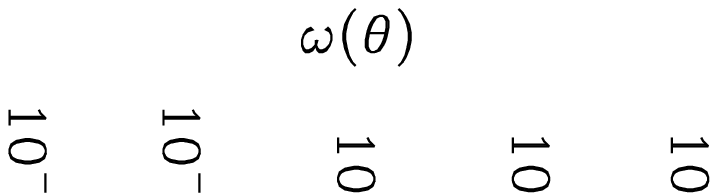}
 \end{center}
 \caption{Angular auto-correlation functions for LAEs selected by the NB
 color-magnitude criteria noted in the panels. 
 The circles with error-bars (red) are the observational results with
 HSC (Ouchi et al.~2018) and the filled data are compared to the models.
 The dotted (black) lines are the models in the completely transparent IGM. 
 The diamonds (green), triangles (blue), and squares (cyan) 
 with error-bars show the IGM-transmitted models with
 different Ly$\alpha$ line profiles in a uniform outflowing gas of a
 velocity of 150 km s$^{-1}$ and an H~{\sc i} column density of 
 $\log_{10}(N_{\rm HI}/{\rm cm^{-2}})=20$, 19, and 18, respectively. 
 The model error-bars, which includes only Poisson errors, 
 are scaled to the survey areas noted in the panels. 
 The thin triple-dot-dashed (magenta) lines are the cases fit to
 Santos et al.~(2016) at $z=5.7$ instead of the HSC data of Konno et
 al.~(2018) and $\log_{10}(N_{\rm HI}/{\rm cm^{-2}})=20$. 
 The $mid$ history is adopted.}
 \label{ACF_mid}
\end{figure*}

\begin{figure*}
 \begin{center}
  \includegraphics[width=4.2cm,angle=90]{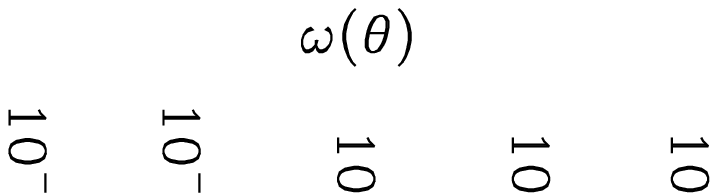}
  \includegraphics[width=4.2cm,angle=90]{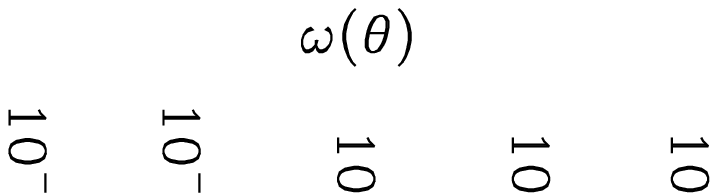}
  \includegraphics[width=4.2cm,angle=90]{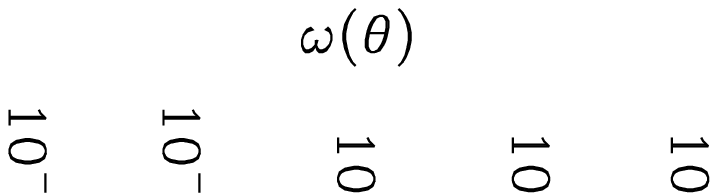}
  \includegraphics[width=4.2cm,angle=90]{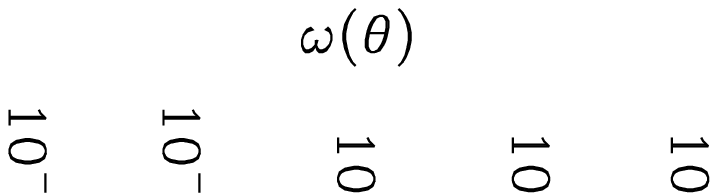}
  \includegraphics[width=4.2cm,angle=90]{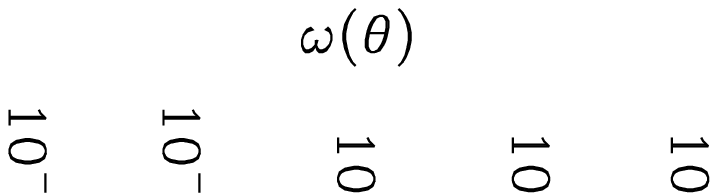}
  \includegraphics[width=4.2cm,angle=90]{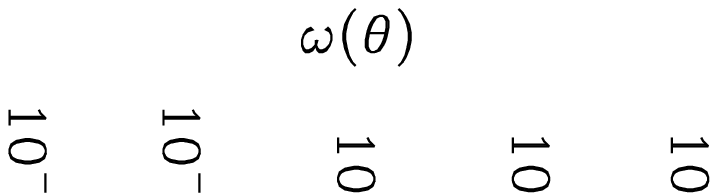}
  \includegraphics[width=4.2cm,angle=90]{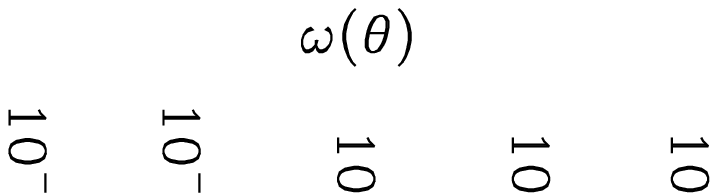}
  \includegraphics[width=4.2cm,angle=90]{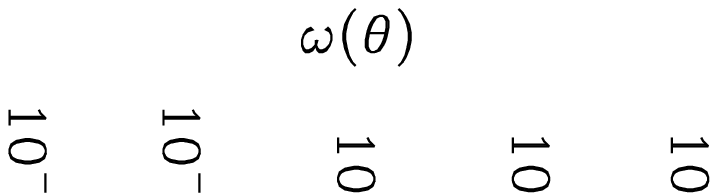}
 \end{center}
 \caption{Same as Fig.\ref{ACF_mid} but for the $early$ history.}
 \label{ACF_early}
\end{figure*}

\begin{figure*}
 \begin{center}
  \includegraphics[width=4.2cm,angle=90]{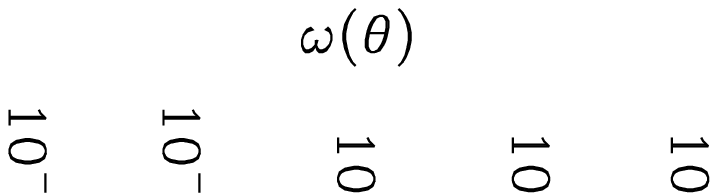}
  \includegraphics[width=4.2cm,angle=90]{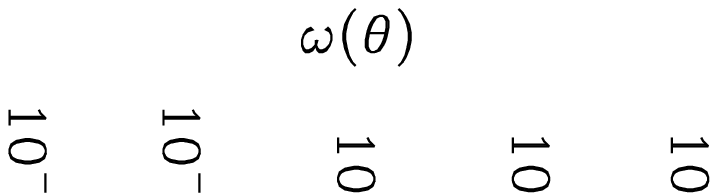}
  \includegraphics[width=4.2cm,angle=90]{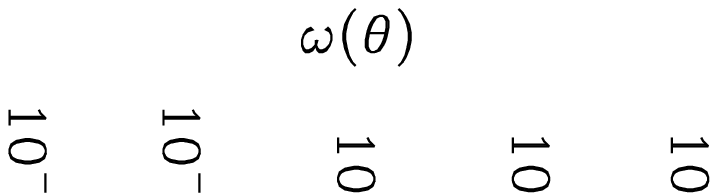}
  \includegraphics[width=4.2cm,angle=90]{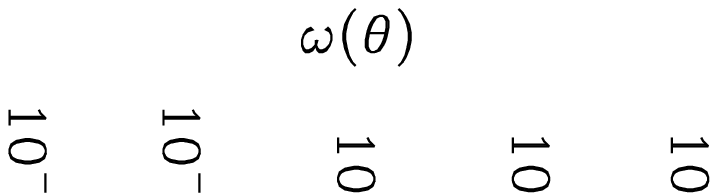}
  \includegraphics[width=4.2cm,angle=90]{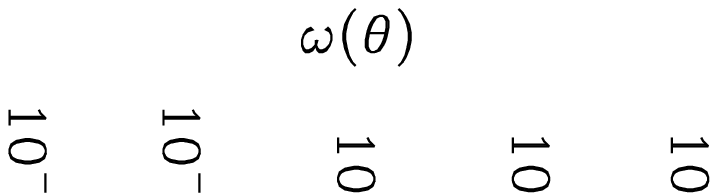}
  \includegraphics[width=4.2cm,angle=90]{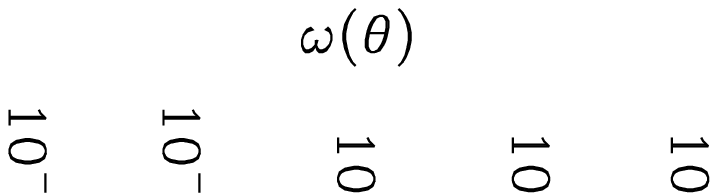}
  \includegraphics[width=4.2cm,angle=90]{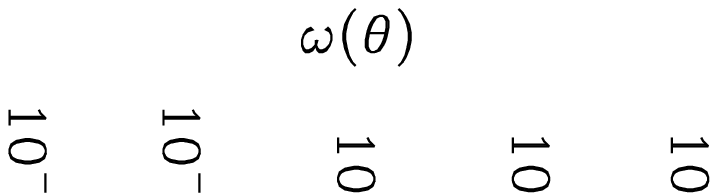}
  \includegraphics[width=4.2cm,angle=90]{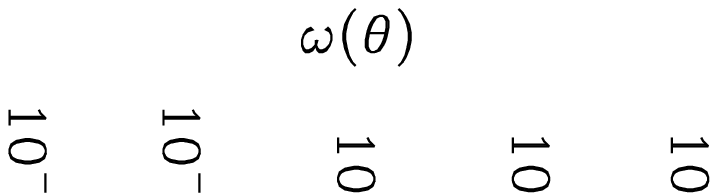}
 \end{center}
 \caption{Same as Fig.\ref{ACF_mid} but for the $late$ history.}
 \label{ACF_late}
\end{figure*}

We compare the model ACFs with the SILVERRUSH early results by
\citet{Ouchi2017}. The model ACFs are calculated by the same way as
\citet{Ouchi2017}; we select mock LAEs by the same color-magnitude
criteria via virtual observations (\S3.6), count the numbers of pairs of
LAE--LAE, LAE--random position, and random--random positions, and
estimate the ACFs according to \citet{LS93}. The virtual observations
are repeated 300- and 500-times to secure sufficiently large mock LAE
catalogs for $z=5.7$ and $6.6$, respectively.
Figs.~\ref{ACF_mid}, \ref{ACF_early} and \ref{ACF_late} show the
comparisons at $z=5.7$ and 6.6 for the $mid$, $early$ and $late$
reionization histories, respectively. 
The areas of the current SILVERRUSH also noted in the panels are
used to scale the model error-bars. However, the model error-bars are
probably underestimated because we account only for Poisson errors. The
effect of the cosmic variance can be estimated by a Jackknife
method observationally from ACFs of \citet{Ouchi2017}. We have found
that $d\omega/\omega\sim0.4$ for a survey area of 20~deg$^2$ 
as the combination of the Poisson errors and the cosmic variance. 
Since the Poisson errors are much smaller than this and the cosmic
variance is the dominant source of uncertainties for the current
SILVERRUSH results.

The dotted (black) lines show the model ACFs through the completely ionized IGM
(i.e. no IGM effect). In these cases, the numbers of the mock LAEs are
the largest in each model and the ACFs up to about 1000 arcsec are
derived successfully as the observations have done, except for Model~F
at $z=6.6$, where the number of bright LAEs satisfying the NB magnitude
criterion is limited and we could derive the ACF only $<100$ arcsec. 
Therefore, Model~F is disqualified here.

The diamonds (green), triangles (blue) and squares (cyan) with
error-bars are the model ACFs through the IGM transmission depending on
the different H~{\sc i} column density for the line profile. At $z=5.7$,
the model ACFs are almost overlapped on the dotted (black) lines
(i.e. the fully-ionized IGM) irrespective of the models 
and the reionization histories because of the high ionization degree at
the redshift in our simulations. Each model predicts a different
correlation amplitude. Comparing with the observed data at $z=5.7$,
Models~C, D, G and H are very consistent with the data within the
error-bars in the angular separation between 60 and 1000 arcsec, while
Models~A, B and E seems marginally overpredict the amplitude. 

At a smaller angular separation ($<60$ arcsec), the observed amplitudes
are smaller than the models which continue to increase up to $\sim10$ arcsec. 
Since we do not identify sub-halos in a halo (\S2.1), 
the models do not include the correlation within a single halo, 
so-called one halo term.
But this becomes important only at $<10$ arcsec. Therefore, the
increase of the model ACF at 10--60 arcsec is caused by a different
thing. This is probably the non-linear halo bias effect 
\citep{Reed2009,Jose2016,Jose2017}. 
\citet{Jose2017} argue that a model with the non-linear effect can
better reproduce the LBG ACFs obtained from the Canada-France-Hawaii
Telescope Legacy Survey data \citep{Hildebrandt2009}. Recently,
\citet{Harikane2017} has also found this feature in the LBG ACFs from
the Subaru/HSC survey called GOLDRUSH \citep{Ono2018} thanks to a huge
number of the sample ($\sim500,000$). 
If the number of the LAE sample increases in future, 
a similar non-linear effect may be found in the observed LAE ACFs 
as predicted by the models.

At $z=6.6$, the difference due to the IGM ionization degree appears,
especially in the $late$ reionization history. The largest impact is the
decrease of the numbers of the mock LAEs. As a result, we have failed to
obtain the ACFs, for example, in Models A, E and F of the $mid$
history, in Model~F of the $early$ history, and many models in the
$late$ history. Since the observations provide us with the firm ACF
measurements even at $z=6.6$, the $late$ history is disfavored. For the
$mid$ and $early$ histories, Models~B, C, D, G, and H are consistent
with the observations between 100 and 1000 arcsec. At a smaller angular scale, 
the model ACFs are again higher than the observations.

In Fig.\ref{ACF_mid}, we also show the ACFs in the calibration with
\citet{Santos2016} instead of \citet{Konno2017}. The results are quite 
similar to those with \citet{Konno2017} and it is hard to distinguish 
the ACFs with different LF calibrations.

\subsection{LAE fraction in LBGs}

\begin{figure*}
 \begin{center}
  \includegraphics[width=7cm]{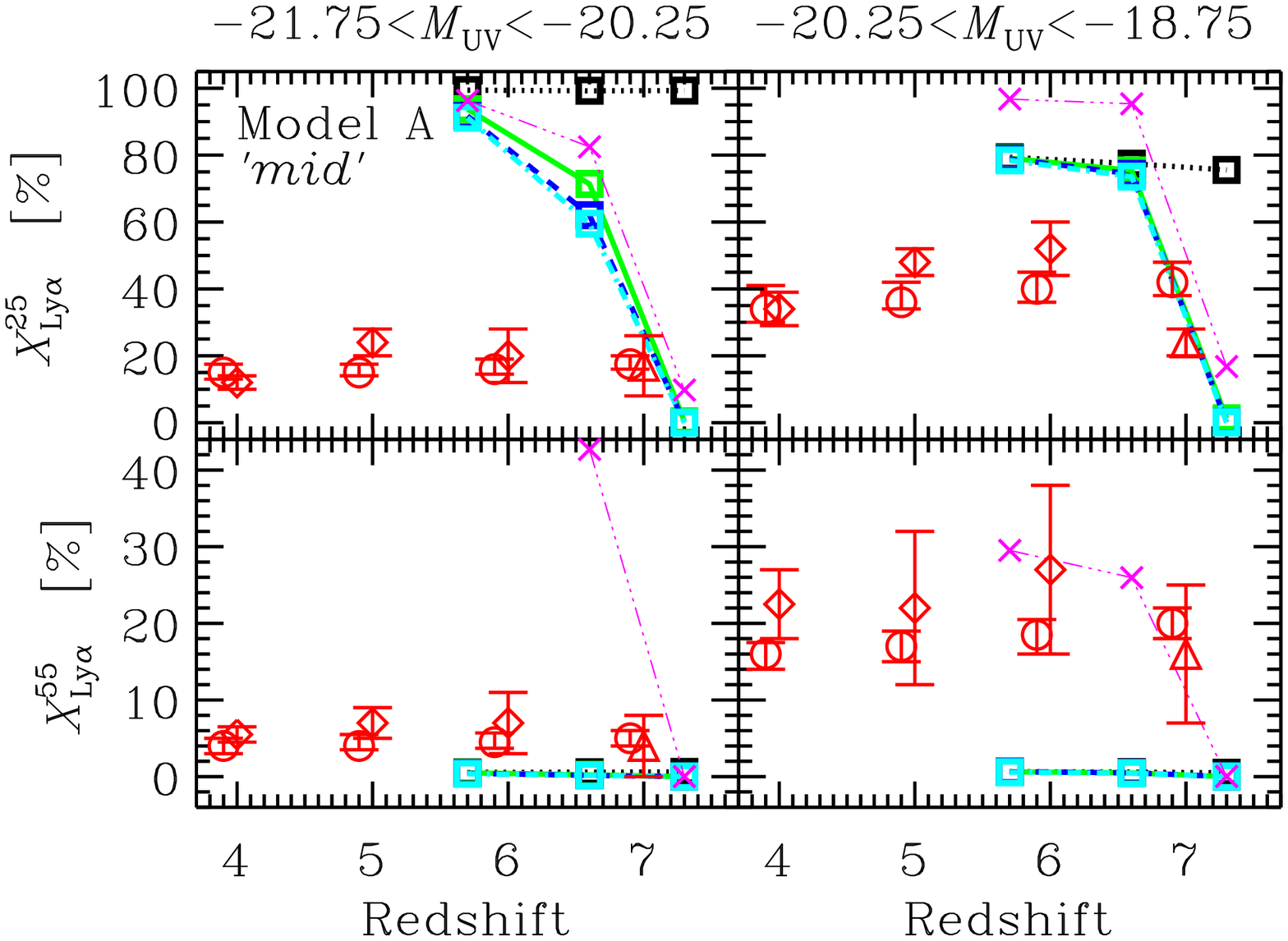}
  \includegraphics[width=7cm]{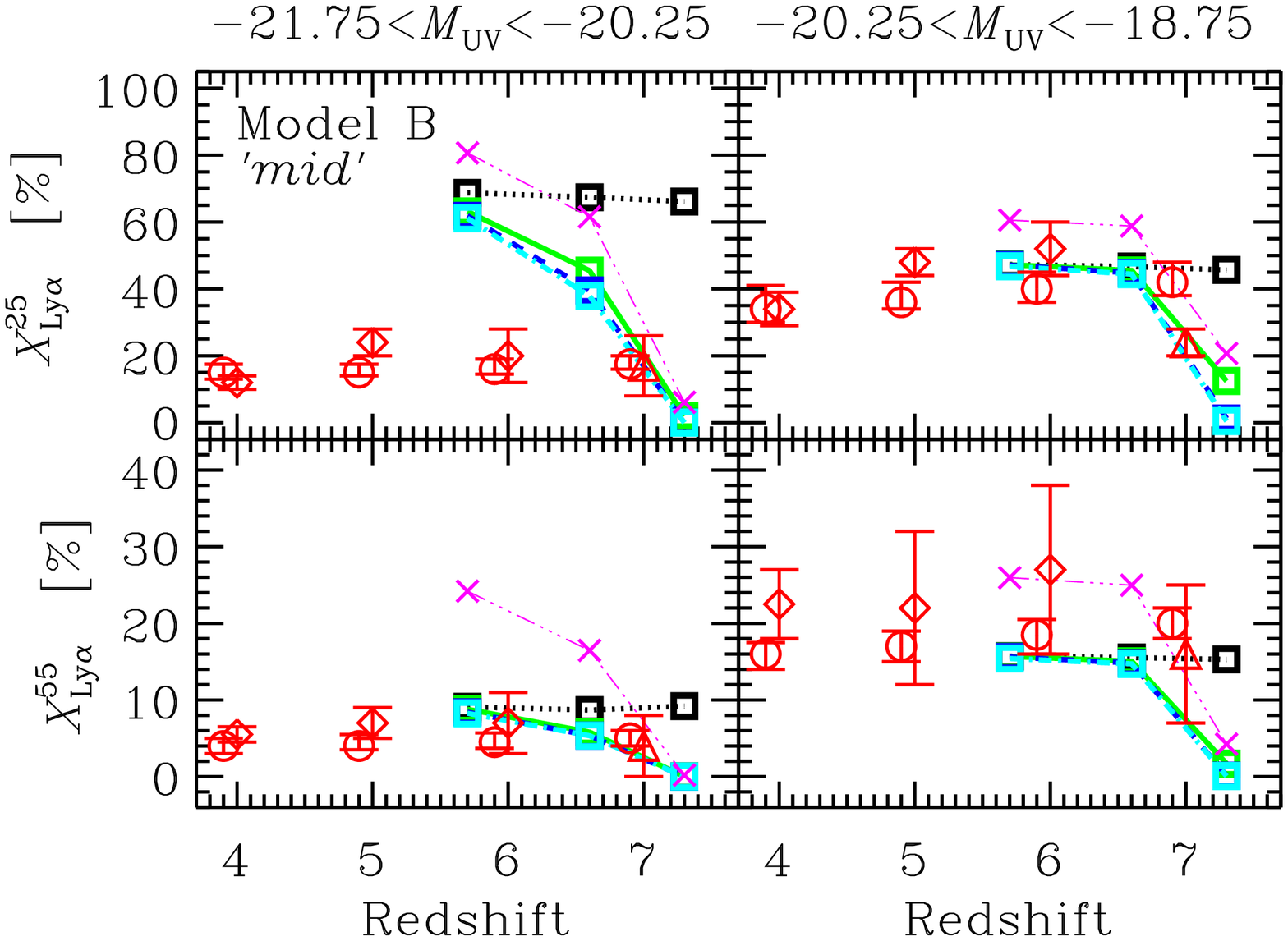}
  \includegraphics[width=7cm]{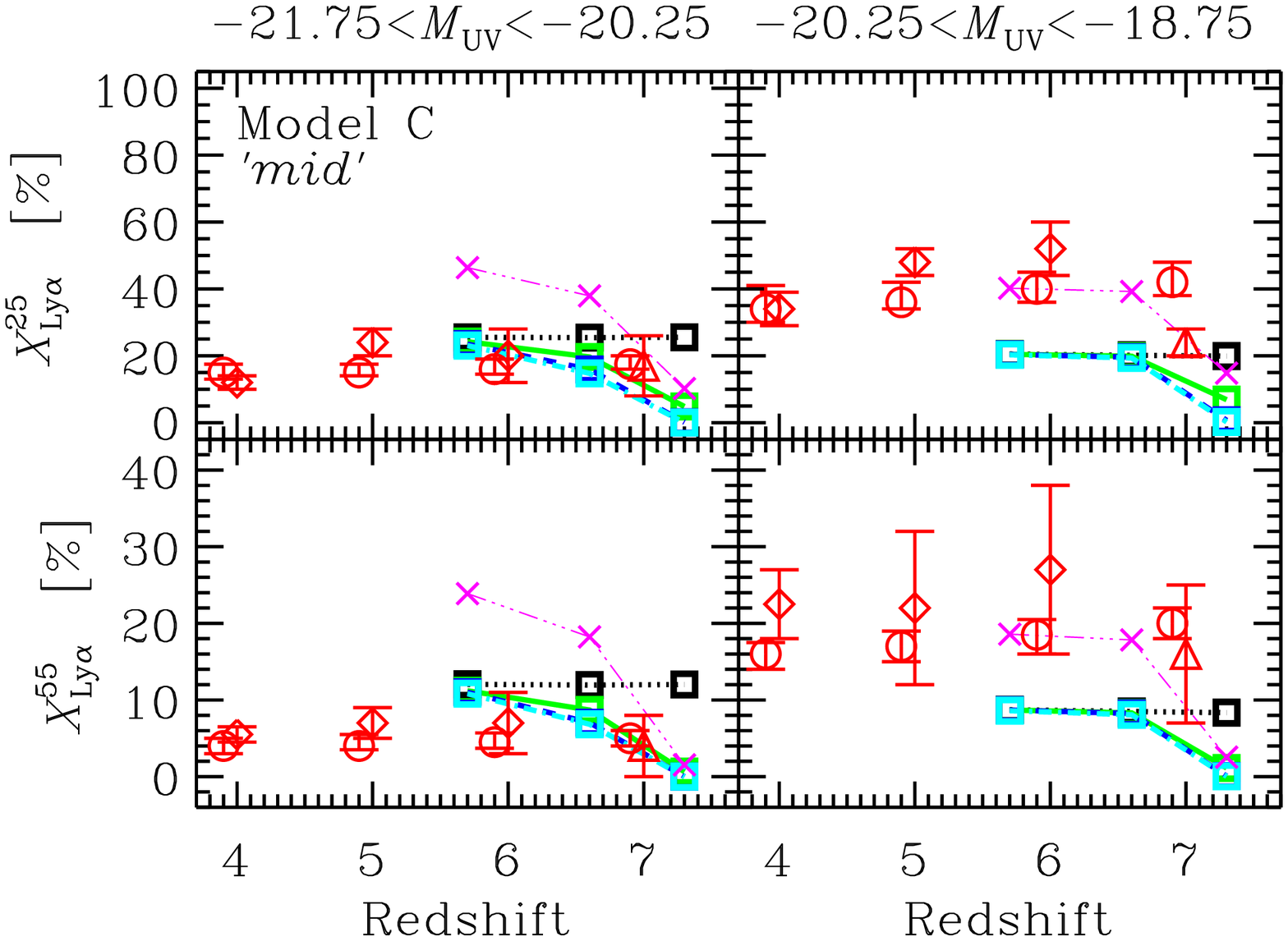}
  \includegraphics[width=7cm]{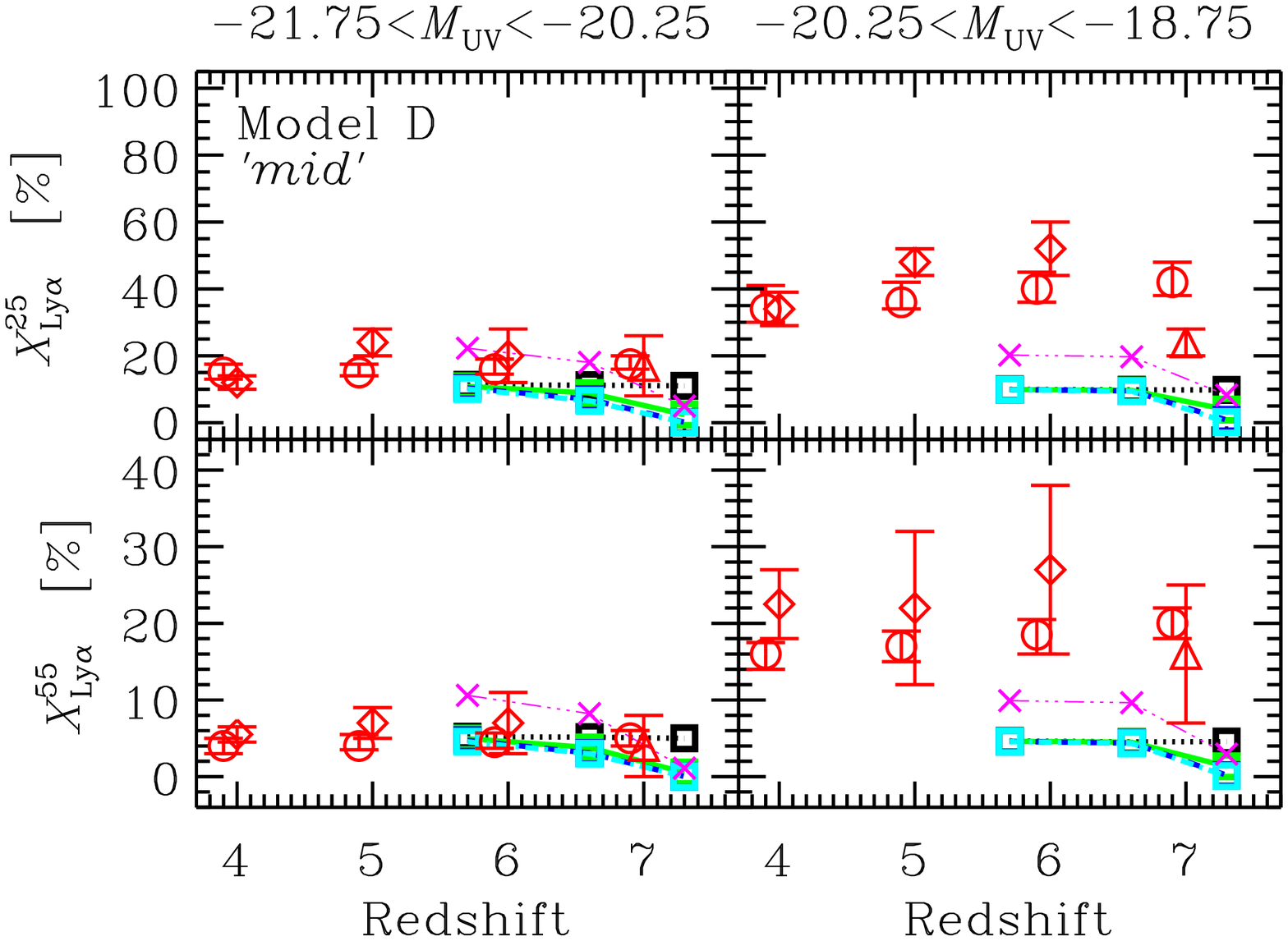}
  \includegraphics[width=7cm]{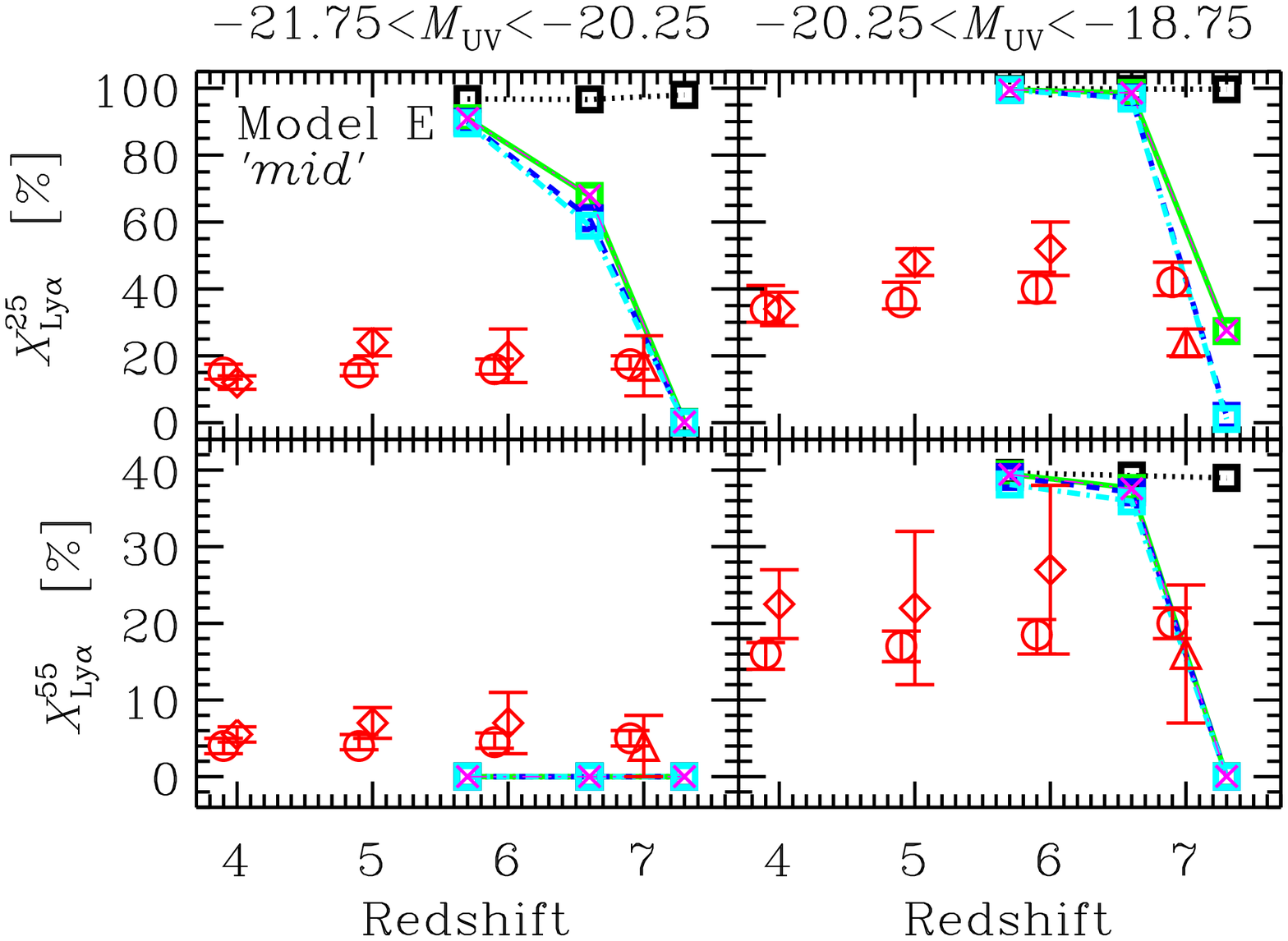}
  \includegraphics[width=7cm]{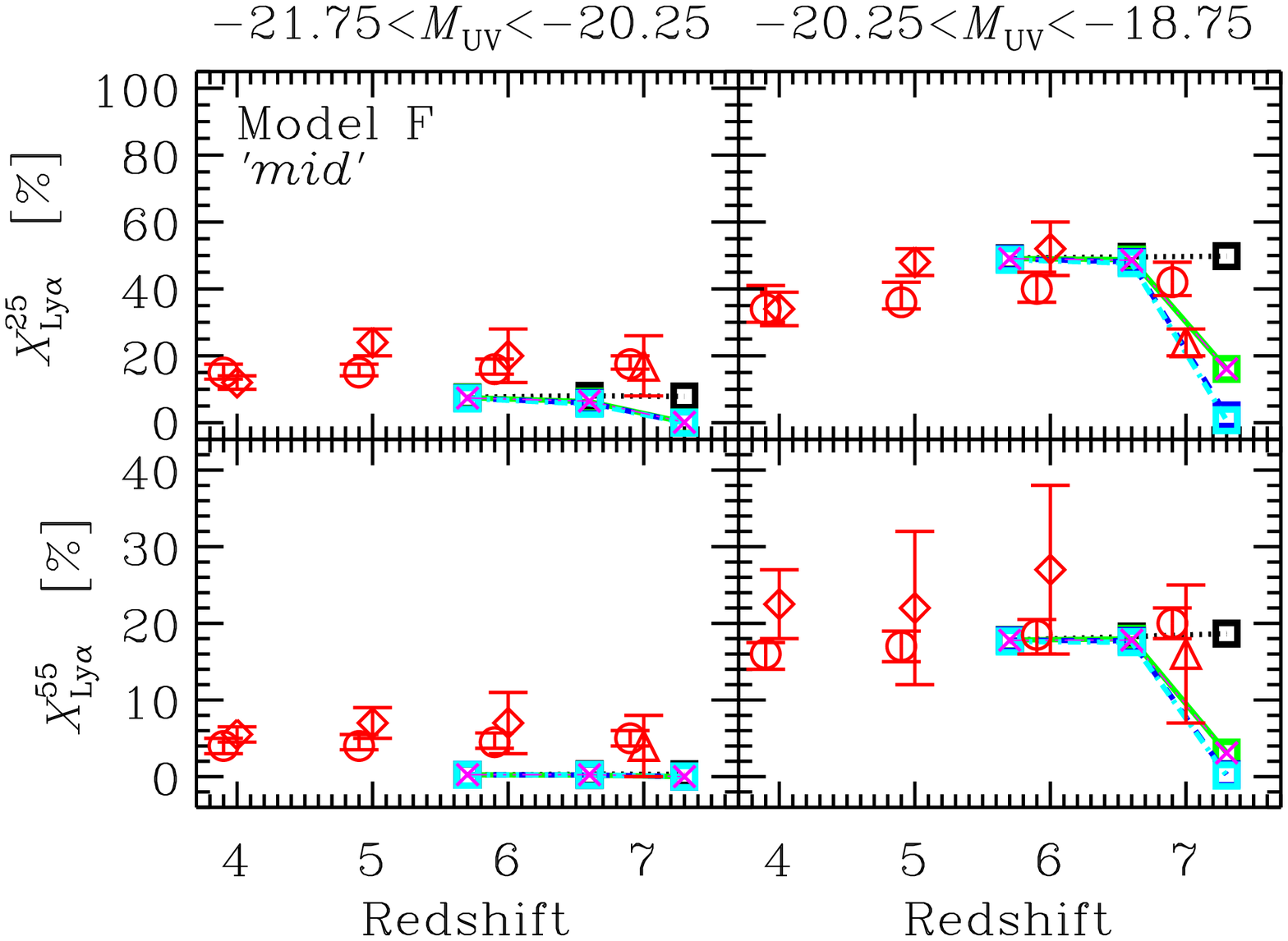}
  \includegraphics[width=7cm]{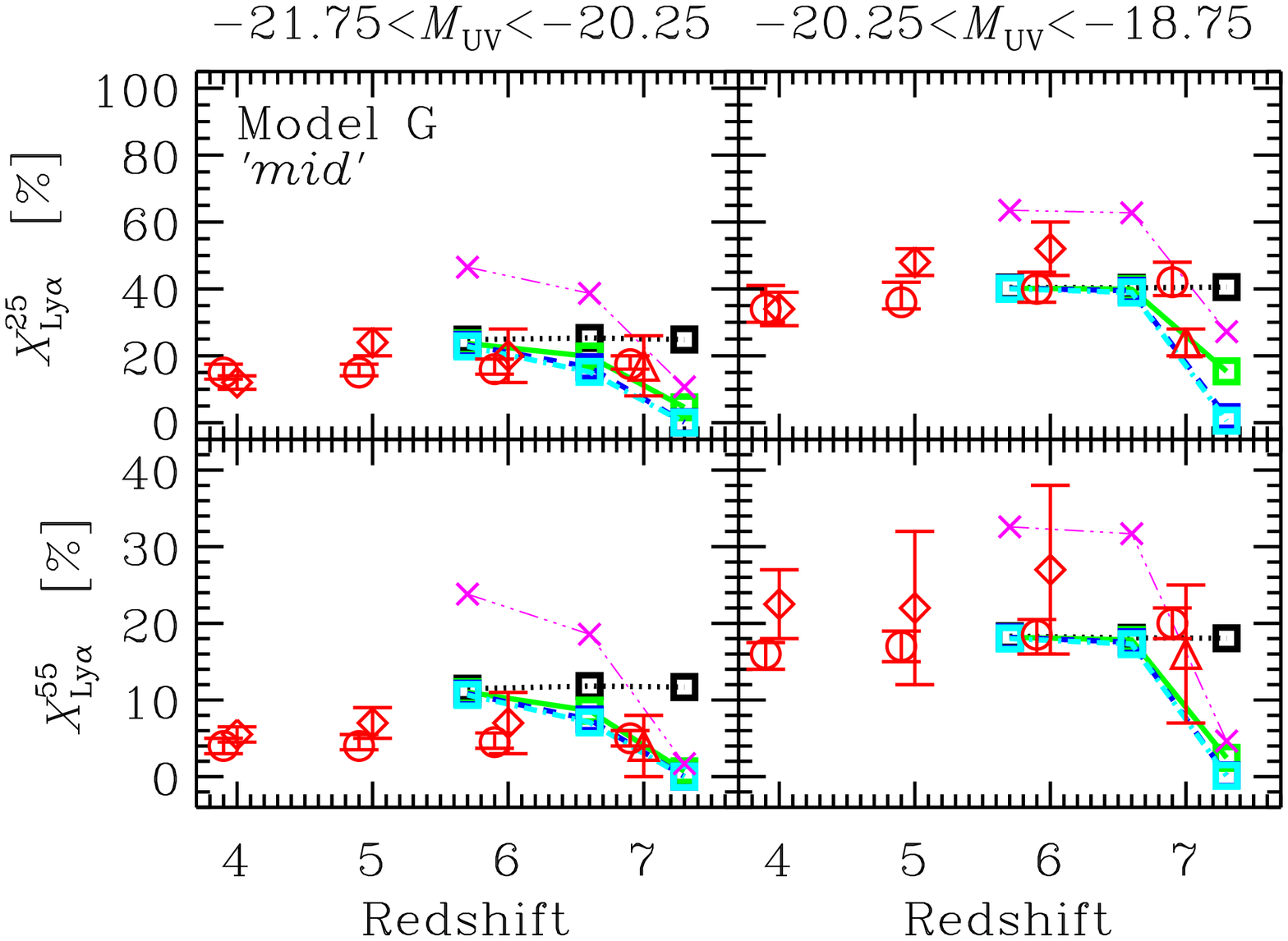}
  \includegraphics[width=7cm]{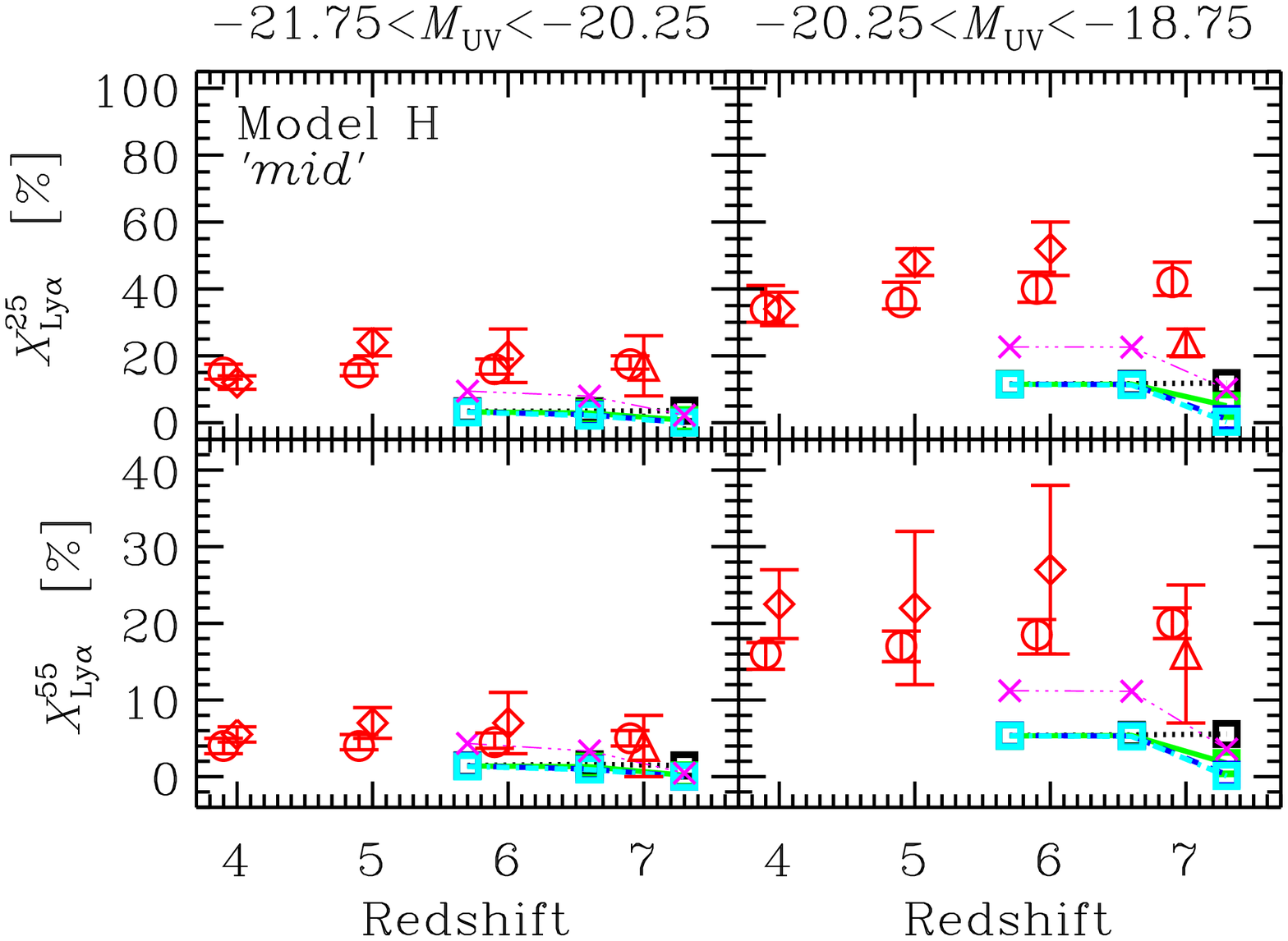}
 \end{center}
 \caption{LAE fractions as a function of redshift for the $mid$ history. 
 Each panel show the number fraction of galaxies having a Ly$\alpha$ 
 equivalent width larger than 25~\AA\ ($X_{\rm Ly\alpha}^{25}$) or 55~\AA\
 ($X_{\rm Ly\alpha}^{55}$) in a sample of galaxies with a UV absolute 
 magnitude in the range noted above the panel.
 The dotted (black) lines are the completely transparent IGM
 cases. The solid (green), dashed (blue), and dot-dashed
 (cyan) lines show the cases with different Ly$\alpha$
 line profiles in a uniform outflowing gas of a velocity of 150 km
 s$^{-1}$ and an H~{\sc i} column density of 
 $\log_{10}(N_{\rm HI}/{\rm cm^{-2}})=20$, 19, and 18, respectively. The
 thin triple-dot-dashed (magenta) lines are the cases fit to Santos et al.~(2016)
 at $z=5.7$ instead of the HSC data of Konno et al.~(2018) and 
 $\log_{10}(N_{\rm HI}/{\rm cm^{-2}})=20$. The diamonds with error-bars
 (red) are direct observations by Stark et al.~(2011). The triangle with
 error-bars (red) is the compilation of direct observations at $z\sim7$ by Ono
 et al.~(2012). The circles with error-bars (red) are indirect estimates for
 the fully-ionized Universe by Oyarz{\'u}n et al.~(2017).}
 \label{LAEfrac_mid}
\end{figure*}

\begin{figure*}
 \begin{center}
  \includegraphics[width=7cm]{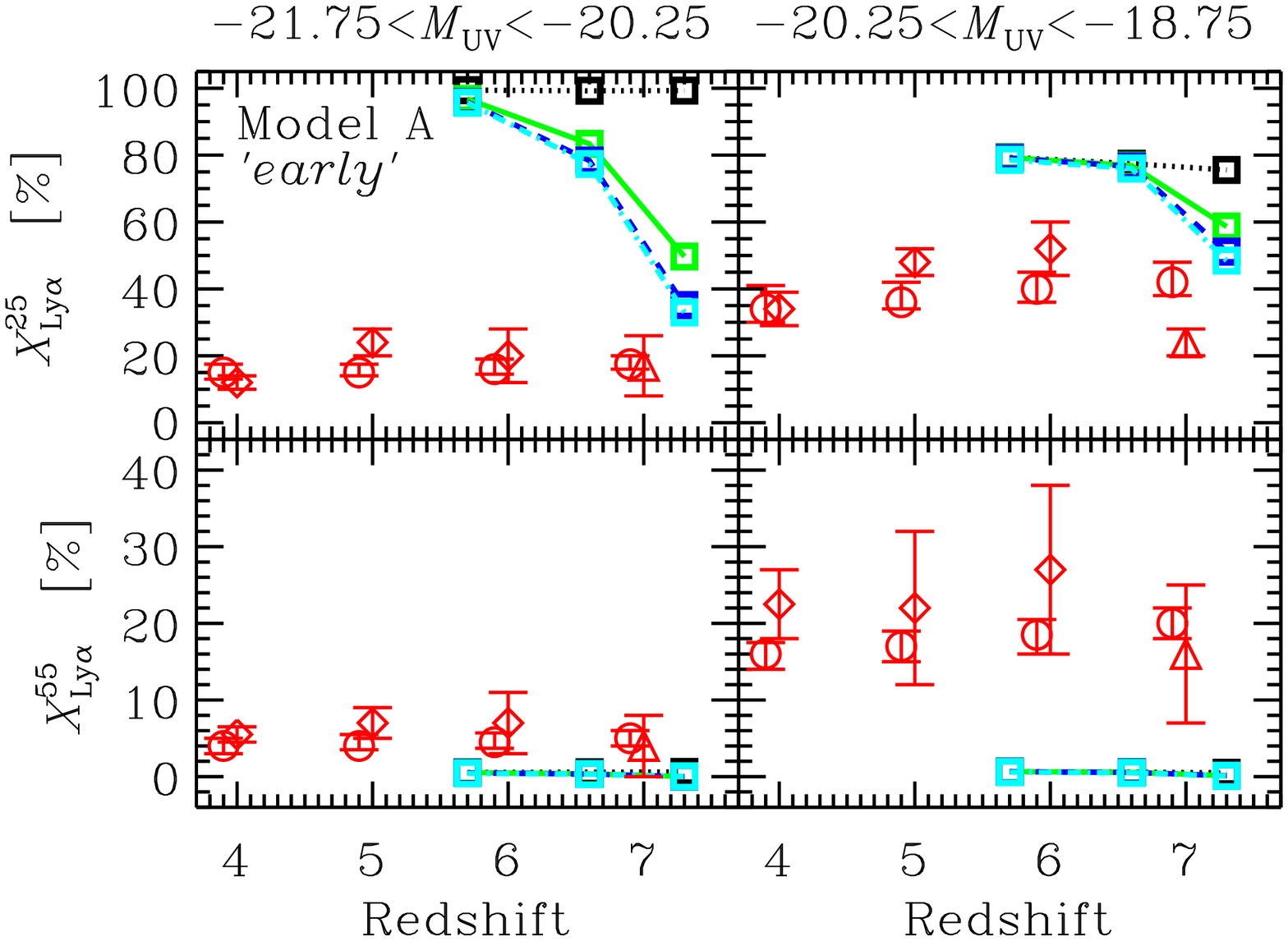}
  \includegraphics[width=7cm]{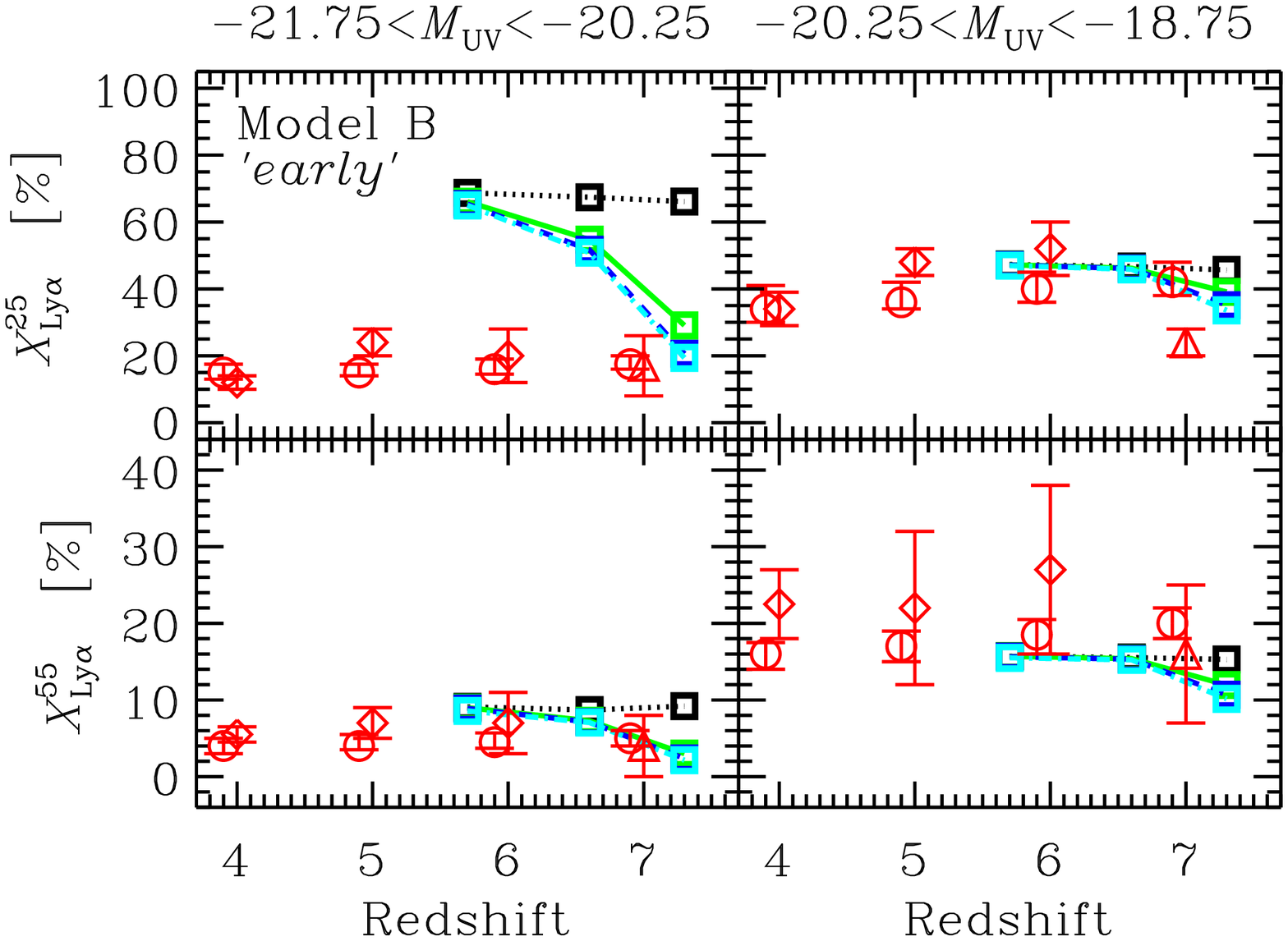}
  \includegraphics[width=7cm]{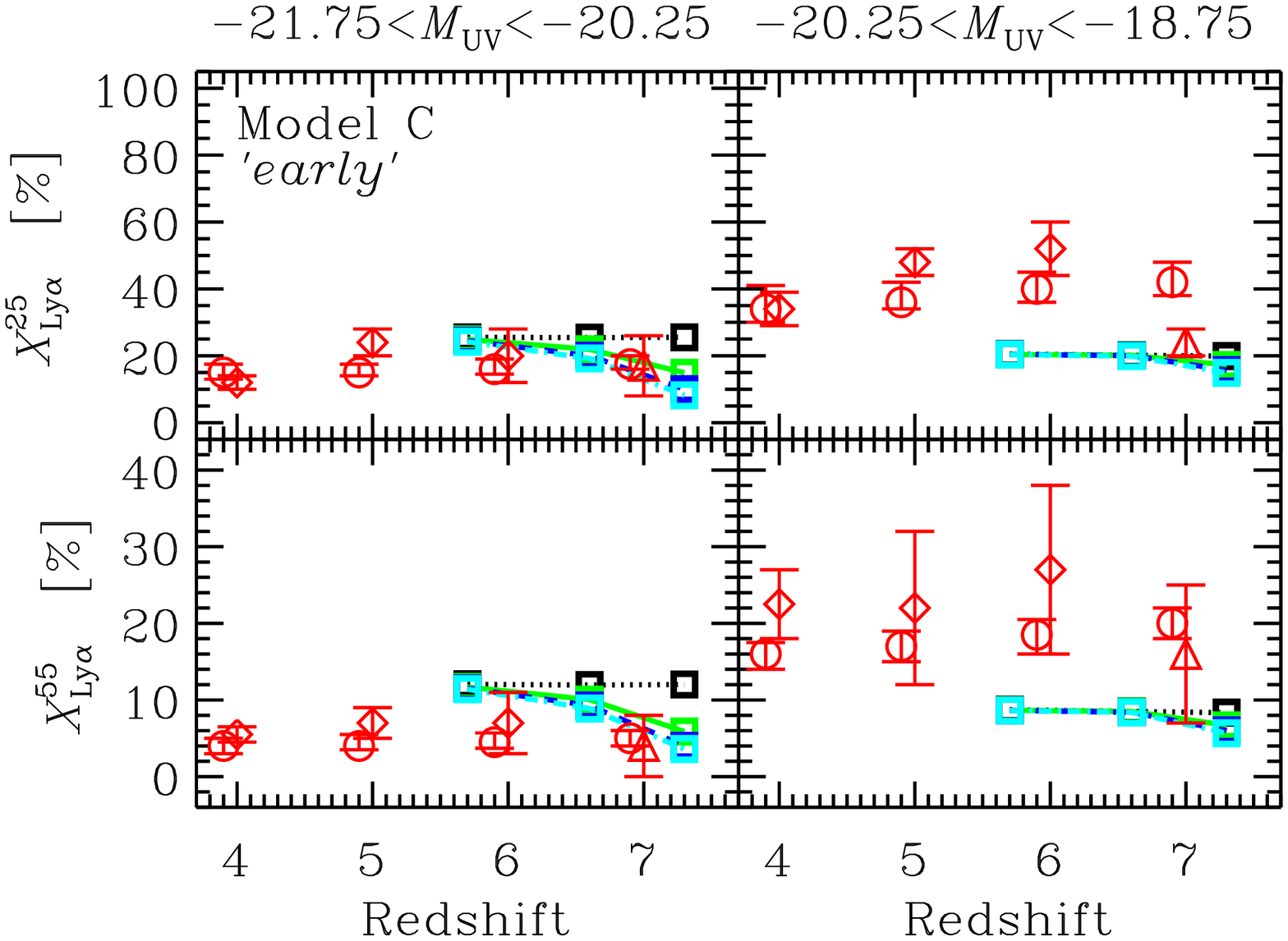}
  \includegraphics[width=7cm]{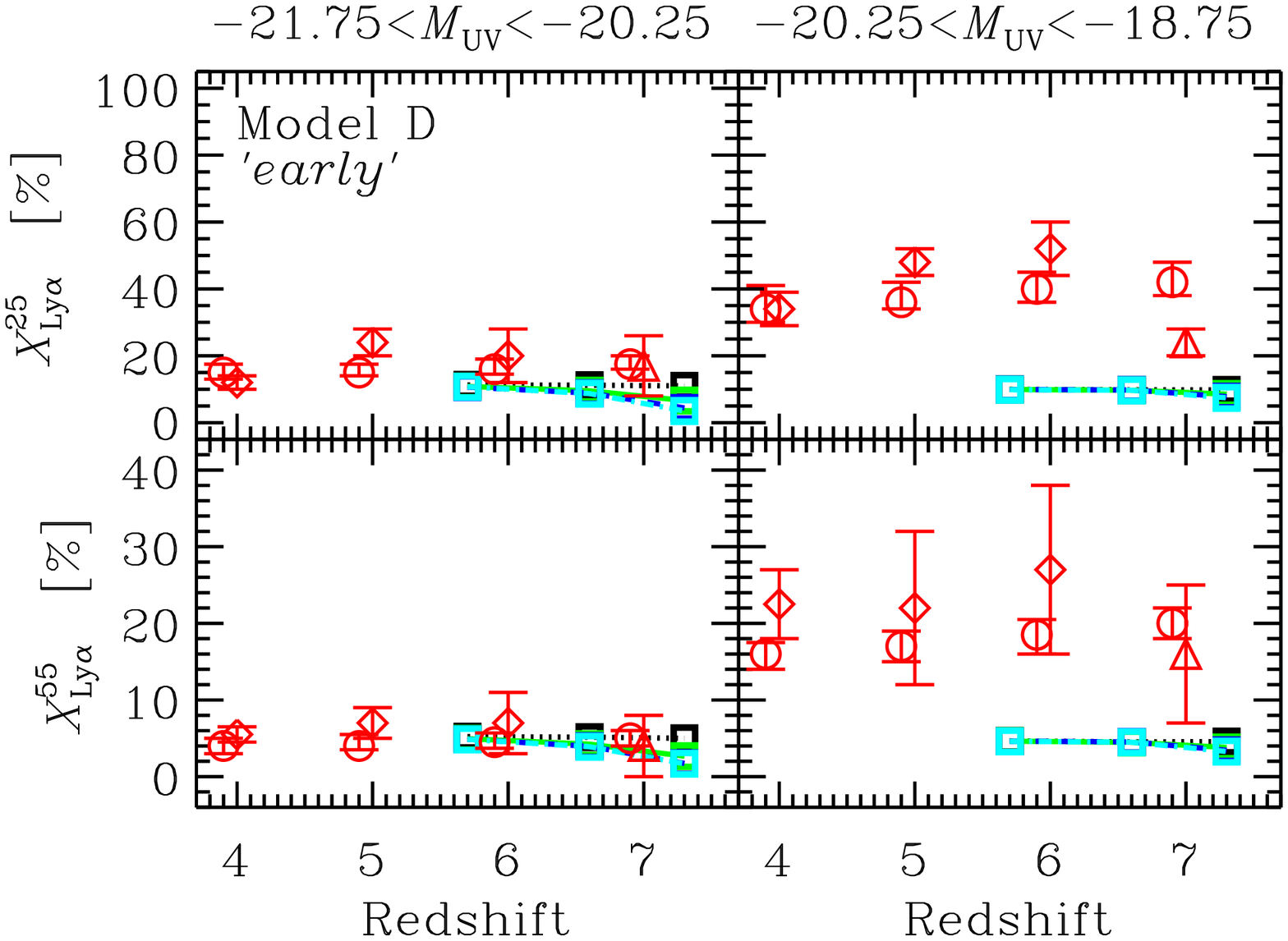}
  \includegraphics[width=7cm]{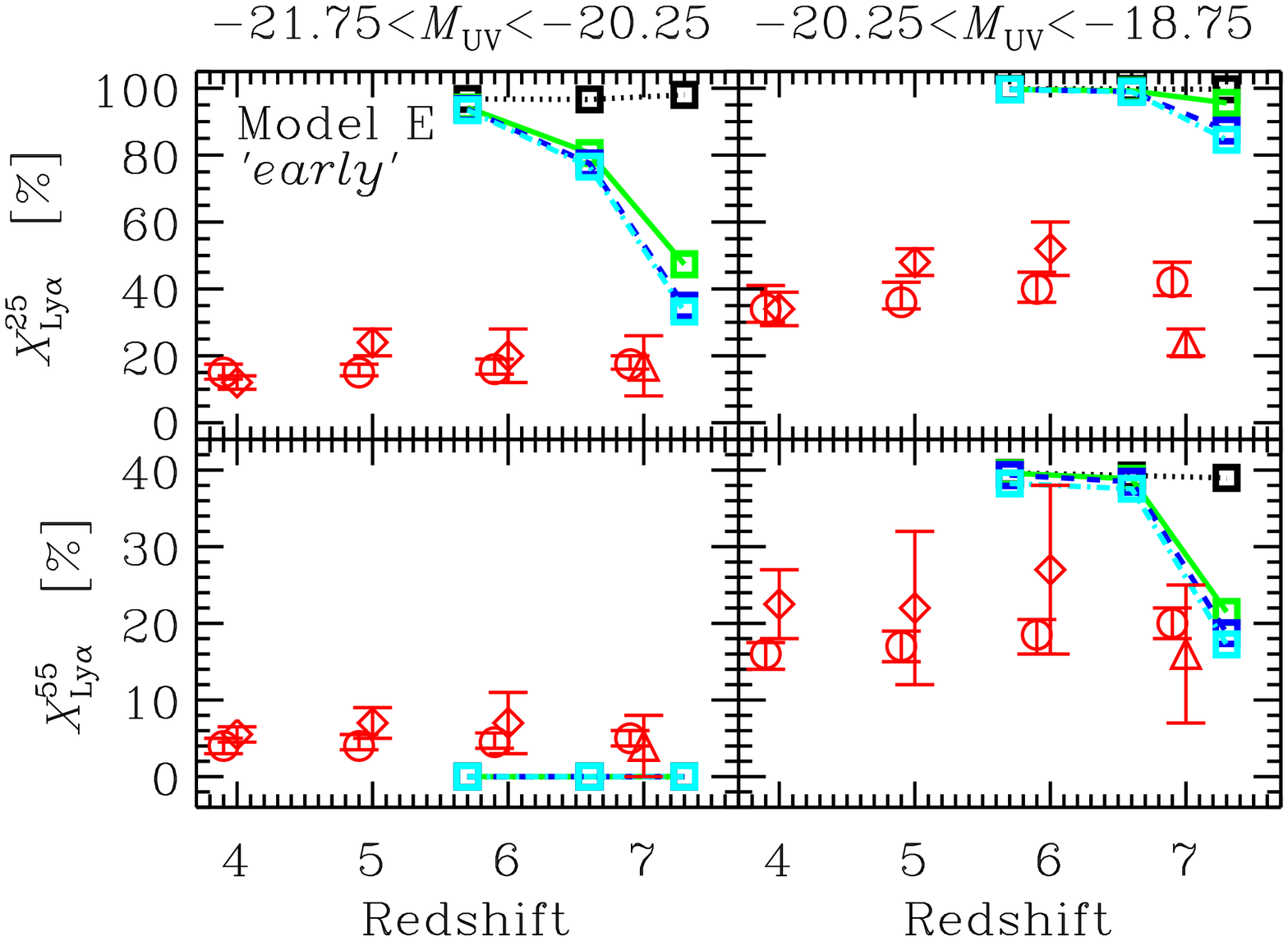}
  \includegraphics[width=7cm]{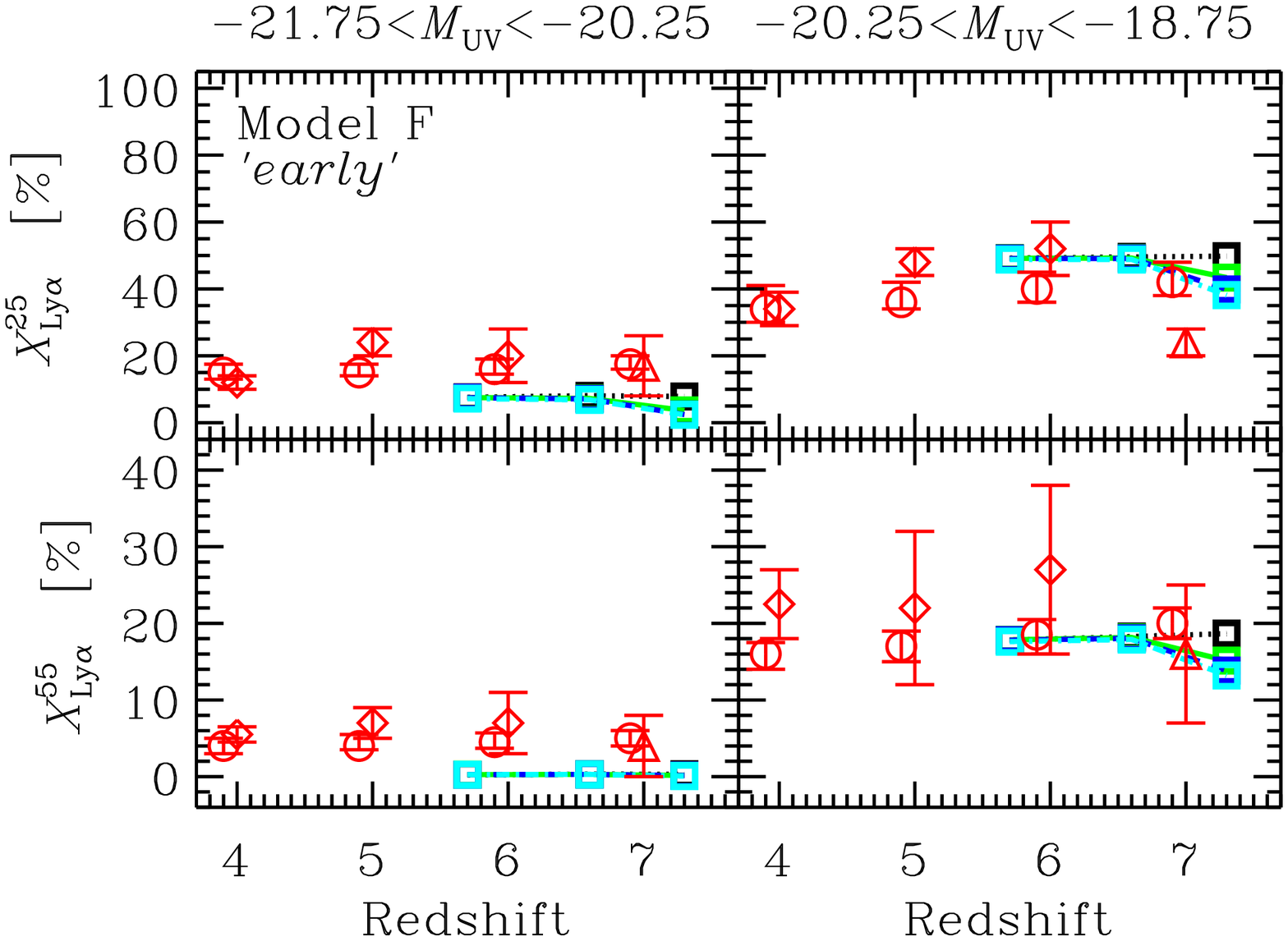}
  \includegraphics[width=7cm]{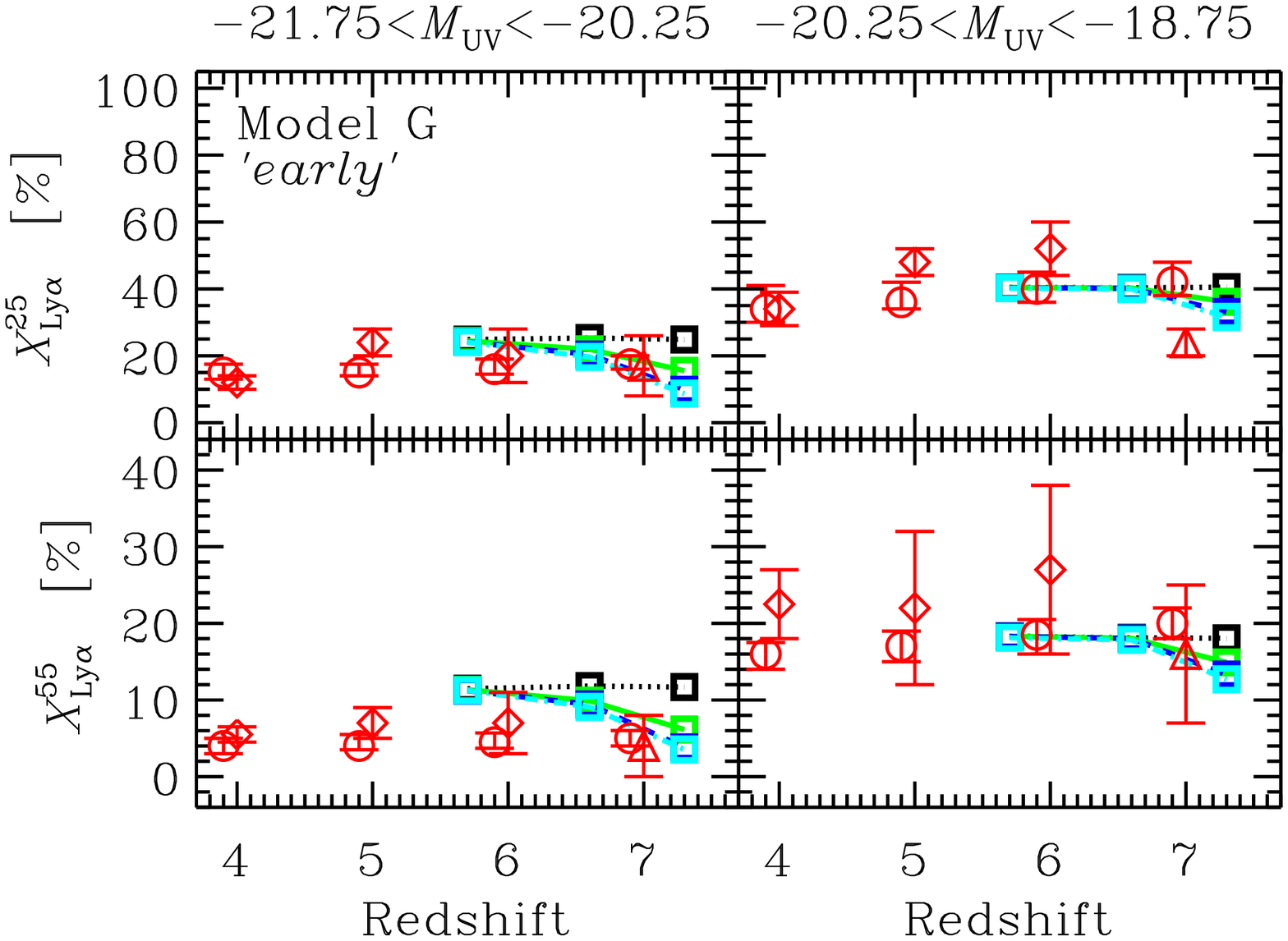}
  \includegraphics[width=7cm]{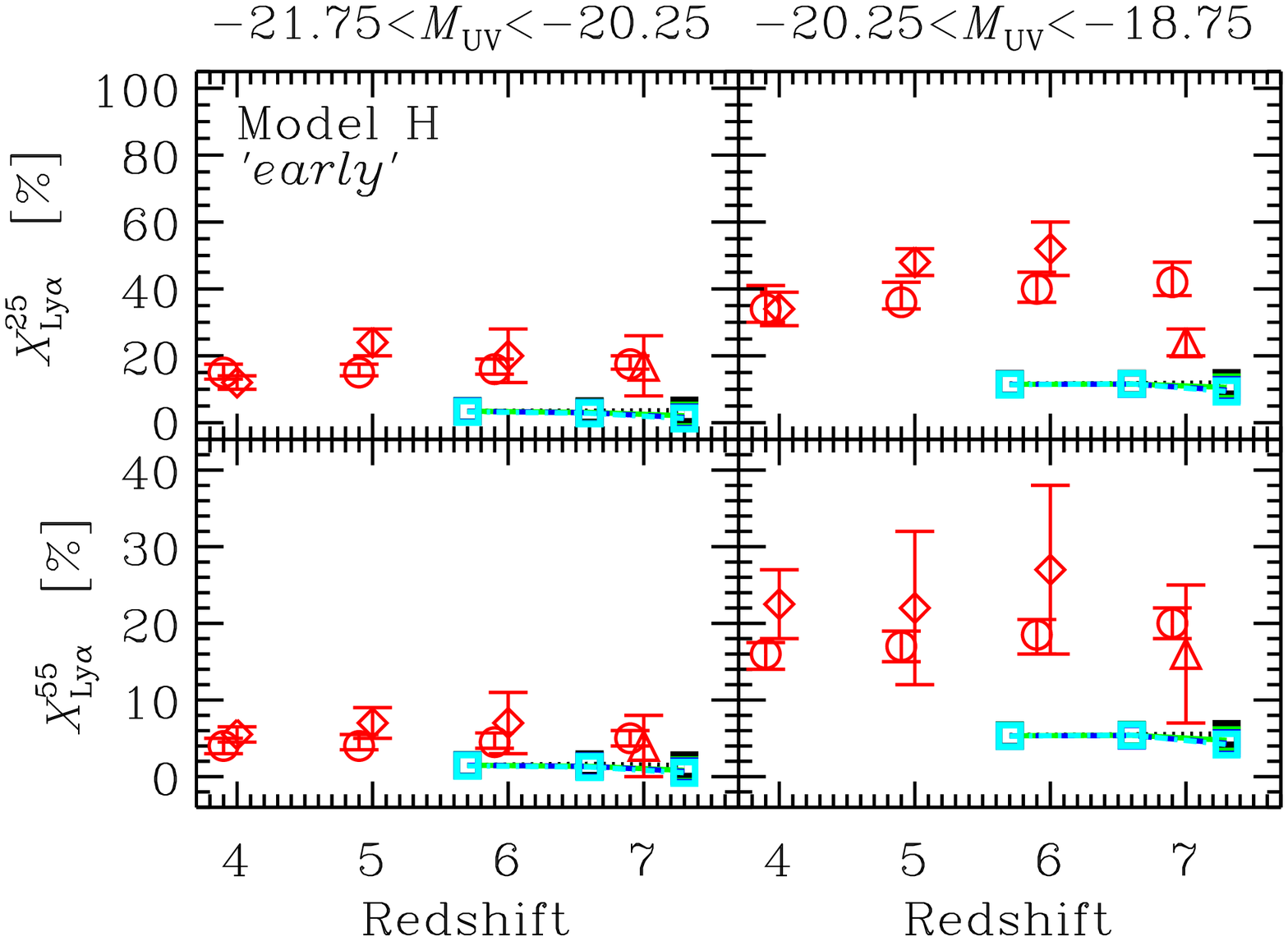}
 \end{center}
 \caption{Same as Fig.\ref{LAEfrac_mid} but for the $early$ history.}
 \label{LAEfrac_early}
\end{figure*}

\begin{figure*}
 \begin{center}
  \includegraphics[width=7cm]{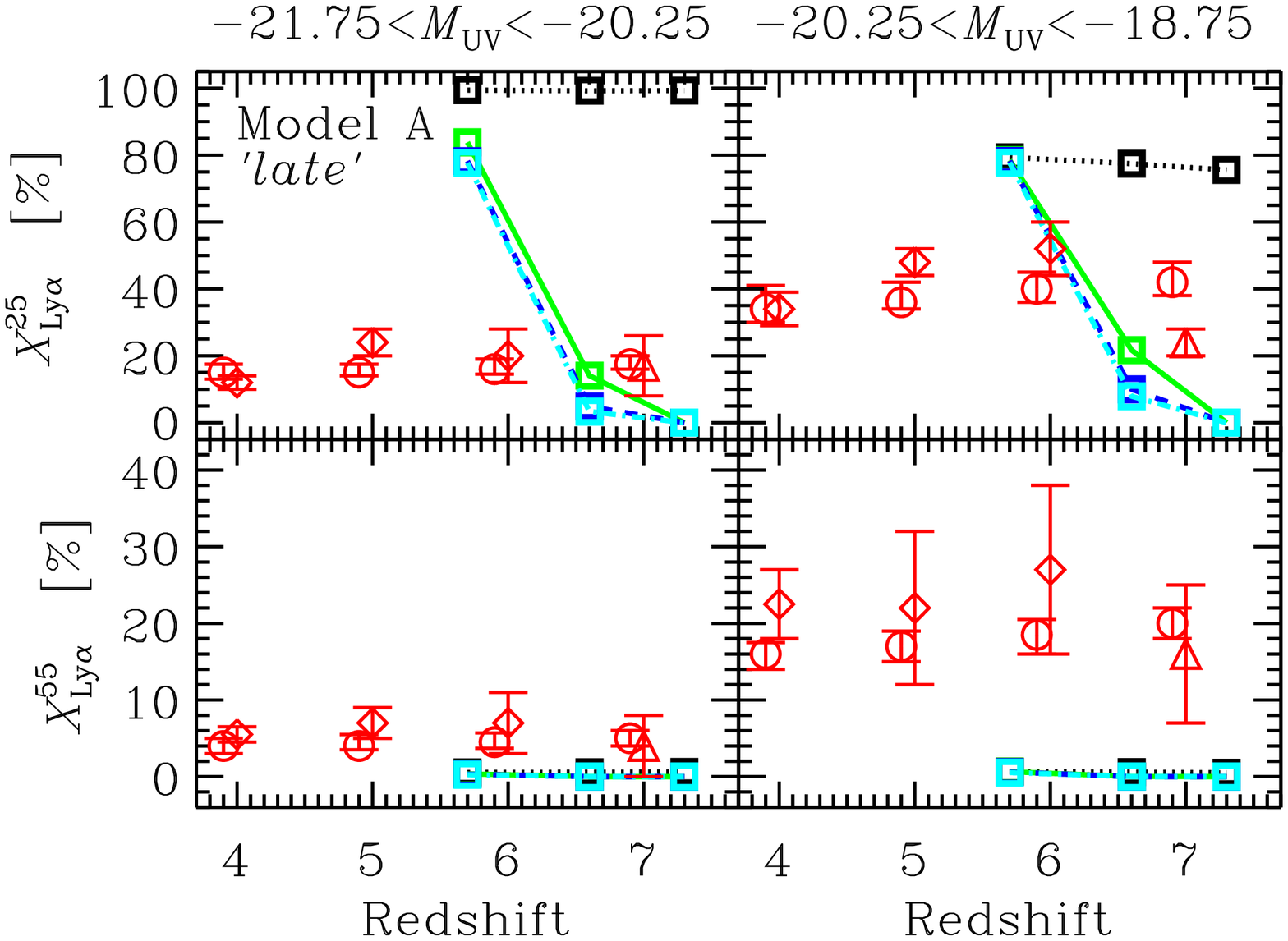}
  \includegraphics[width=7cm]{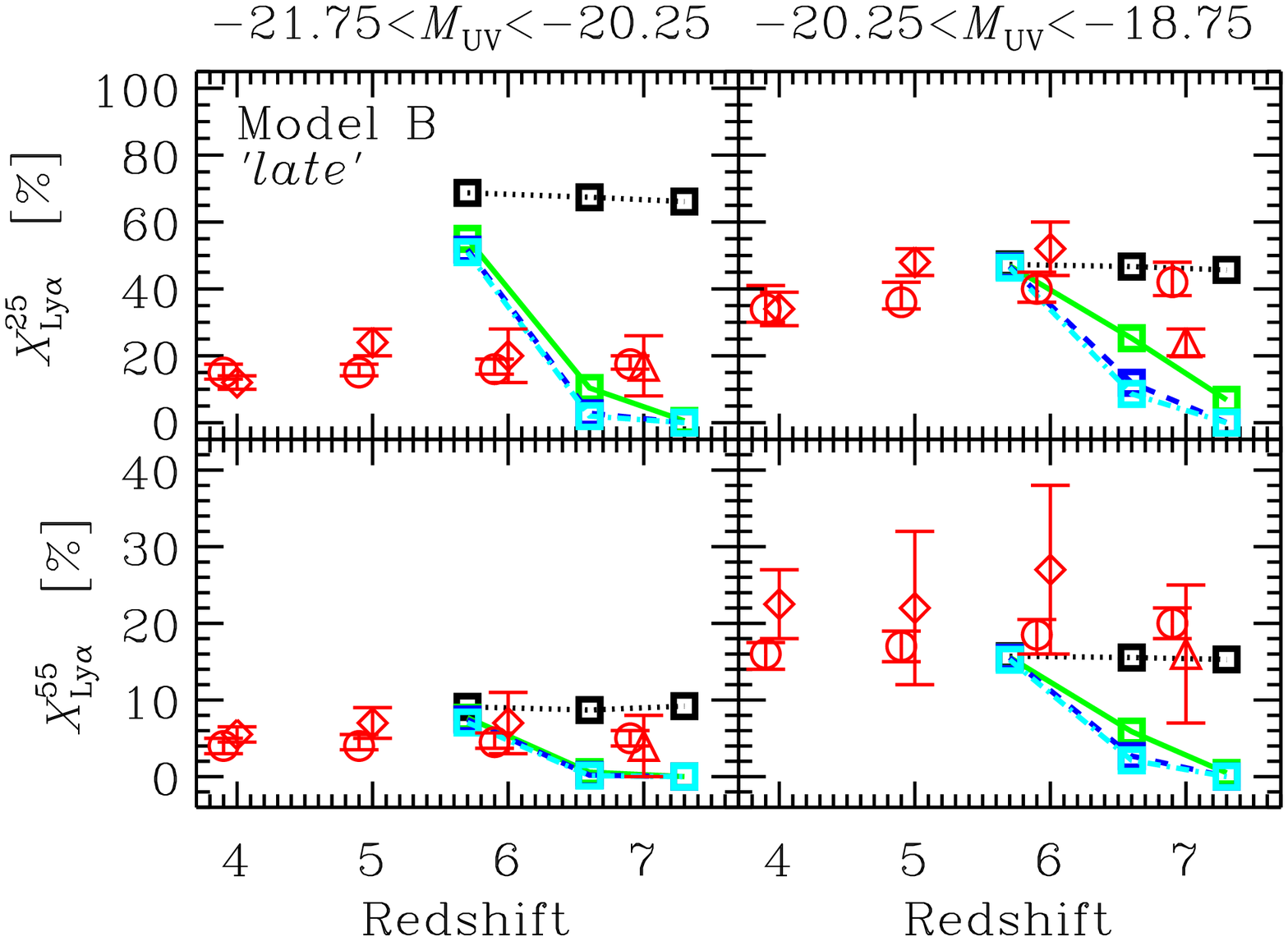}
  \includegraphics[width=7cm]{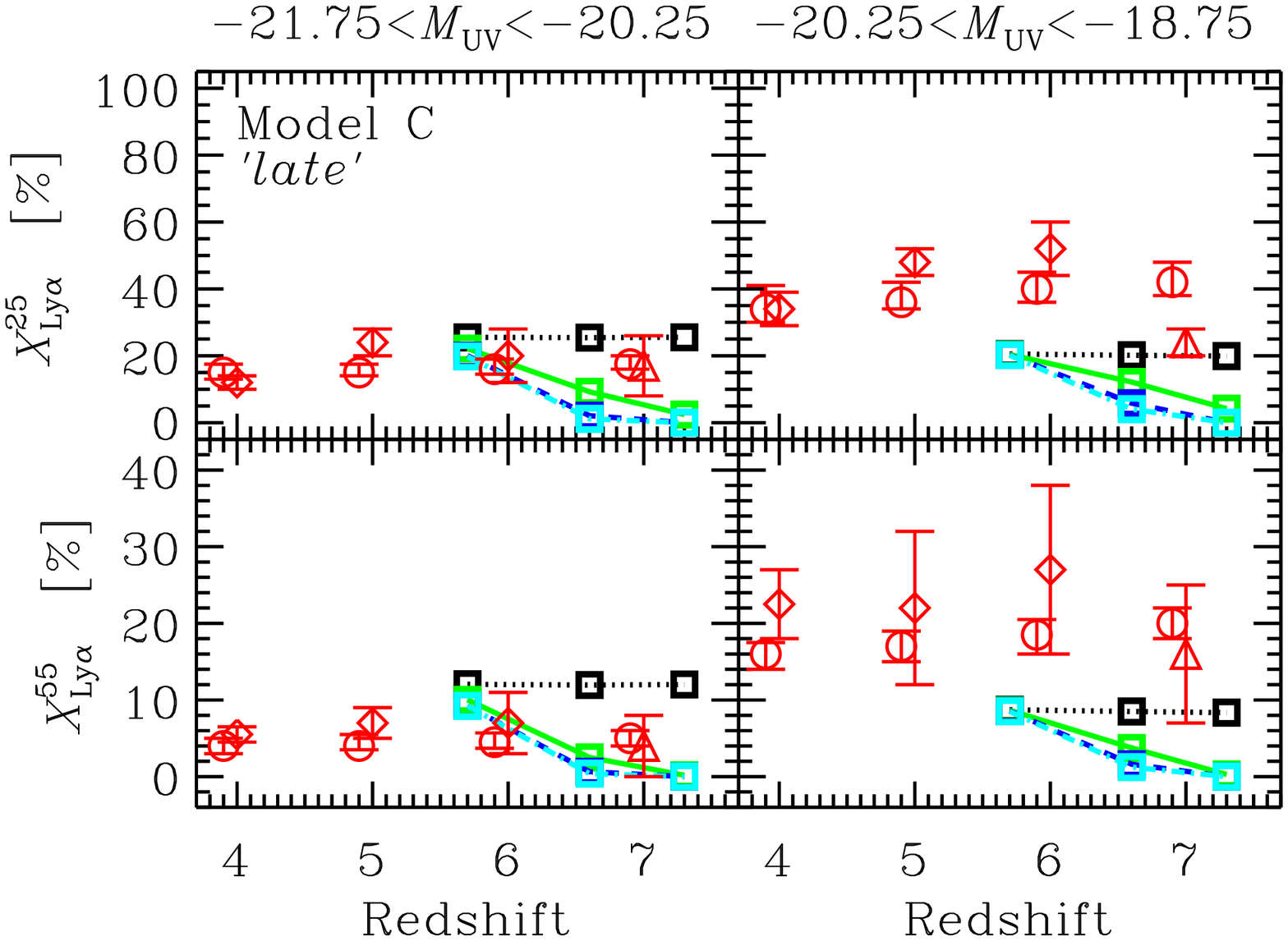}
  \includegraphics[width=7cm]{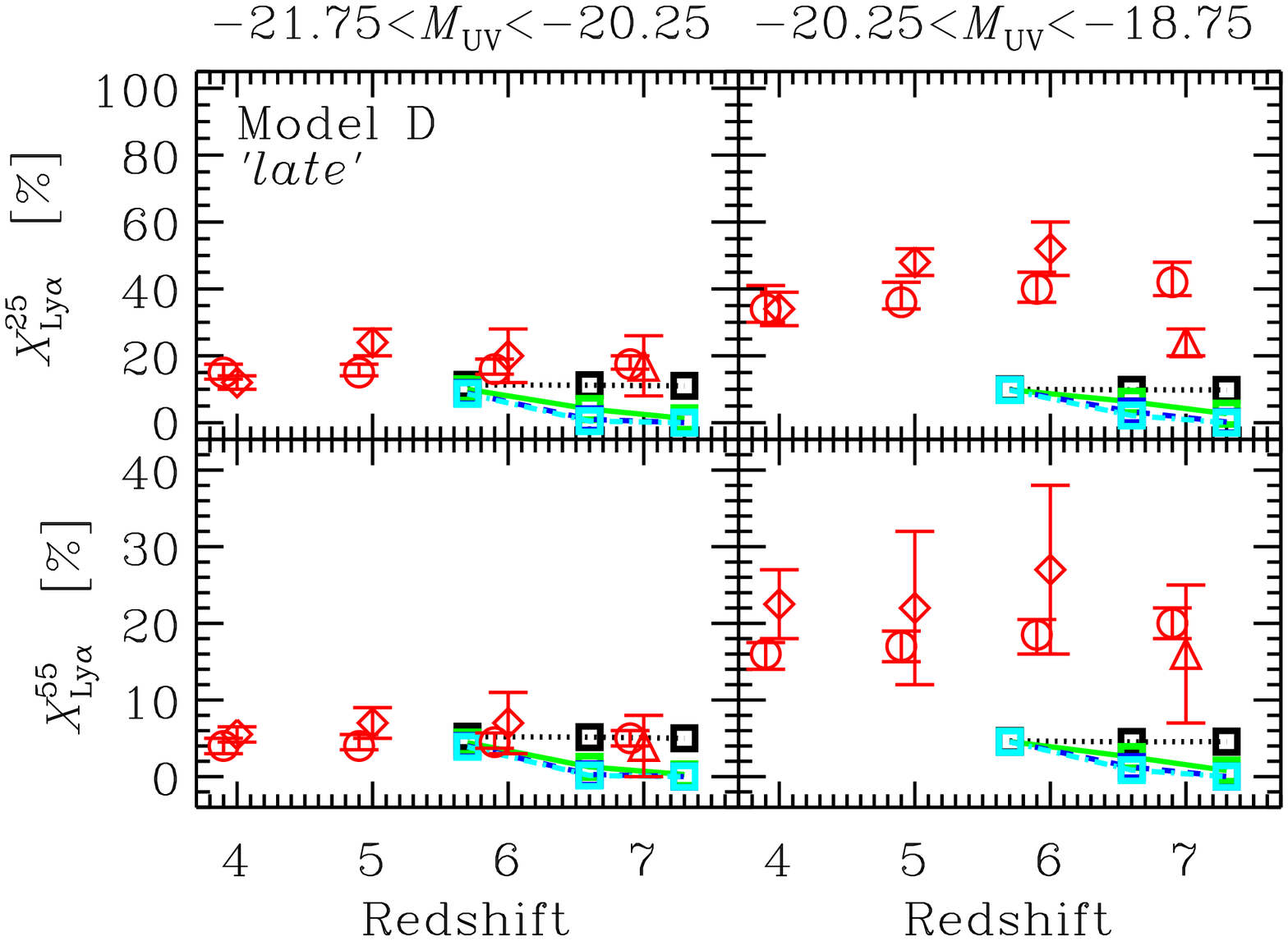}
  \includegraphics[width=7cm]{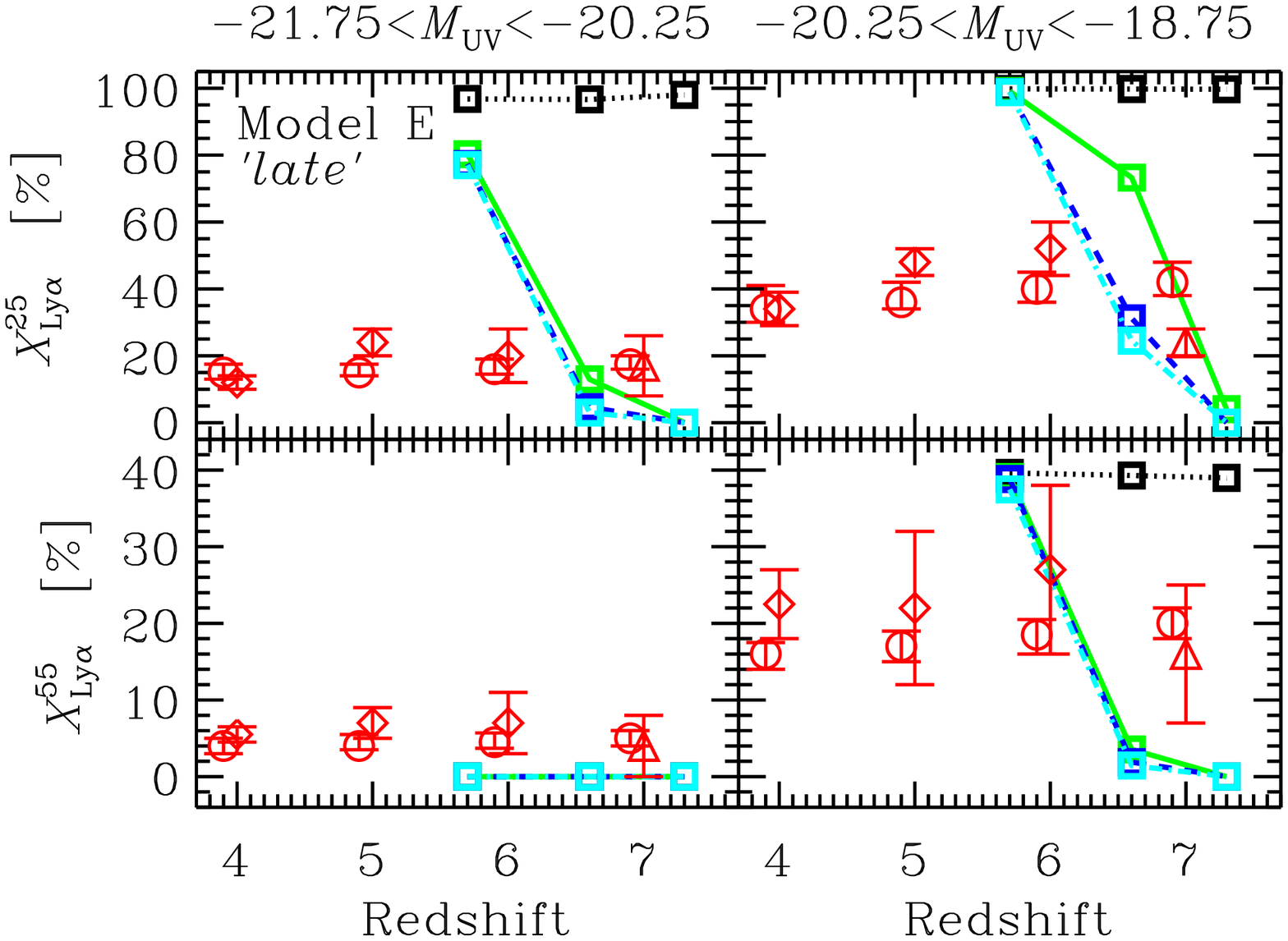}
  \includegraphics[width=7cm]{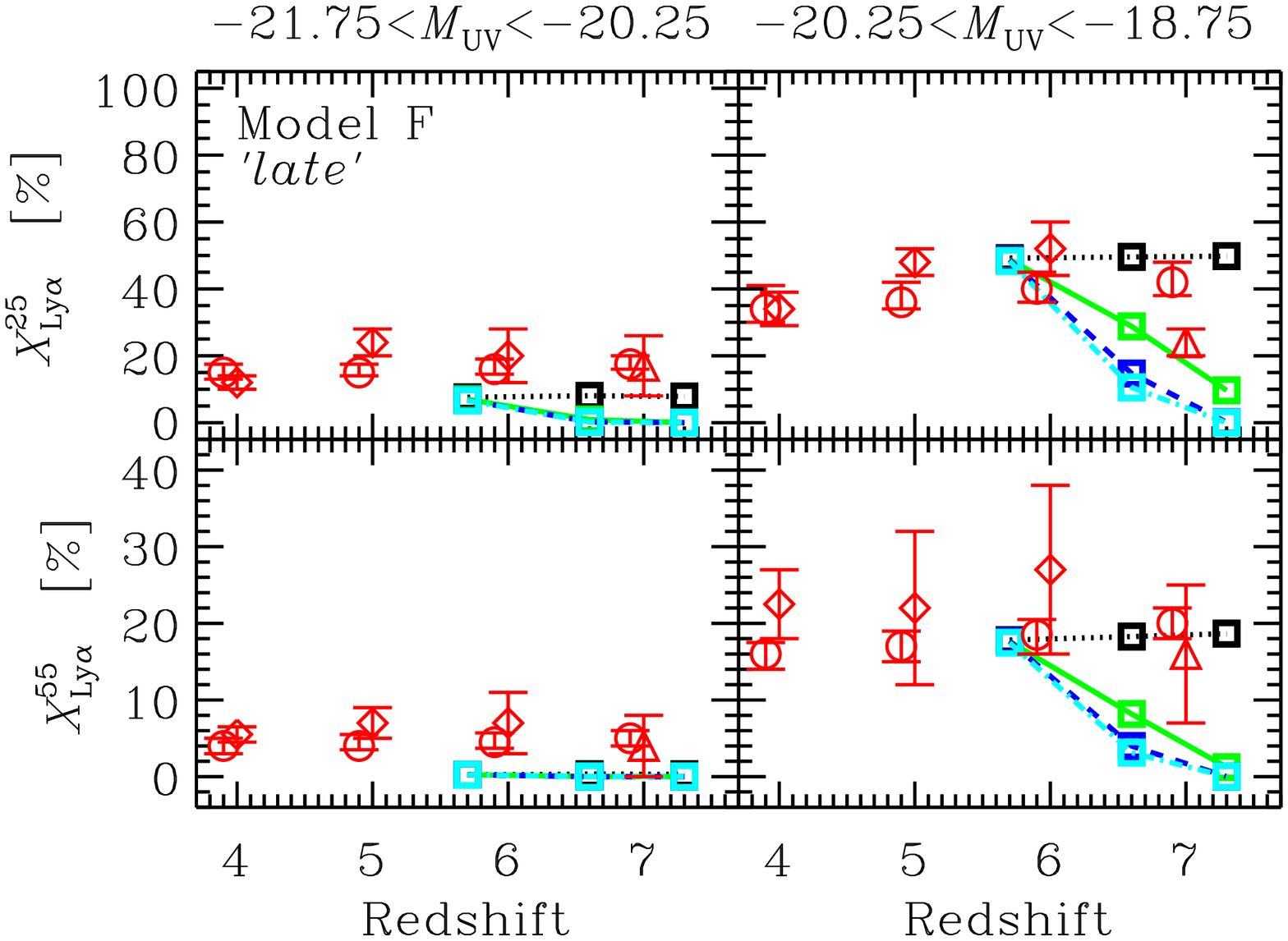}
  \includegraphics[width=7cm]{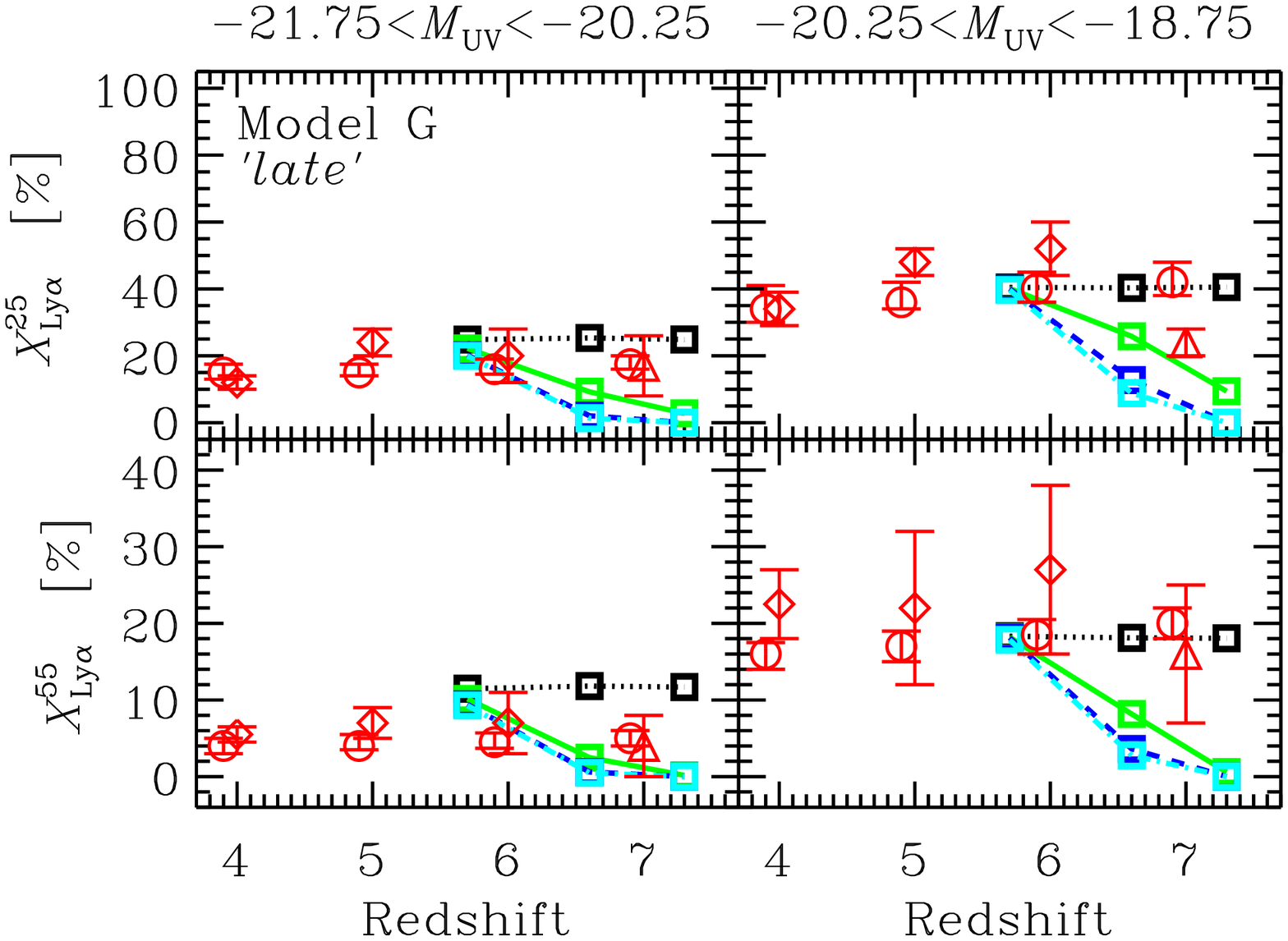}
  \includegraphics[width=7cm]{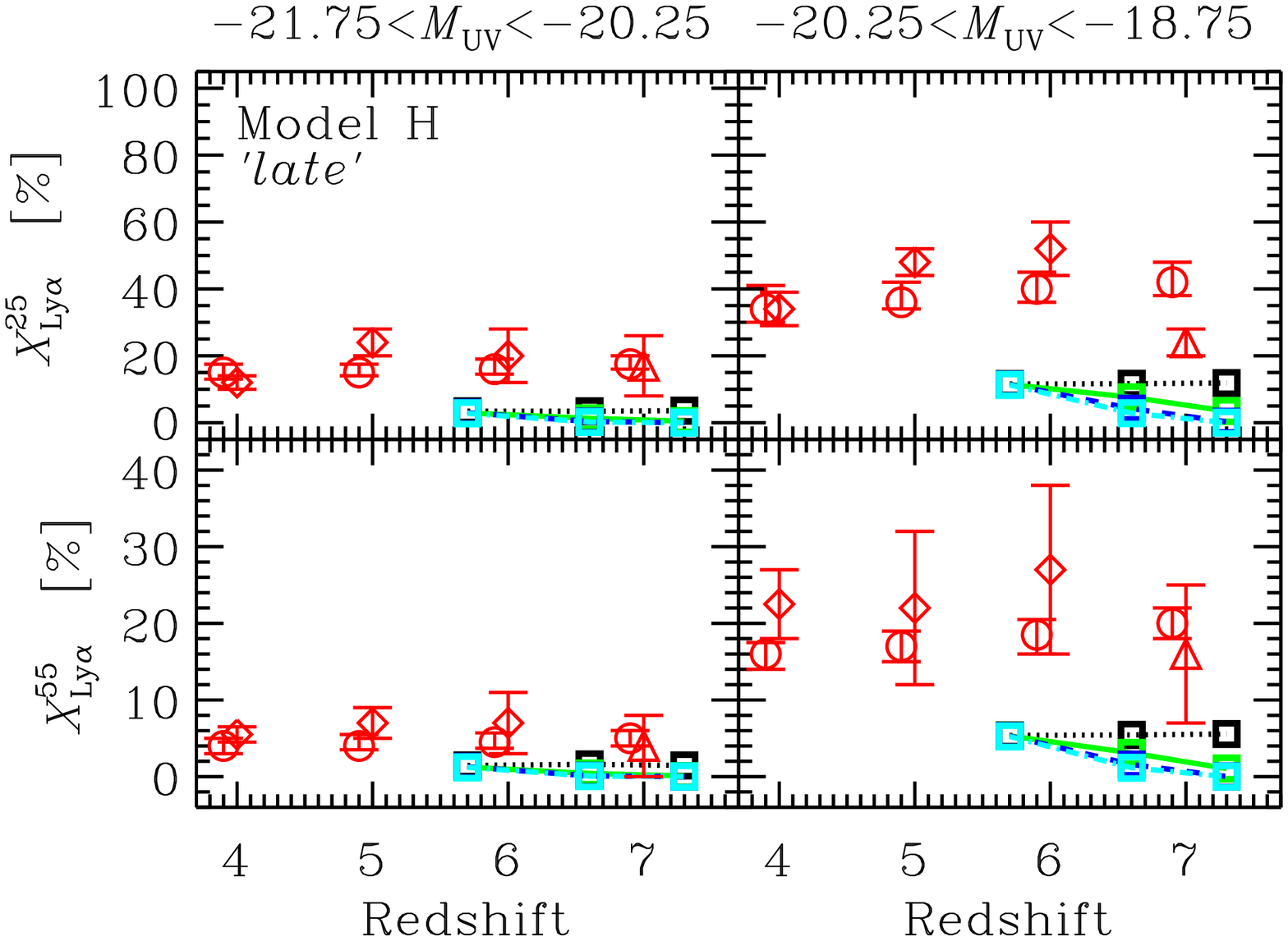}
 \end{center}
 \caption{Same as Fig.\ref{LAEfrac_mid} but for the $late$ history.}
 \label{LAEfrac_late}
\end{figure*}

We compare the observed LAE fractions in LBGs with those predicted by
our models. Since the observational LAE fractions were usually obtained
from spectroscopy, their EWs can be regarded as true ones
(i.e. not-estimated by NB photometry). Therefore, we will use the true
EWs in our models. However, there might be systematic bias in the
observed data because only a small number of $z\gtrsim6$ LBGs has been
targeted and the Ly$\alpha$ fraction is sensitive to the inhomogeneity
of the spectroscopic detection limit (e.g., \cite{Ono2012}).

Figs.~\ref{LAEfrac_mid}, \ref{LAEfrac_early} and \ref{LAEfrac_late}
show the comparisons of the observed LAE fraction in LBGs with the
models for the $mid$, $early$ and $late$ reionization histories,
respectively. For each model, there are four sets: the smaller or 
larger threshold of the Ly$\alpha$ equivalent width (EW) and the 
brighter or fainter UV luminosity range. We have found that this 
comparison is the most critical one. Only Model~G shows a reasonable 
agreement with all four sets. Model~B is the second best because it 
reproduces the three sets but it fails in the lower EW and the 
brighter UV. Models~C and D reproduce the two brighter UV sets.
However, they significantly underpredict the LAE fraction in the 
fainter UV galaxies because there is no halo mass (or UV luminosity 
in our model) dependence of the Ly$\alpha$ optical depth which is a key
ingredient to reproduce a higher LAE fraction in a fainter galaxy 
sample. Models~E and F can reproduce one or two sets in the fainter 
sample but not in the brighter sample. Models~A and H fail in all 
four sets.

It is interesting to look into the reason why Models~A and H fail.
Model~A has no stochastic process in Ly$\alpha$ production and
Ly$\alpha$ transmission in a halo but does have in UV luminosity and
slope. Neglecting the latter effect, we obtain the rest-frame EW 
$\approx30$~\AA~$M_{\rm h,10}^{0.1}T_\alpha^{\rm IGM}$ for 
$\beta=-2$ and $f_{\rm esc,\alpha}^{\rm ISM}=0.2$ with 
$\tau_\alpha=1.6$. As a result, $\sim100$\% UV bright galaxies 
satisfy EW$>25$~\AA\ in an ionized Universe (i.e. 
$T_\alpha^{\rm IGM}\sim1$), but there is no LAEs with EW$>55$~\AA.
The halos with $M_{\rm h,10}\lesssim10$ corresponding to 
$M_{\rm UV}\gtrsim-20$ are affected by the stochastic processes in 
UV luminosity and slope and the EWs fluctuate, resulting in an  
LAE fraction $<100\%$ in the UV fainter samples. 
Model~H, on the other hand, has all stochastic processes. Due to 
the fluctuation in Ly$\alpha$ production, less massive halos have 
a significant chance to get a bright Ly$\alpha$ luminosity. Since 
they are numerous, a larger pivot value of the Ly$\alpha$ optical depth
in a halo is required to keep the Ly$\alpha$ LF consistent with the 
observations. Model~H also has a halo mass dependency in the Ly$\alpha$ 
optical depth, extinguishing Ly$\alpha$ in more massive halos. 
The halo mass is directly connected to the UV luminosity in our 
modeling. As a result, a dearth of LAEs in massive halos bright 
in UV emerges, which is inconsistent with the observations.

The effect of the reionization (or IGM neutrality) can be found 
in Figs.~\ref{LAEfrac_mid}, \ref{LAEfrac_early} and \ref{LAEfrac_late}. 
For example, in the $mid$ history (Fig.~\ref{LAEfrac_mid}), 
we can find significant decrements of the LAE fraction at $z=7.3$ 
where $x_{\rm HI}\sim0.5$ (see Fig.~1) and even at $z=6.6$ 
($x_{\rm HI}\sim0.01$) in some cases. In the $late$ history 
(Fig.~\ref{LAEfrac_late}), the LAE fraction decrements are significant 
already at $z=6.6$ ($x_{\rm HI}\sim0.5$). Even in the $early$ history 
(Fig.~\ref{LAEfrac_early}), the decrements can be found at $z=7.3$ 
($x_{\rm HI}\sim0.2$) and also at $z=6.6$ ($x_{\rm HI}\sim0.001$) 
in few cases. Therefore, the reionization signature is indeed imprinted 
in the redshift evolution of the LAE fraction 
\citep{Stark2010,Stark2011,Ono2012} 
and the decrements become significant when $x_{\rm HI}\gtrsim0.1$.
This point will be discussed again in \S5.2.

In Fig.~\ref{LAEfrac_mid}, we also show the LAE fractions in the 
LF calibration with \citet{Santos2016}. Unlike the calibration
with \citet{Konno2017}, we can not find any model reproducing all 
four sets of the LAE fraction evolution. In the \citet{Santos2016} 
calibration, the LAE fractions are often overestimated, 
indicating possible overestimation of their LF.

\section{Discussion}

From the comparisons in the previous section, we have identified 
Model~G as the best model in this paper (Table~1). 
In this section, we will discuss the physical properties of the LAEs in
Model~G (\S5.1) and implications how to discuss cosmic reionization with
LAEs (\S5.2).

\subsection{Nature of LAEs in the reionization epoch}

\begin{figure*}
 \begin{center}
  \includegraphics[width=6cm]{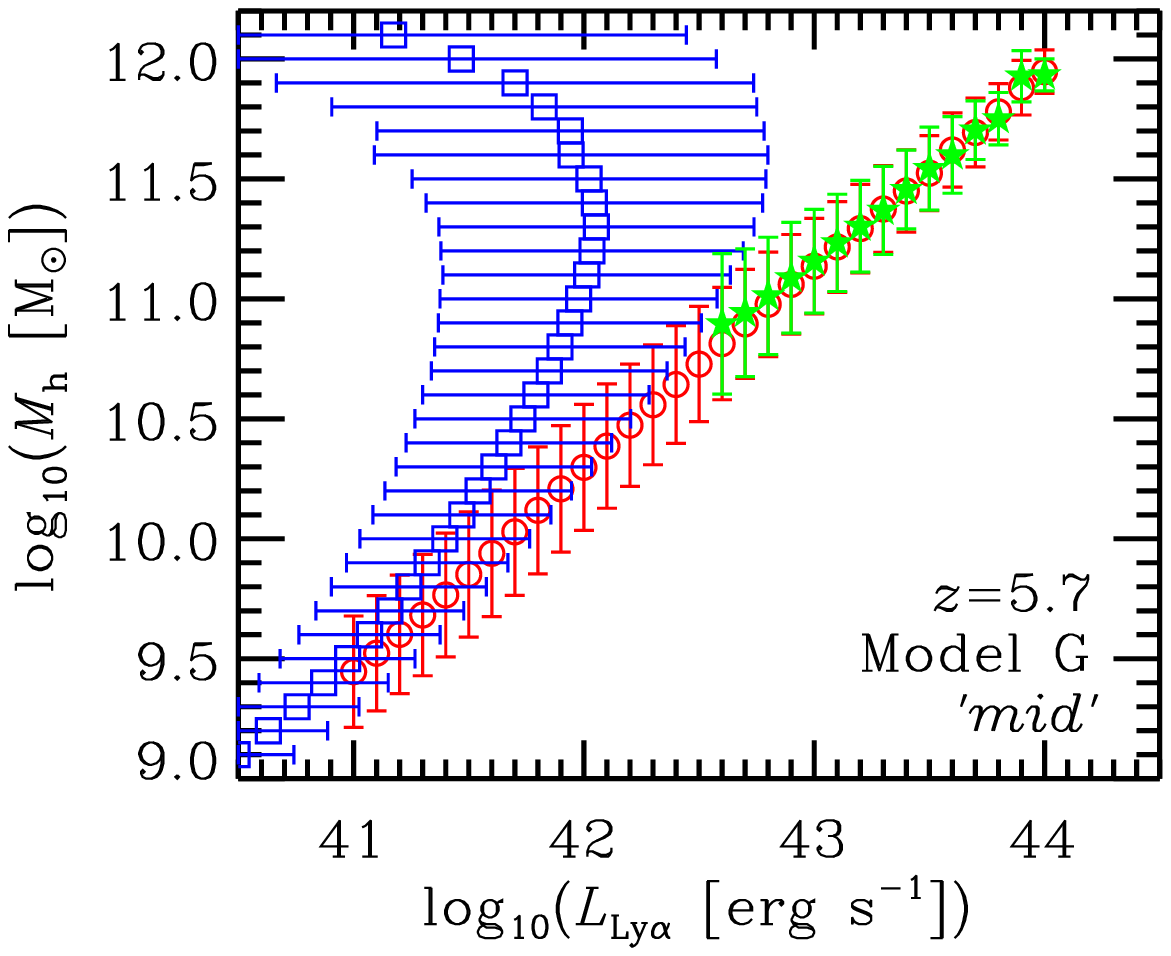}
  \includegraphics[width=6cm]{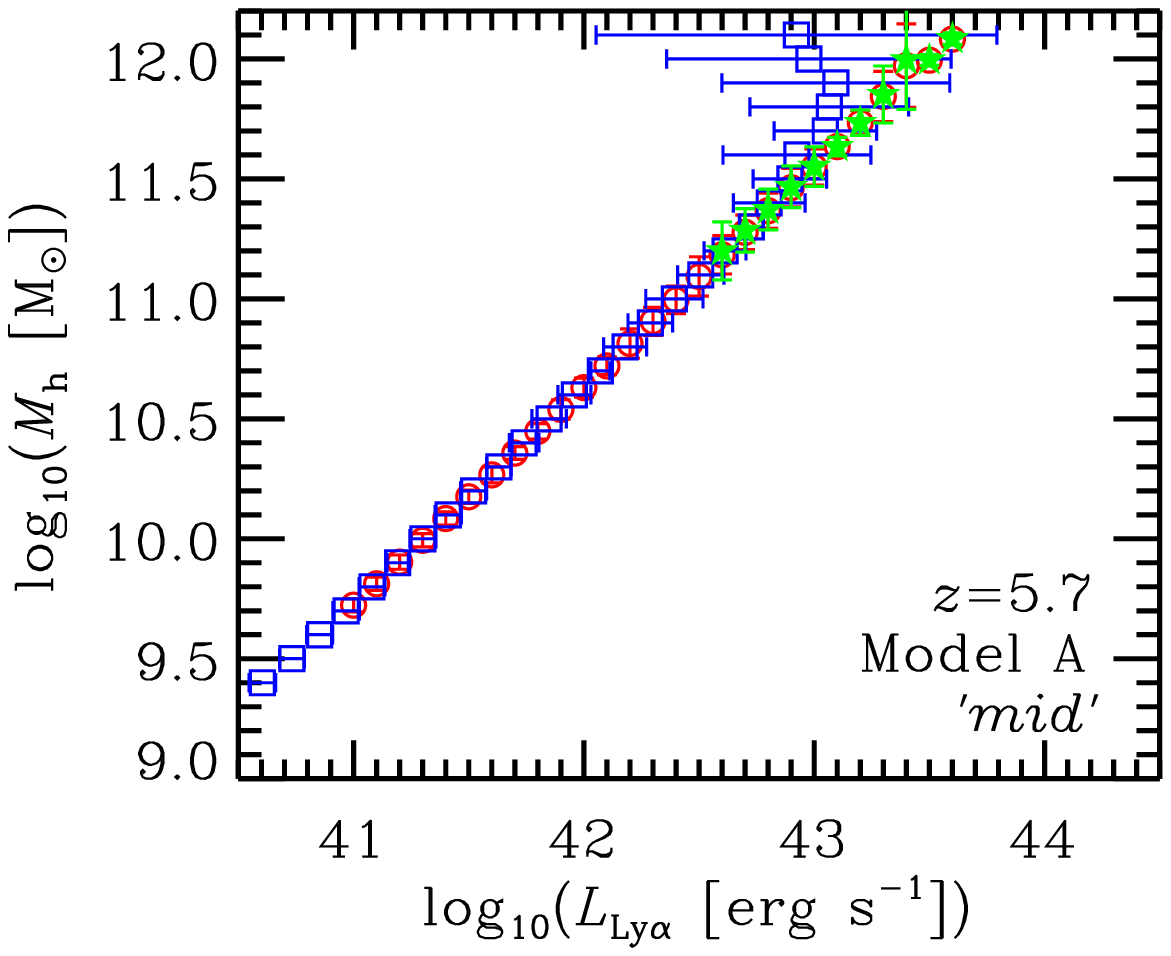}
  \includegraphics[width=6cm]{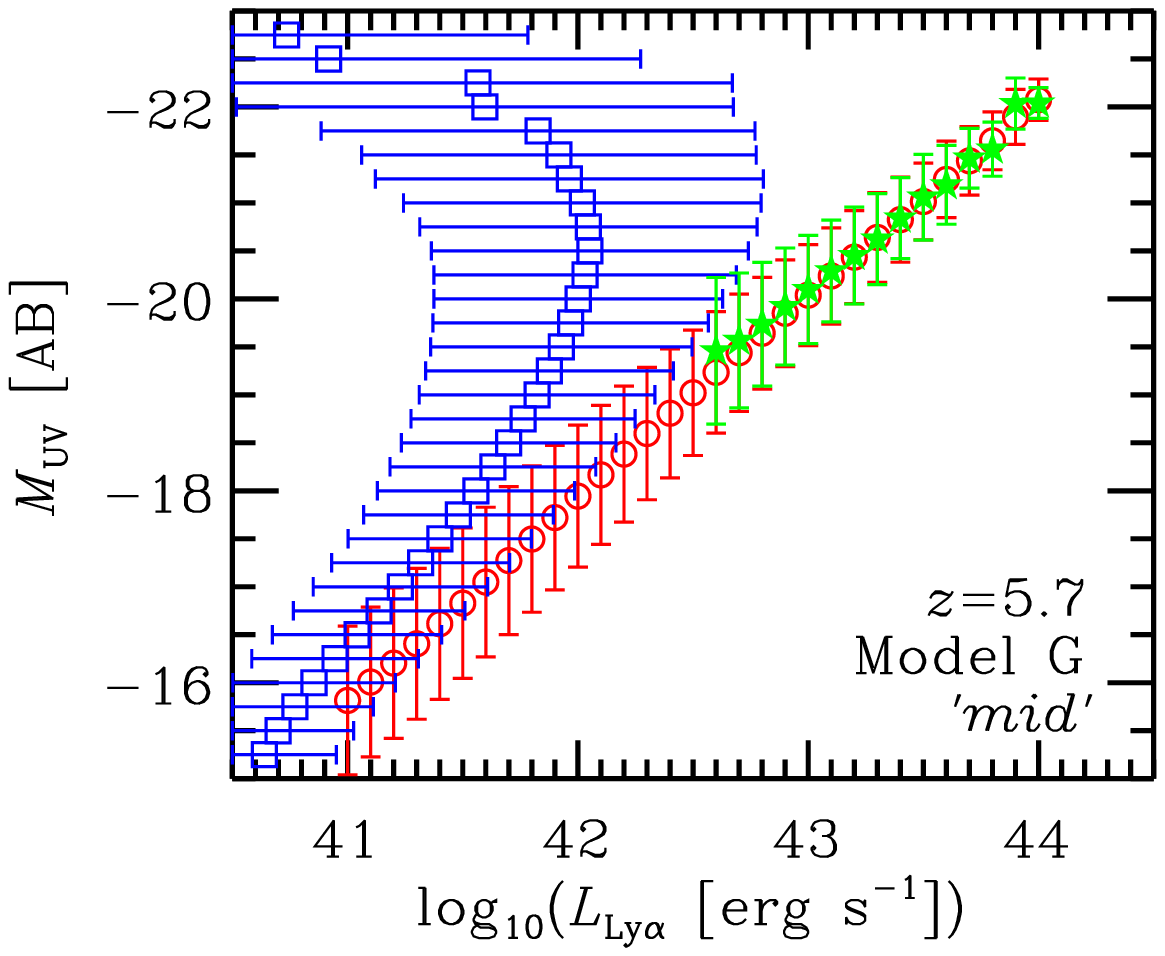}
  \includegraphics[width=6cm]{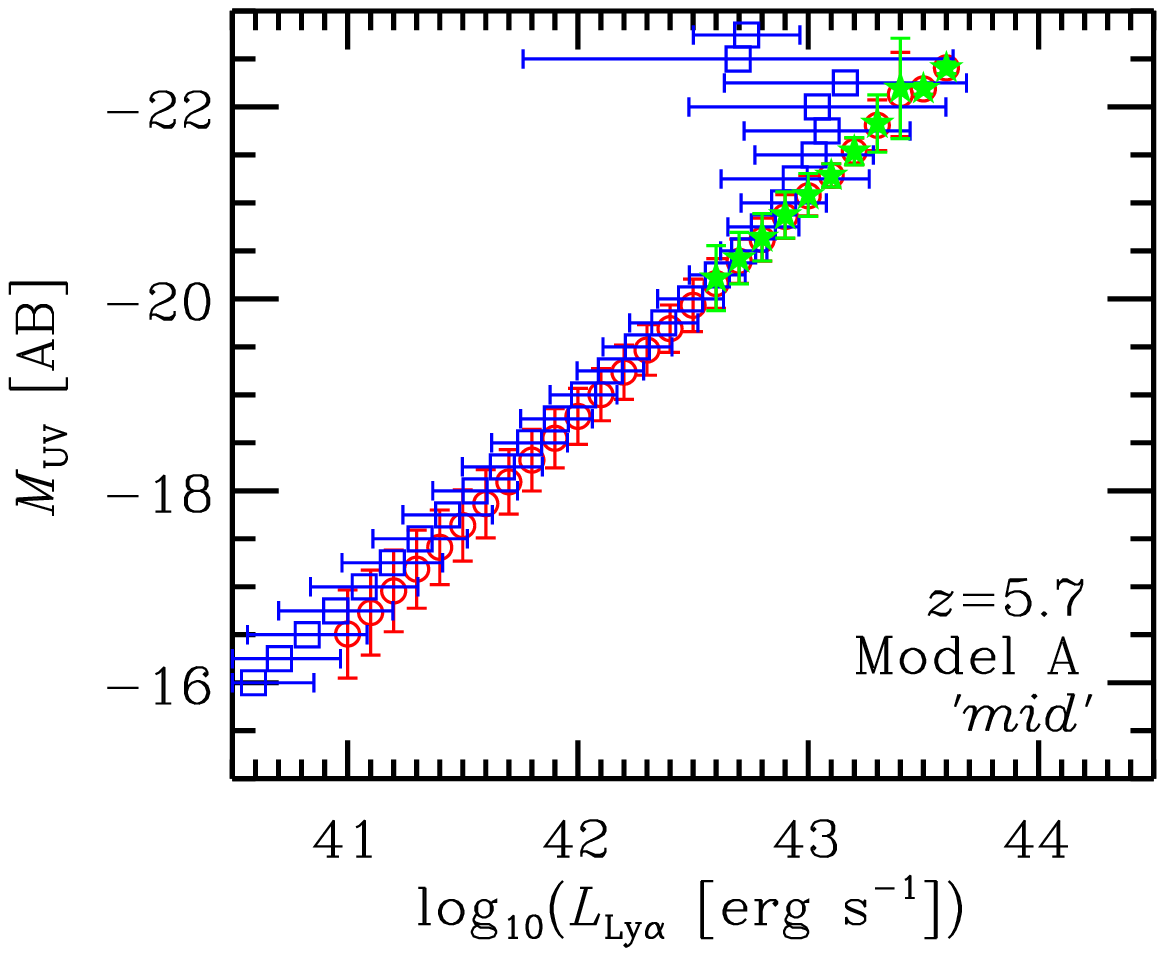}
  \includegraphics[width=6cm]{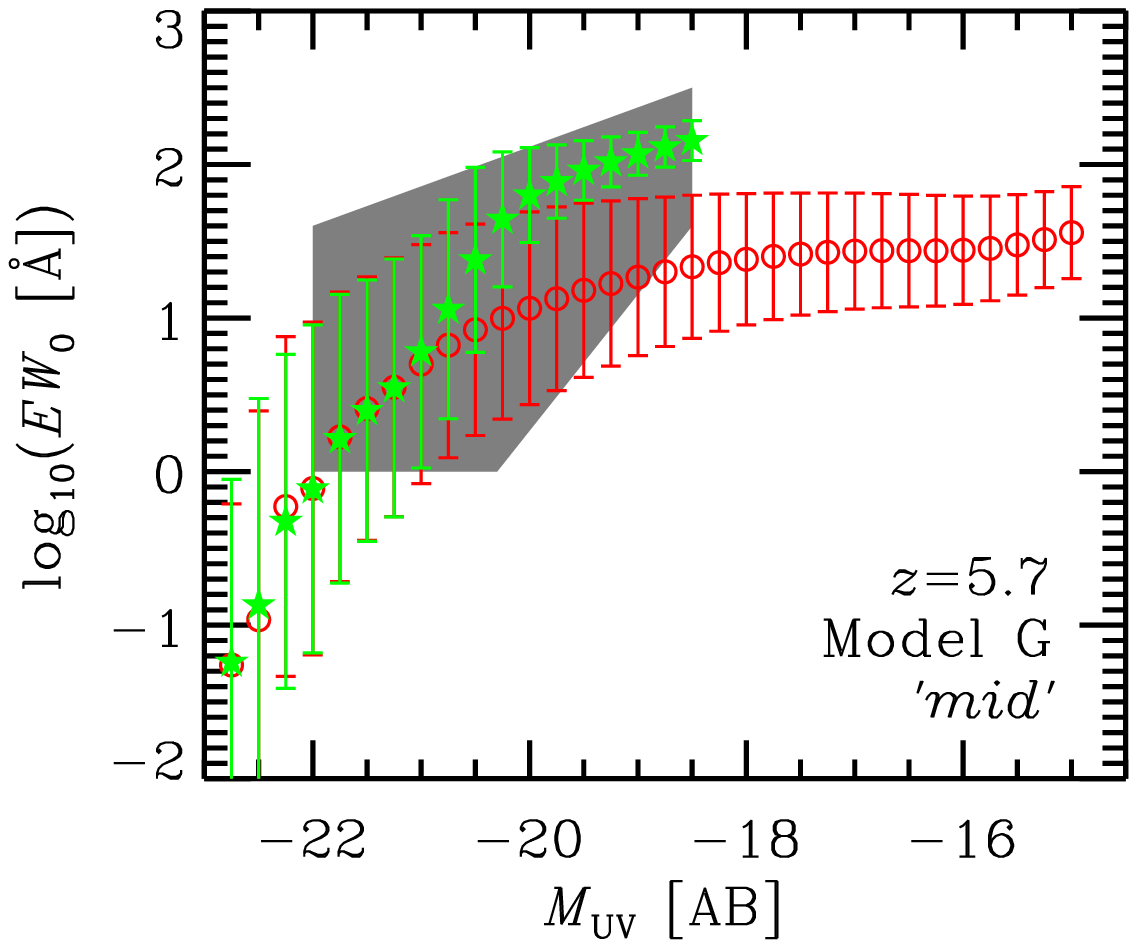}
  \includegraphics[width=6cm]{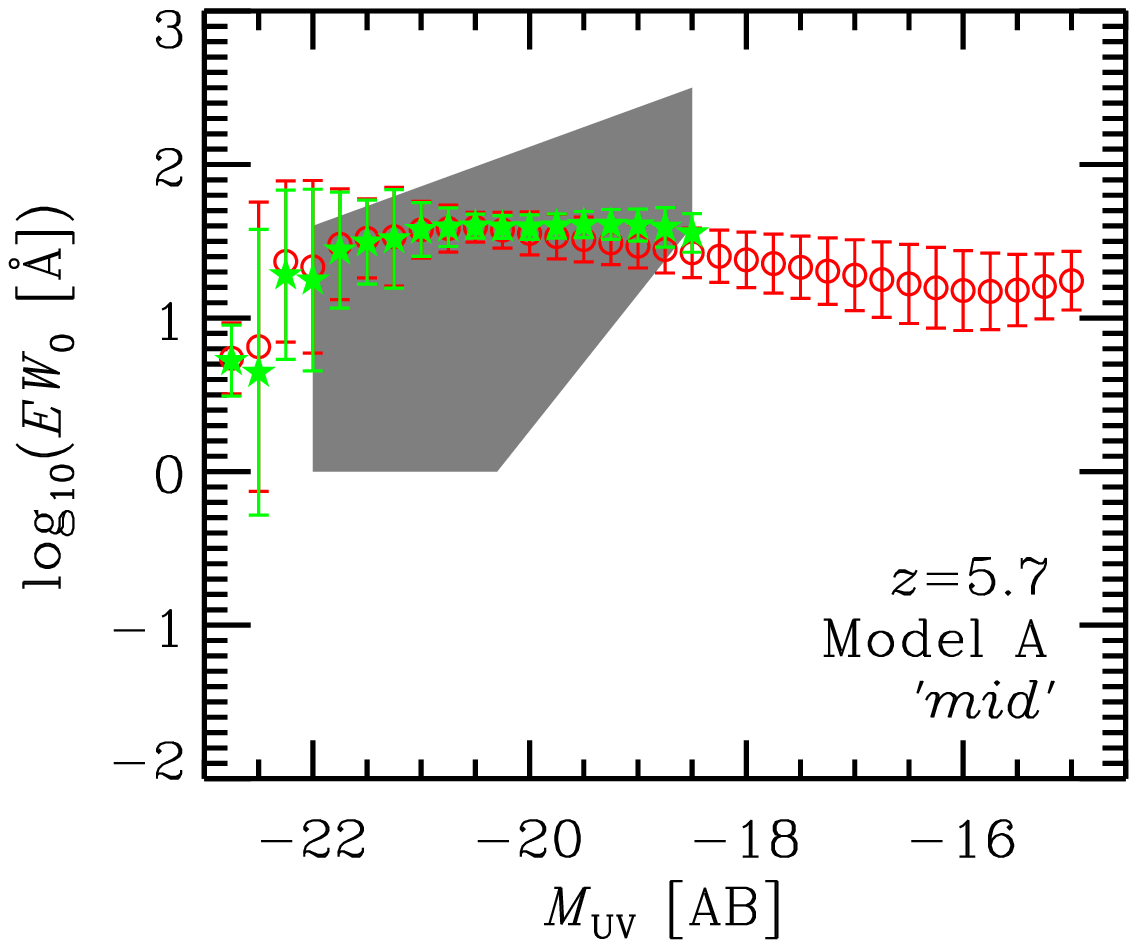}
 \end{center}
 \caption{Some correlations between halo mass, UV magnitude, 
 Ly$\alpha$ luminosity and equivalent width for the best-model, 
 Model~G (left panels) and the simplest model, Model~A (right panels),
 as a reference. The $mid$ reionization history and the redshift $z=5.7$
 are adopted. 
 The circles and error-bars (red) show the averages and standard
 deviations for the simulated galaxies binned along the horizontal axis. 
 On the other hand,
 the squares and error-bars (blue) are those binned along the vertical axis.
 The five-pointed-stars with error-bars (green) are the same as the
 circles but for the simulated galaxies which would be selected as LAEs
 by the color-magnitude 
 selection of the real observations (Shibuya et al.~2018). In the bottom
 panel, we also show the observed area from Furusawa~et~al.~(2016) 
 as the shaded area. The simulated
 galaxies selected as LAEs but having a small equivalent width in the
 bright-end are contaminants but its number fraction is as small as 3\%.}
 \label{correlation}
\end{figure*}

Fig.~\ref{correlation} shows some physical properties of galaxies 
predicted from the best-model, Model~G. We also show the predictions
from the simplest model, Model~A, as a reference. The top panels show the
predicted relations between halo mass, $M_{\rm h}$, and observable
Ly$\alpha$ luminosity, $L_{\rm Ly\alpha}$. The circles and error-bars
(red) represent the average $M_{\rm h}$ and its standard deviation of the
halos having a $L_{\rm Ly\alpha}$ within a 0.1-dex range around the
value of the horizontal axis. The squares and error-bars (blue), on the other
hand, represent the average $L_{\rm Ly\alpha}$ and its standard
deviation of the halos having a $M_{\rm h}$ within a 0.1-dex range
around the value of the vertical axis. For Model~A, the circles and
squares are completely overlapped, except for $L_{\rm Ly\alpha}>10^{43}$
erg~s$^{-1}$ and $M_{\rm h}>10^{11.5}$ M$_\odot$ because of the
one-to-one correspondence of $M_{\rm h}$ and $L_{\rm Ly\alpha}$ in the
model without any stochastic process. The deviation for the massive
or luminous halos is caused by a fluctuation of the IGM transmission. In
our reionization simulation, the neutral fraction around massive halos
tends to be higher and the transmission varies significantly, whereas
that around less massive halos tends to be lower and the transmission
keeps high (Hasegawa et al.\ in prep.).

Interestingly, Model~G is very different from Model~A. In Model~G,
luminous LAEs ($>10^{43}$~erg~s$^{-1}$) tend to have a massive halo 
mass ($>10^{11}$~M$_\odot$) as shown by the circles. However, such a 
massive halo on average has much smaller Ly$\alpha$ luminosity (see 
the squares) and only a small fraction of these massive halos is 
observed as luminous LAEs. This is very consistent with a duty cycle 
hypothesis of the LAE population (e.g., \cite{Ouchi2010,Ouchi2017}).
The middle panels of Fig.~\ref{correlation} show correlations 
between the UV magnitude, $M_{\rm UV}$, and $L_{\rm Ly\alpha}$ and 
indicate a very similar thing. According to the LAE fraction in UV 
luminous halos (Figs.~\ref{LAEfrac_mid}, \ref{LAEfrac_early} and 
\ref{LAEfrac_late}), the LAE fraction is 5--20\% depending on the 
equivalent width ($EW_0$) criterion at $M_{\rm UV}\simeq-20.5$.

The five-pointed stars with error-bars (green) are the same as the circles but
for the halos which would be selected as LAEs by real observations 
\citep{Shibuya2017}. We set the NB limiting magnitude corresponding to 
$L_{\rm Ly\alpha}=10^{42.5}$ erg s$^{-1}$. We have found a typical halo
mass of $M_{\rm h}\simeq10^{11}$~M$_\odot$ which excellently agrees with
that estimated by \citet{Ouchi2017} through the clustering analysis.
This is a natural consequence because we have reproduced the observed 
ACFs of \citet{Ouchi2017} in \S4.2.

The bottom panels of Fig.~\ref{correlation} show the relation between 
the UV magnitude, $M_{\rm UV}$, and the Ly$\alpha$ equivalent width in 
the source rest-frame, $EW_0$. The shaded area is the observed range 
in the diagram \citep{Furusawa2016}. There is an observational trend
that UV bright galaxies have a smaller $EW_0$, so-called ``Ando
relation'' \citep{Ando2006,Ando2007}. Note that the sample galaxies for
the observation of \citet{Furusawa2016} 
are mainly NB-selected LAEs and a comparison with our
models is fair. We have applied the same $M_{\rm UV}$ limit as the 
observation to the color-magnitude selected halos in the model
(five-pointed stars). The offset between the five-pointed stars and the
circles in Model~G is caused by the NB excess selection in the mock
observation. In Model~G, there are some NB-selected ``LAEs'' with
$EW_0<1$ \AA\ at $M_{\rm UV}<-22$. These halos were selected by a Lyman
break mimicking an NB excess. However, the number fraction of such bright
NB-selected halos among the NB-selected halos of $M_{\rm UV}<-18.5$ is
as small as 3\%. The best-model, Model~G, excellently reproduces the
Ando relation although the simplest, Model~A, does not. The origin 
of the Ando relation in Model~G is the halo mass dependence of the
Ly$\alpha$ optical depth. Therefore, the Ando relation can be
understood as a consequence from a simple scaling of halo mass:  
more H~{\sc i} in more massive halos. 

\begin{figure*}
 \begin{center}
  \includegraphics[width=5cm]{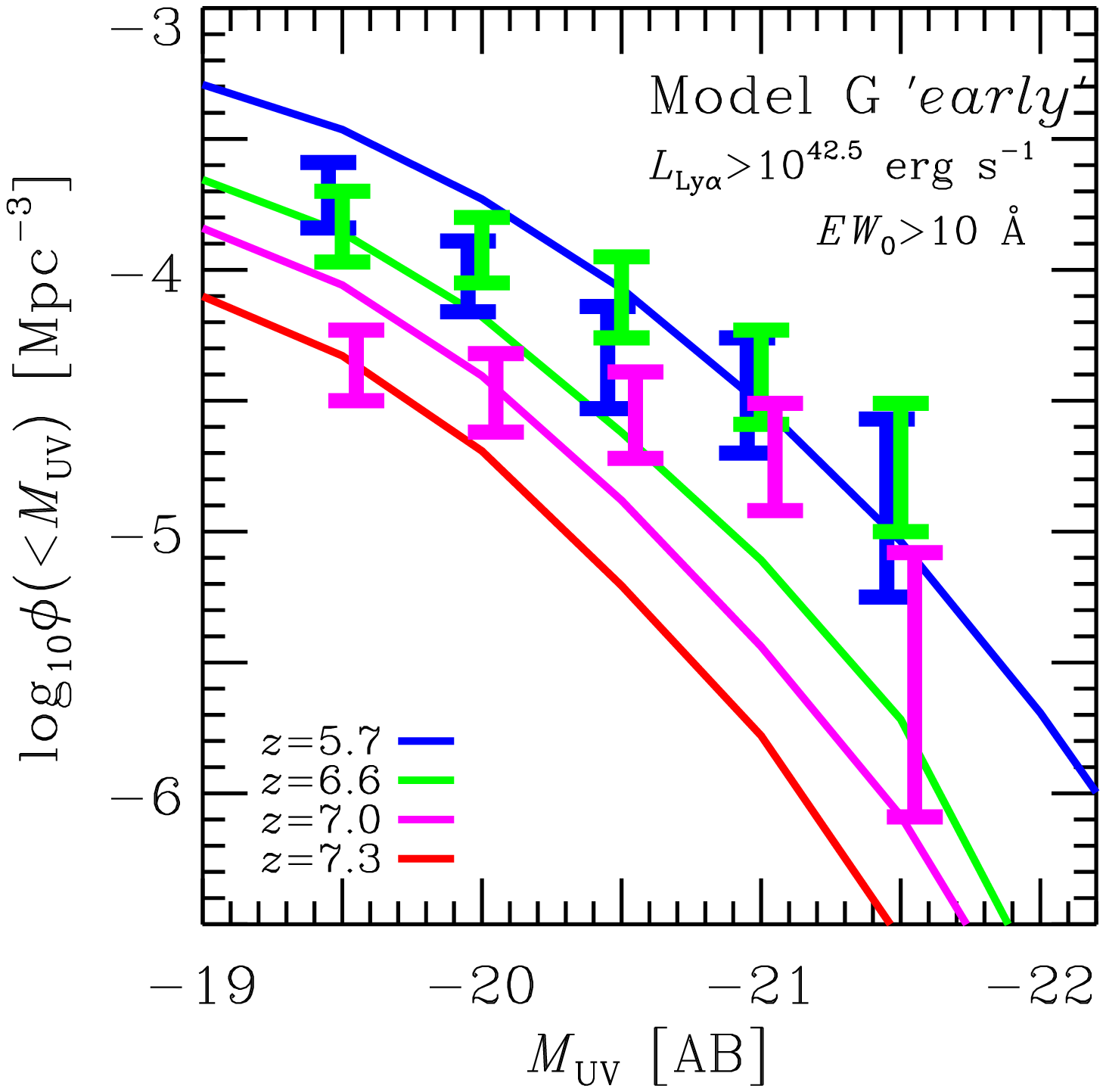}
  \includegraphics[width=5cm]{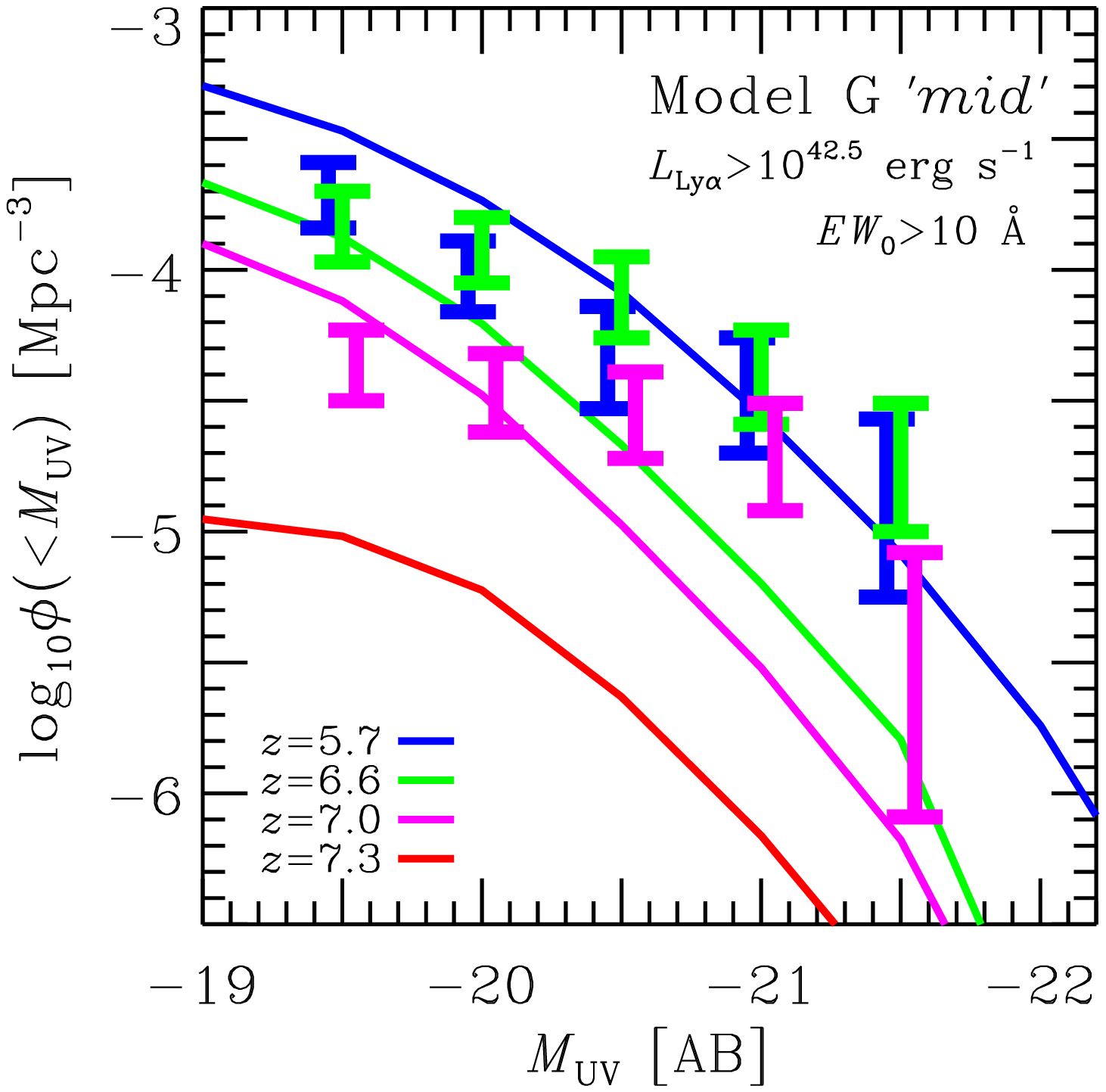}
  \includegraphics[width=5cm]{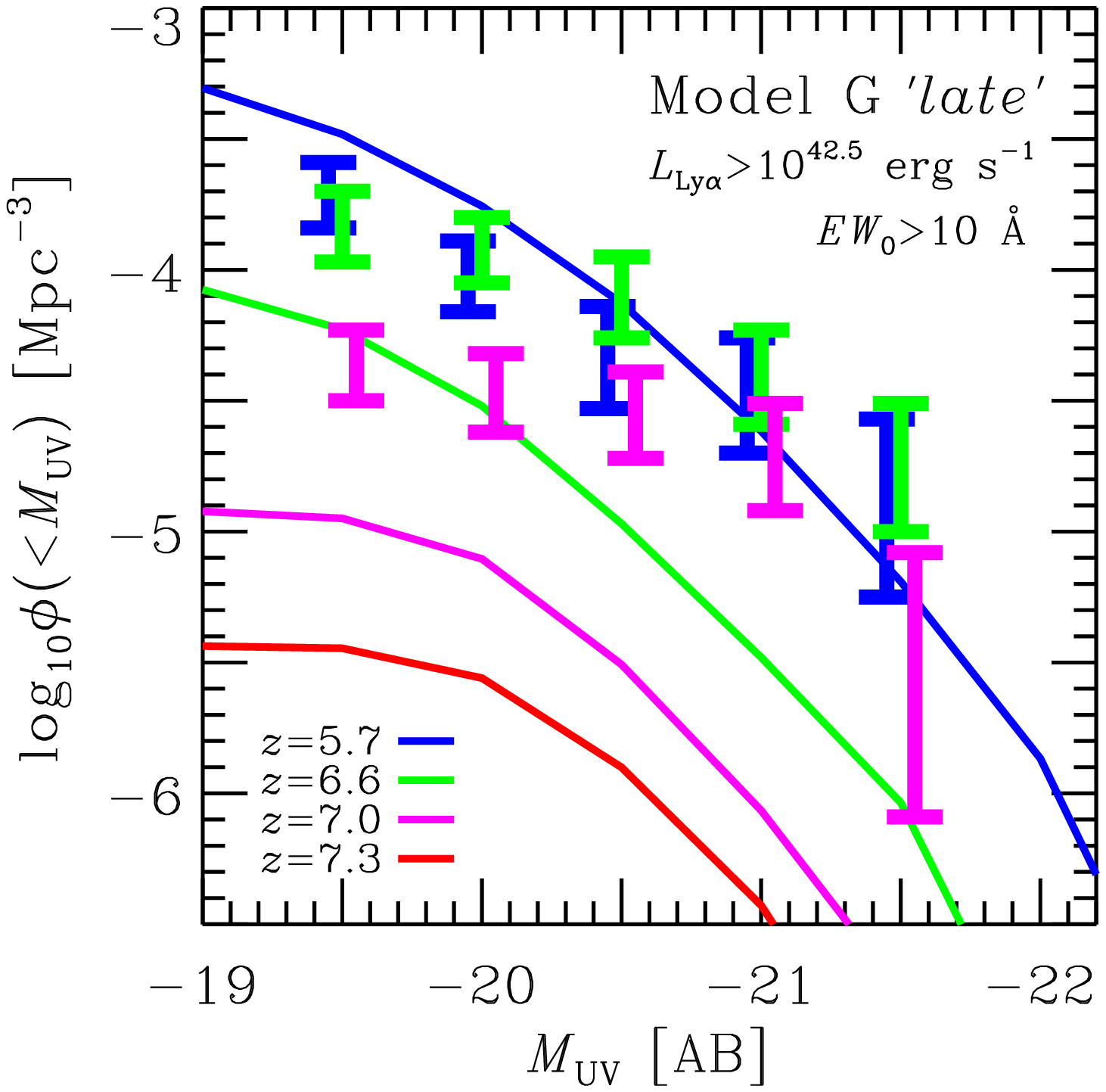}
  \includegraphics[width=5cm]{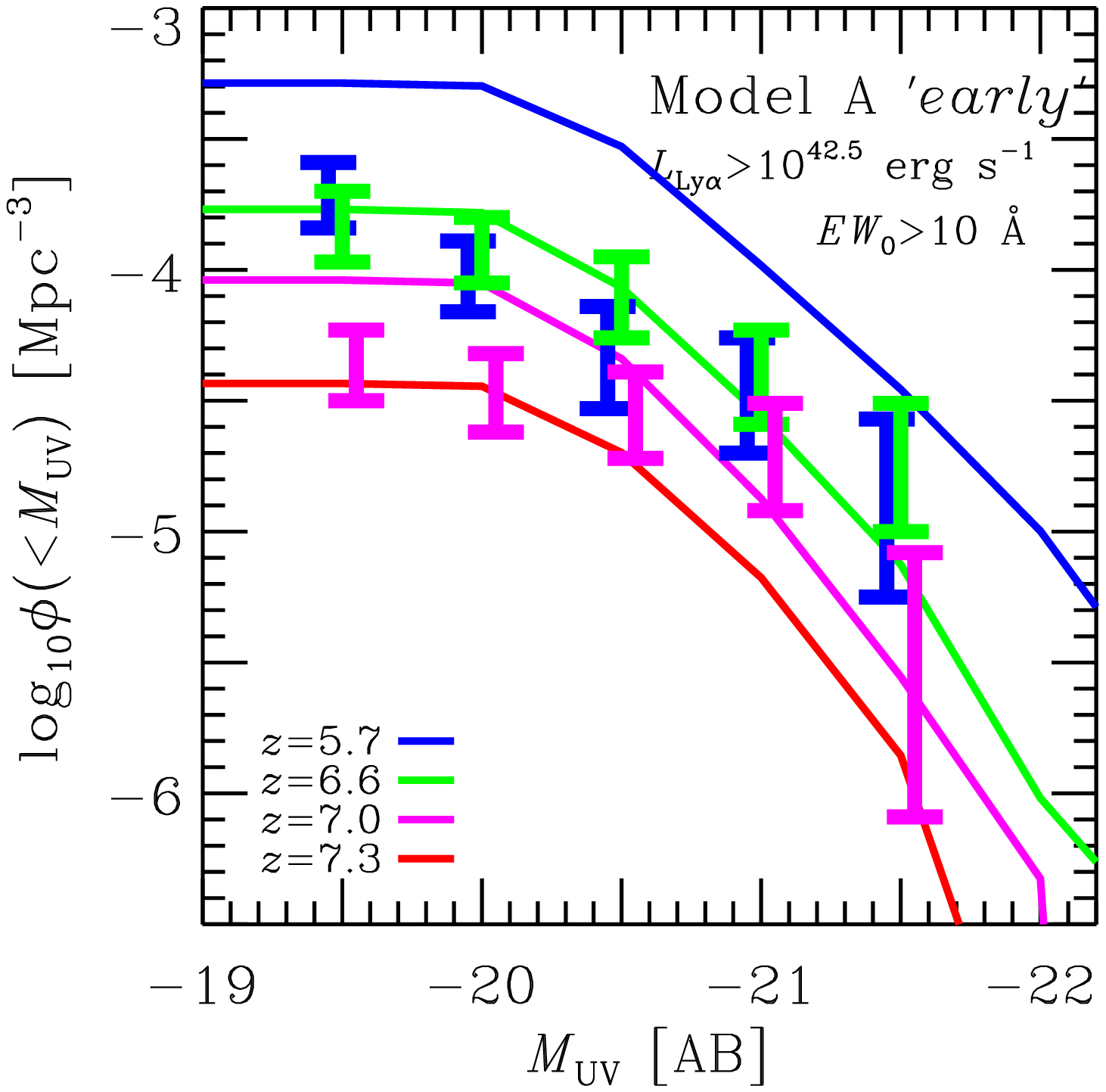}
  \includegraphics[width=5cm]{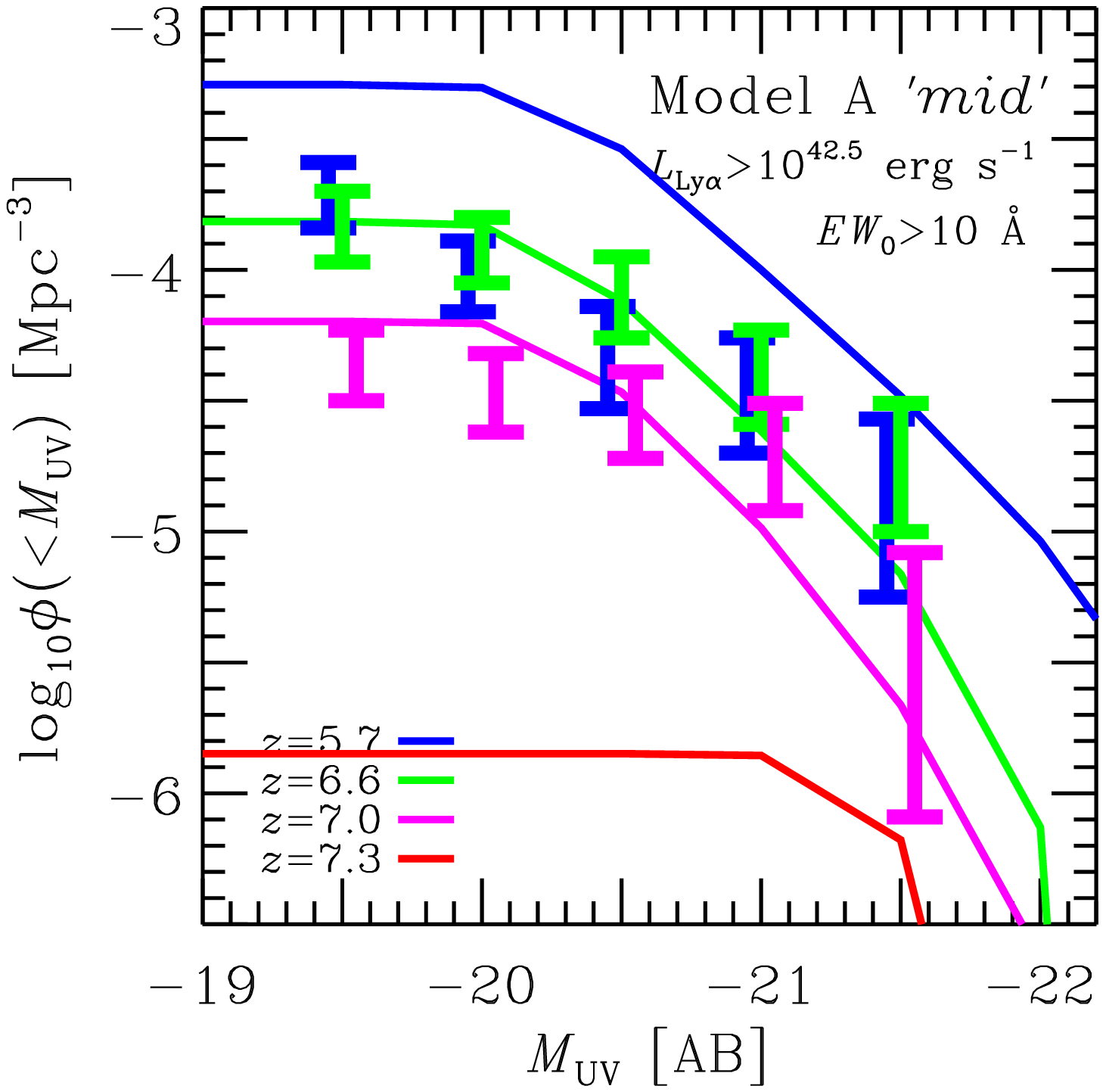}
  \includegraphics[width=5cm]{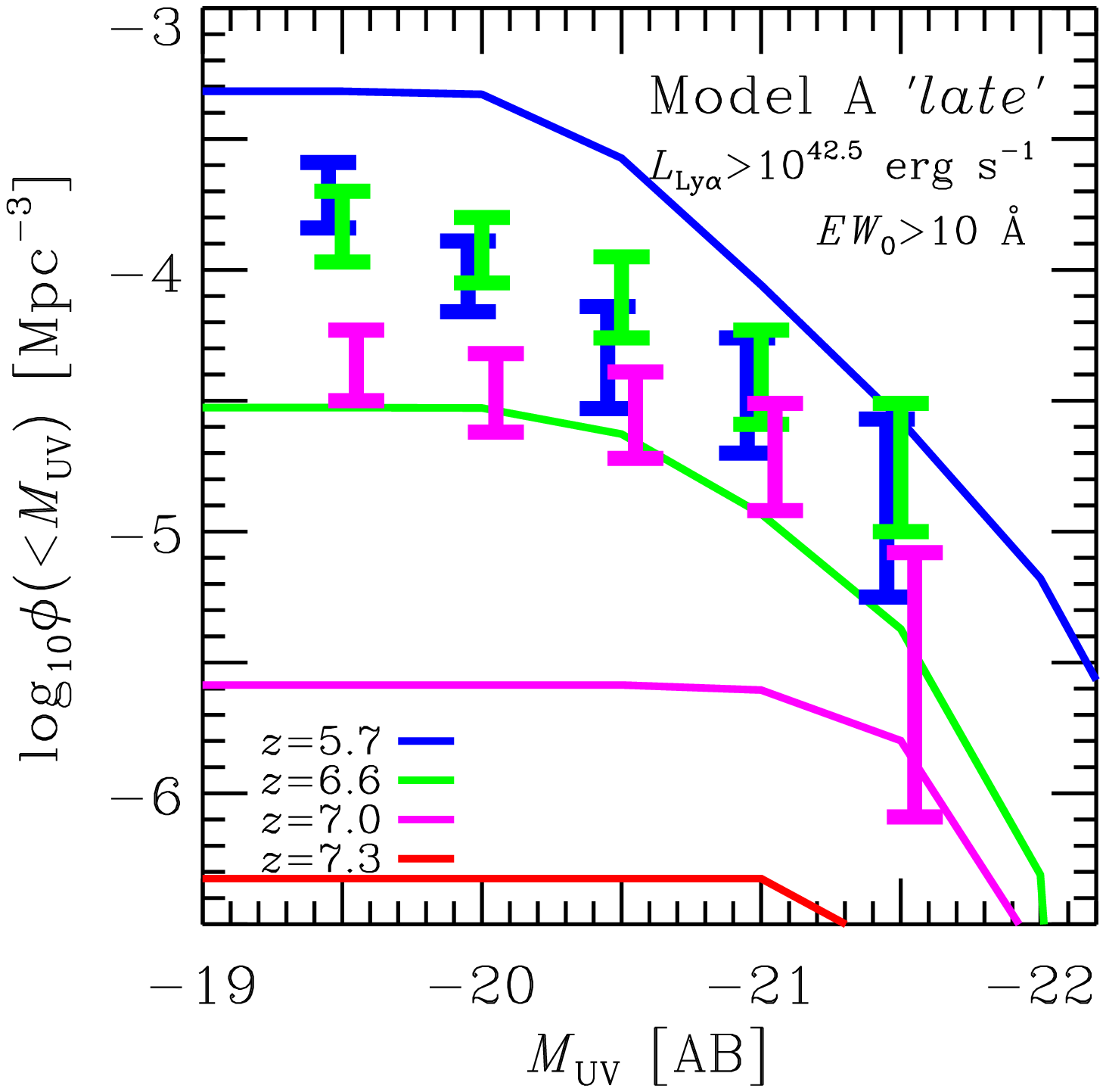}
 \end{center}
 \caption{Cumulative UV luminosity functions of LAEs 
 at $z=5.7$ (blue), 6.6 (green), 7.0 (magenta) and 7.3 (red) for the best-model, 
 Model~G (upper panels) and for the simplest model, Model~A, as a
 reference (lower panels). The LAEs are selected if their Ly$\alpha$
 luminosity is $>10^{42.5}$~erg~s$^{-1}$ and Ly$\alpha$ equivalent width
 in the source rest-frame is $>10$~\AA. The $early$, $mid$, and $late$
 reionization histories are shown in the left, middle and right panels,
 respectively. The observational data are taken from Ota~et~al.(2017).}
 \label{LAEUVLF}
\end{figure*}

Fig.~\ref{LAEUVLF} shows the UV luminosity functions for LAEs at 
$z=5.7$, $6.6$, $7.0$ and $7.3$. The observational data are taken 
from \citet{Ota2017} and references therein. We show the best-model, 
Model~G, and the simplest model, Model~A as a comparison. Both models 
agree with the observations in some cases but do not in other cases. 
The $early$ or $mid$ reionization histories show better agreement than 
the $late$ cases. However, the observed little evolution between 
$z=5.7$ and $z=6.6$ is not consistent with the models which predict 
a significant evolution. The physical reason of this disagreement is 
unclear. On the other hand, we have found that the model UV luminosity 
function is sensitive to the selection criteria for the LAEs. In this 
comparisons, we have adopted a set of Ly$\alpha$ luminosity and 
equivalent width cuts similar to those in \citet{Ota2017}. If we 
change the Ly$\alpha$ luminosity limit, however, the UV LF easily moves 
vertically. Therefore, we need to be careful in the Ly$\alpha$ 
luminosity limit in such comparisons in future.

\subsection{To derive the reionization history with LAEs}

\subsubsection{Ly$\alpha$ luminosity function and density}

\begin{figure*}
 \begin{center}
  \includegraphics[width=5cm]{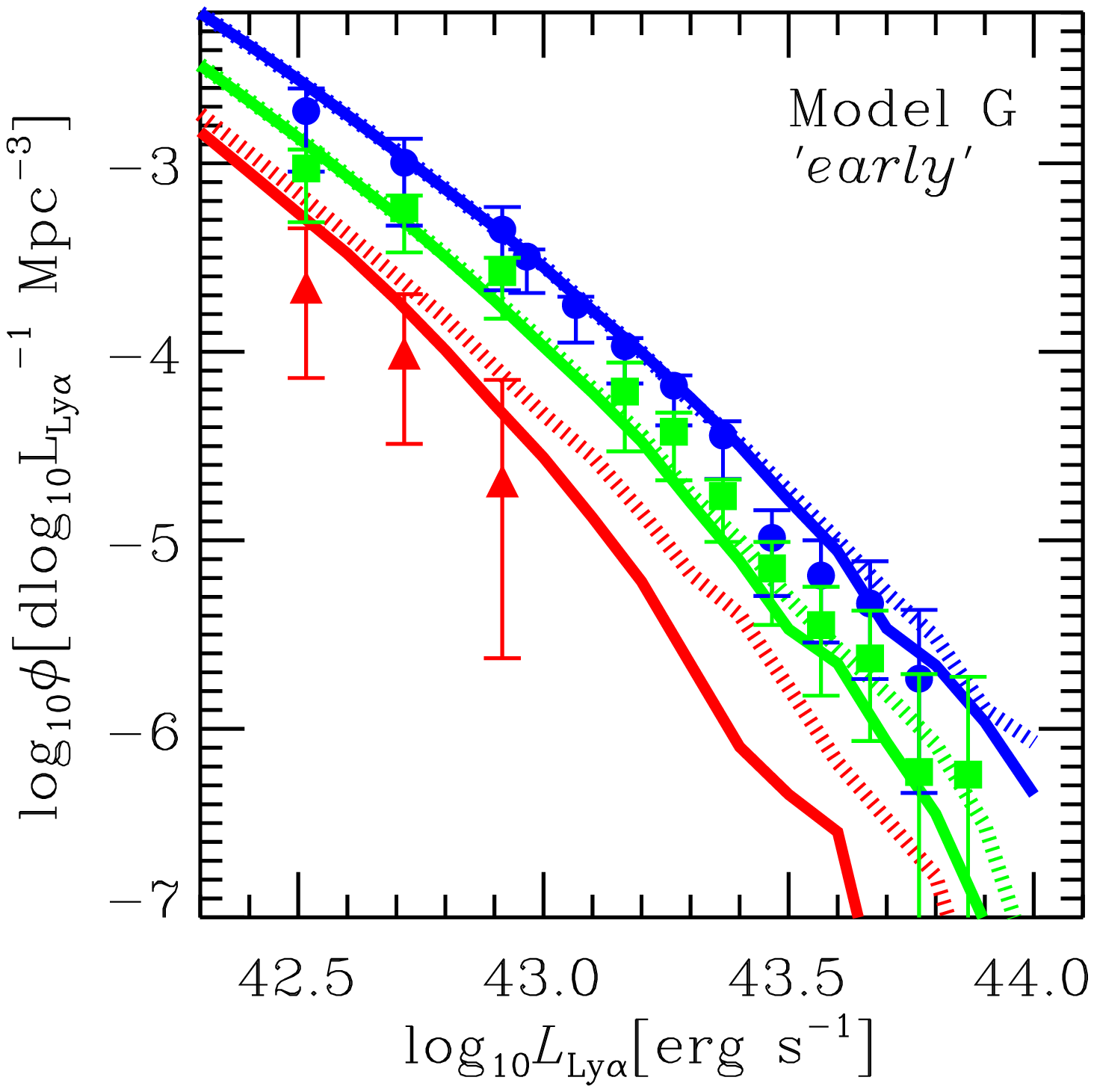}
  \includegraphics[width=5cm]{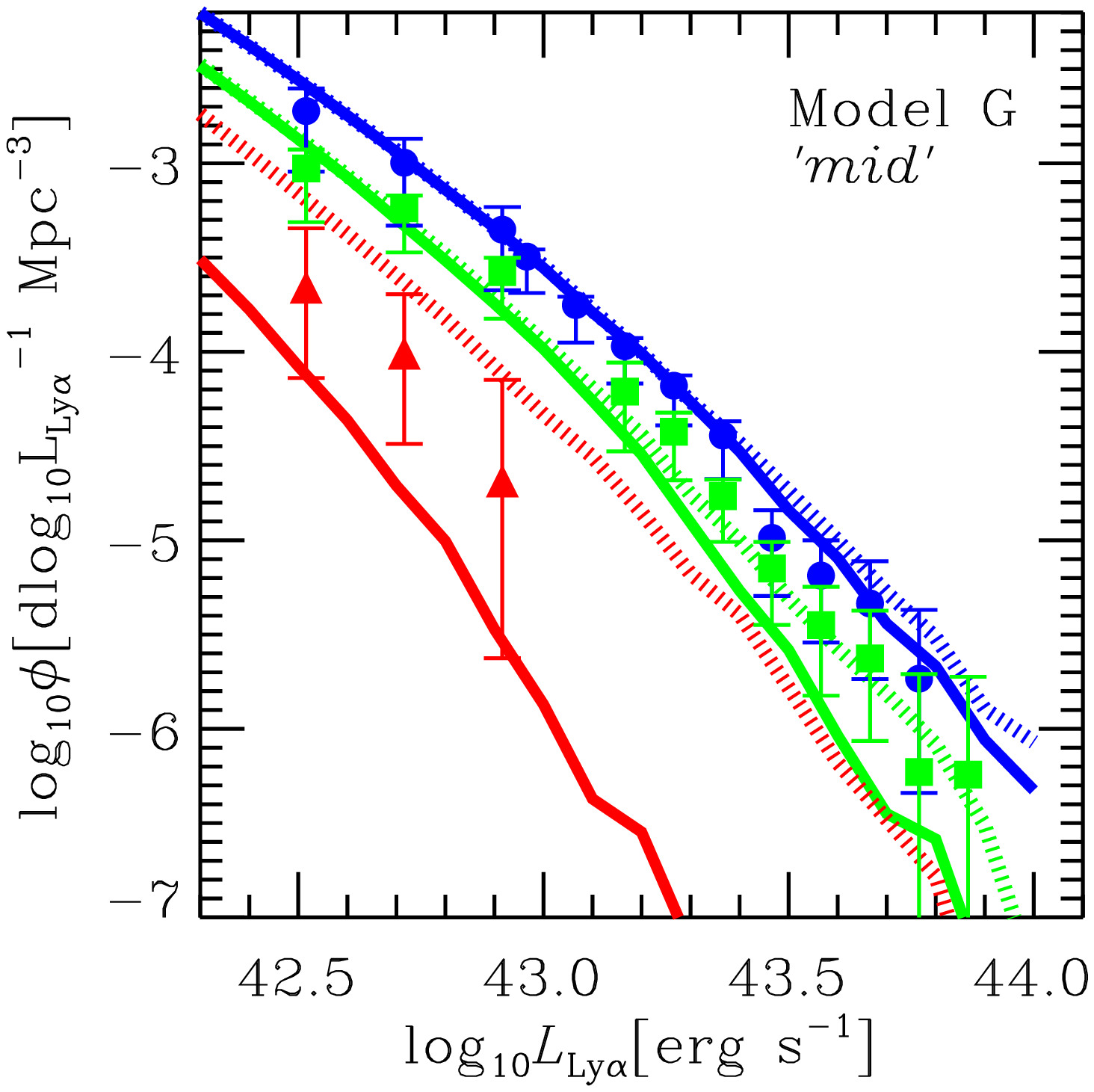}
  \includegraphics[width=5cm]{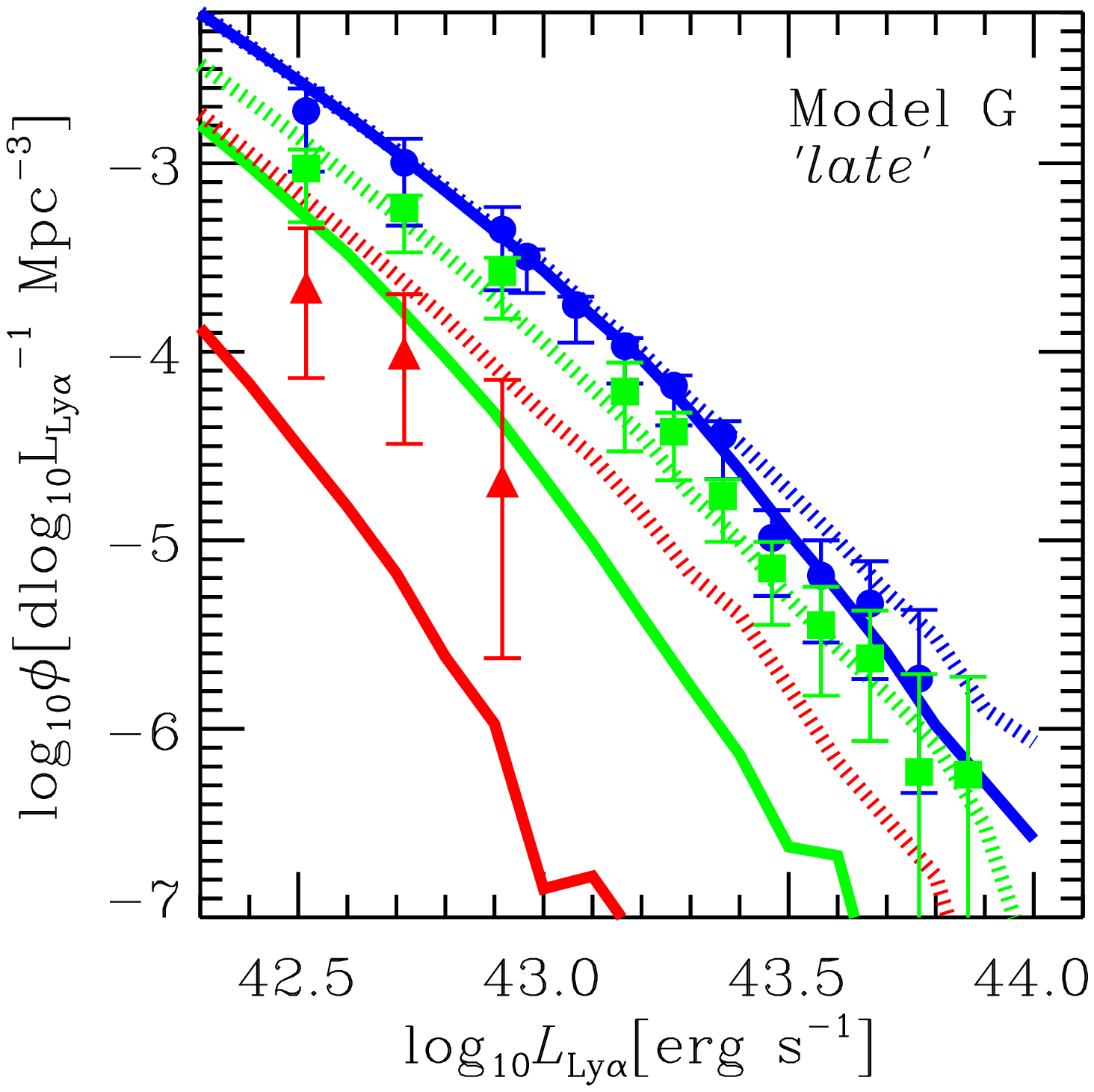}
 \end{center}
 \caption{Redshift evolution of Ly$\alpha$ luminosity functions 
 at $z=5.7$ (blue), 6.6 (green) and 7.3 (red)
 of the best-model, Model~G. The $early$, $mid$ and $late$ reionization 
 histories are shown in the left, middle and right panels, 
 respectively. The dotted lines are the fully-ionized IGM cases and 
 the solid lines are the cases with the Ly$\alpha$ transfer in the IGM.
 The observational data are taken from Konno~et~al.~(2014,2018).
 We have assumed the Ly$\alpha$ line profile through an outflowing gas 
 with $v=150$ km s$^{-1}$ and $\log_{10}(N_{\rm HI}/{\rm cm^2})=20$.}
 \label{LaLFzevol}
\end{figure*}

Fig.~\ref{LaLFzevol} shows the redshift evolution of the Ly$\alpha$ 
LFs and comparisons with the best-model, Model~G. The three panels 
correspond to the $early$, $mid$ and $late$ reionization histories 
from left to right, respectively. The solid and dotted curves are 
the cases with/without the IGM Ly$\alpha$ transfer effect, respectively.
Namely, the dotted curves are the fully-ionized IGM cases. At $z=5.7$, 
the models reproduce the observations very well independent of the 
reionization model because the neutral fraction is very low at the 
redshift in any model. On the other hand, the $late$ model predicts 
much fewer LAEs than the observations at $z=6.6$ and $7.3$. The $mid$ 
model also predicts somewhat fewer LAEs but the $early$ model expects 
a bit more LAEs at $z=7.3$. Therefore, the real Universe may have 
evolved through a middle history of the $mid$ and $early$ models.

In this way, we may derive the reionization history from the Ly$\alpha$ 
LFs. Such attempts have been made in literature (e.g., 
\cite{Kashikawa2006,Ouchi2010,Kashikawa2011,Konno2014,Ota2017,Konno2017}).
The most serious difficulty in this analyses, however, is the degeneracy 
between the evolution of the LAE population and the IGM neutrality. 
Some authors have assumed the evolution (or no evolution) in the LAE 
population estimated from the LAE UV LFs and other authors have used 
theoretical model predictions. We take the latter approach here. 
The advantage compared to the previous works is that the best-model, 
Model~G, in this paper can reproduce all of the Ly$\alpha$ LFs, ACFs, 
and LAE fractions very well. Therefore, the reliability of the model 
would be high.

\begin{figure}
 \begin{center}
  \includegraphics[width=7cm]{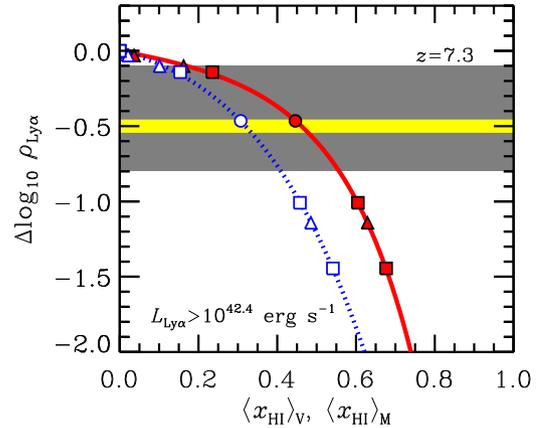}
 \end{center}
 \caption{Decrements of the Ly$\alpha$ luminosity density 
 as a function of the IGM hydrogen neutral fraction, $x_{\rm HI}$. 
 The circles, triangles, and squares are the predictions of the best-model, 
 Model~G, at $z=6.6$ (NB921), $z=7.0$ (NB973) and $z=7.3$ (NB101), 
 respectively. The filled (red) and open (white) symbols show the
 volume-averaged and mass-averaged $x_{\rm HI}$ cases, respectively. The
 solid (red) and dotted (blue) lines are fitting functions for the two
 cases described in equation~(\ref{fiteq}). The same symbols along a
 fitting line show the predictions at the same redshift but in different
 reionization histories (then different $x_{\rm HI}$). 
 The limiting Ly$\alpha$ luminosity is indicated in the panel.
 We also show the observations with their uncertainties at $z=7.3$ 
 as the horizontal line and shade.}
 \label{reionconst}
\end{figure}

We introduce a measure of the IGM neutral fraction: the decrement of 
the observed Ly$\alpha$ luminosity density (LD) compared to the model 
predictions in a fully-ionized IGM. Namely, 
\begin{equation}
 \Delta \log_{10} \rho_{\rm Ly\alpha} \equiv 
  \log_{10} \rho_{\rm Ly\alpha}^{\rm obs} 
  - \log_{10} \rho_{\rm Ly\alpha}^{\rm NoIGM}\,,
\end{equation}
where $\rho_{\rm Ly\alpha}$ is the integrated Ly$\alpha$ LD.
Fig.~\ref{reionconst} shows these decrements obtained from Model~G.
We have found that these decrements at different redshifts 
and in different reionization histories can be 
described by the following single function of the neutral fraction, 
$x_{\rm HI}$, excellently: 
\begin{equation}
 y = a x \exp(b x^c)\,,
 \label{fiteq}
\end{equation}
where $y$ is the decrement, $\Delta\log_{10}\rho_{\rm Ly\alpha}$, 
and $x$ is the volume-averaged or mass-averaged 
neutral fraction. The fitting parameters of $a$, $b$ and $c$ are listed 
in Table~\ref{fitparam} for a limiting Ly$\alpha$ luminosity. In 
Table~\ref{xHIestimate}, we list the model predictions of the Ly$\alpha$ 
LDs in a fully-ionized IGM, the observed Ly$\alpha$ LDs, the decrements 
of the observed LDs and the estimated hydrogen neutral fractions.
The observed decrements at $z\leq7.0$ are consistent with zero within 
uncertainties and we have obtained only upper limits of $x_{\rm HI}$. 
At $z=7.3$, the 
decrement is significantly negative, indicating a non-zero value of 
$x_{\rm HI}$. The resultant values are $0.5$ and $0.3$ for the 
volume-averaged and mass-averaged $x_{\rm HI}$, respectively. These 
$x_{\rm HI}$ at $z=7.3$ are consistent with the estimation by 
\citet{Konno2014} who assumed an evolution of the LAE population 
estimated from the LAE UV LFs and an analytic IGM transmission model 
by \citet{Santos2004}.

There is a precaution for using equation~(19). These LD decrements are 
derived from the cases with a Ly$\alpha$ line profile through an 
outflowing gas with its velocity $v=150$~km~s$^{-1}$ and H~{\sc i} 
column density $\log_{10}(N_{\rm HI}/{\rm cm^2})=20$. As we saw in 
Figs.~\ref{LaLF_mid}, \ref{LaLF_early} and \ref{LaLF_late}, the 
H~{\sc i} column density has an effect on the LF decrements. 
Therefore, if the typical H~{\sc i} column density in LAEs at the 
redshift interested is different from that assumed here, the fitting 
equation~(19) will change. In addition, there may be systematic
uncertainty in $\rho_{\rm Ly\alpha}^{\rm NoIGM}$ caused by the model
calibration of LFs which we assume 0.1-dex (Table~\ref{xHIestimate}).

\begin{table}
 \tbl{Fitting parameters of equation~(\ref{fiteq}) for the case of $\log_{10}(L_{\rm Ly\alpha}~[{\rm erg~s^{-1}}])>42.4$.}{%
 \begin{tabular}{llll}
  \hline
  & $a$ & $b$ & $c$ \\
  \hline
  $\langle x_{\rm HI} \rangle_{\rm V}$ & $-0.515$ & $2.77$ & $1.70$ \\
  $\langle x_{\rm HI} \rangle_{\rm M}$ & $-0.639$ & $2.46$ & $0.883$ \\
  \hline
 \end{tabular}}\label{fitparam}
 \begin{tabnote}
  These parameters are derived from Model G assuming the Ly$\alpha$ line 
  profile through an outflowing gas with $v=150$~km~s$^{-1}$ and 
  $\log_{10}(N_{\rm HI}/{\rm cm^2})=20$.
 \end{tabnote}
\end{table}

\begin{table*}
 \tbl{A summary of the estimations of the hydrogen neutral fraction.}{%
 \begin{tabular}{lllllll}
  \hline
  $z$ & $\log_{10}\rho_{\rm Ly\alpha}^{\rm NoIGM}$ $^\dag$ 
  & $\log_{10}\rho_{\rm Ly\alpha}^{\rm obs}$ $^\dag$ 
  & $\Delta\log_{10}\rho_{\rm Ly\alpha}$ 
  & $\langle x_{\rm HI} \rangle_{\rm V}$ 
  & $\langle x_{\rm HI} \rangle_{\rm M}$ & References$^\ddag$ \\
  \hline
  5.7 & $39.7\pm0.1$ & $39.54_{-0.09}^{+0.07}$ & $-0.2\pm0.2$ 
	      & $<0.4$ & $<0.3$ & (1) \\
  6.6 & $39.4\pm0.1$ & $39.26_{-0.08}^{+0.07}$ & $-0.1\pm0.2$
	      & $<0.4$ & $<0.2$ & (1) \\
  7.0 & $39.2\pm0.1$ & $39.08_{-0.25}^{+0.14}$ & $-0.1\pm0.3$
	      & $<0.4$ & $<0.3$ & (2) \\
  7.3 & $39.0\pm0.1$ & $38.49_{-0.17}^{+0.27}$ & $-0.5_{-0.3}^{+0.4}$
	      & $0.5_{-0.3}^{+0.1}$ & $0.3_{-0.2}^{+0.1}$ 
		      & (3) \\
  \hline
 \end{tabular}}\label{xHIestimate}
 \begin{tabnote}
  $^\dag$ The unit of the luminosity density is
  erg~s$^{-1}$~Mpc$^{-3}$. The lower limit of the luminosity in
  integration is $\log_{10}(L_{\rm Ly\alpha}~[{\rm erg~s^{-1}}])=42.4$. 
  We assume 0.1-dex systematic uncertainty of the model calibration.\\
  $\ddag$ The references of the observed luminosity densities. The
  numbers correspond to the following references: (1) Konno et al.~(2017), 
  (2) Ota et al.~(2017), (3) Konno et al.~(2014).
 \end{tabnote}
\end{table*}

\subsubsection{Other methods}

The LAE ACFs are suggested to be useful to constrain the reionization 
history as well as the reionization topology (e.g., \cite{mcquinn07}).
On the other hand, our model predictions in Figs.~\ref{ACF_mid}, 
\ref{ACF_early} and \ref{ACF_late} do not show very clear differences 
in the ACFs between $z=5.7$ and $6.6$. This is partly because 
$x_{\rm HI}$ is too small even at $z=6.6$ to make differences in ACFs.
If we go to higher redshifts where $x_{\rm HI}$ is sufficiently large, 
however, the observable numbers of LAEs may be too small to obtain 
firm ACF measurements.
Another reason may be the reionization topology in our simulation. 
This is basically ``inside-out'' as found in Fig.~\ref{fig:xHImap}, 
but our simulation seems more modest than previous ones. The IGM 
Ly$\alpha$ transmission at the redshift interested in this paper is 
actually lower for more massive halos. This is caused by two new 
features in our reionization simulation. One is that a smaller 
$f_{\rm esc}$ for more massive halos. The other is the spatially 
different clumping factor (Hasegawa et al.~in prep.\ for more details). 
Therefore, the LAE ACFs may not be a useful probe of reionization 
as previously thought although further investigations in simulations 
are required.

The redshift evolution of the LAE fractions can be useful to derive 
the reionization history (e.g., \cite{Ono2012}). Our model predictions 
shown in Figs.\ref{LAEfrac_mid}, \ref{LAEfrac_early} and \ref{LAEfrac_late} 
support this idea; the LAE fraction decrements become significant 
when $x_{\rm HI}\gtrsim0.1$. On the other hand, \citet{Oyarzun2017} raises 
a possibility that the drop of the LAE fraction is caused by survey 
incompleteness in $M_{\rm UV}$. Future analyses are required to examine 
and correct the incompleteness effect.

The 21~cm--LAE cross-correlation is proposed to be another powerful 
tool to deduce the reionization history, especially the size of the 
ionized bubbles (e.g., \cite{Lidz2009,Kubota2017,Yoshiura2017}). 
So far, the LAE modeling
remains rather simple assuming one-to-one correspondence between halo
mass and Ly$\alpha$ luminosity (e.g., \cite{Kubota2017}). Since we have
shown its break down in this paper, adopting a better LAE model in such
analyses will be an interesting future work.

\section{Conclusion}

We have presented new models of LAEs at $z\gtrsim6$, the reionization era, 
in a large-scale ($162^3$ comoving Mpc$^3$) radiative transfer simulation 
of reionization. Our LAE modeling is based on physically-motivated 
analytic equations, depending on the presence or absence of dispersion 
of Ly$\alpha$ emissivity in a halo, dispersion of the Ly$\alpha$ optical
depth in the halo, $\tau_\alpha$, and the halo mass dependence of
$\tau_\alpha$. We have critically examined the model with one-to-one
correspondence between halo mass and Ly$\alpha$ luminosity before
transmission in the intergalactic medium (IGM), which is often adopted
in previous studies. Comparing with the $z=5.7$ Ly$\alpha$ luminosity
function (LF) from the early data of the Subaru/HSC survey of LAEs
called SILVERRUSH \citep{Ouchi2017,Konno2017}, we have calibrated the
single adjustable parameter in our models: a typical Ly$\alpha$ optical
depth in a halo of $10^{10}$ M$_\odot$. After comparisons of our $2^3$
models with Ly$\alpha$ LFs at $z=6.6$ and $7.3$, angular
auto-correlation functions (ACFs) of LAEs at $z=5.7$ and $6.6$, LAE
fractions in Lyman break galaxies (LBGs) at $5<z<7$, we have identified
the best model which successfully reproduces all these observations. Our
main findings are as follows: 

\begin{itemize}
 \item The Ly$\alpha$ LFs and ACFs are reproduced by multiple models,
 but the LAE fraction is turned out the most critical test. The simplest 
 model adopting one-to-one correspondence between halo mass and Ly$\alpha$ 
 luminosity (Model~A) overpredicts (or underpredicts) the number fraction 
 of LAEs with $EW_0>25~{\rm \AA}$ (or $EW_0>55~{\rm \AA}$) among LBGs. 
 Therefore, this model which many previous studies adopted has been ruled 
 out.

 \item The dispersion of $\tau_\alpha$ and the halo mass dependence of 
 $\tau_\alpha$ are essential to explain all observations reasonably.
 However, a large dispersion of Ly$\alpha$ emissivity assumed in this 
 paper reduces a typical halo mass of LAEs too much and underpredicts 
 the LAE fractions in LBGs with $M_{\rm UV}<-19$. Therefore, the 
 Ly$\alpha$ emissivity dispersion among halos may be small if it exists.

 \item Based on the best model in this paper (Model G), we have discussed 
 the physical properties of LAEs at $z>6$. A typical halo mass of the 
 LAEs is estimated at $\simeq10^{11}$ M$_\odot$. The so-called 
 ``Ando-relation'' is also reproduced and its physical origin is a 
 simple scaling of halo mass: more H~{\sc i} in more massive halos.

 \item We have presented a simple formula to estimate the intergalactic 
 neutral hydrogen fraction, $x_{\rm HI}$, from the observed Ly$\alpha$ 
 luminosity density at $z\gtrsim6$. While $x_{\rm HI}$ at $z=5.7$, $6.6$, 
 and $7.0$ are still consistent with zero within uncertainties and the 
 obtained upper limits are $<0.40$ ($1\sigma$) as a volume-average, 
 we have obtained a non-zero value of $x_{\rm HI}=0.5_{-0.3}^{+0.1}$ 
 as a volume-average at $z=7.3$.
\end{itemize}

Finally, we note a possible direction for future updates of the model.
The best model, Model~G, is characterized by the significant fluctuation
in the halo Ly$\alpha$ optical depth (or the Ly$\alpha$ escape fraction)
and the halo mass dependence on the optical depth.  
However, it may be puzzling that Model~G has less importance of the
fluctuation in the Ly$\alpha$ production which is predicted by the
radiation hydrodynamics (RHD) simulations (see Fig.~3). 
In this respect, our simple analytic treatment may be insufficient. 
Therefore, we will examine a new modeling of the halo Ly$\alpha$
luminosity by using Ly$\alpha$ transfer simulations in galaxies produced
by the RHD simulations \citep{Abe2018}. 
The new recipe will include the production and transfer of Ly$\alpha$
photons in halos as well as the line profile self-consistently. 
Such an updated LAE model will be compared with the full data set of
SILVERRUSH \citep{Ouchi2017} and CHORUS (Inoue et al.~in prep.) in
future.

\begin{ack}
The authors appreciate Taishi Nakamoto, Nobunari Kashikawa, Ken
 Mawatari, and Takuya Hashimoto for discussions and encouragements.

AKI is supported by JSPS KAKENHI Grant Number 23684010, 26287034 and
 17H01114.
KH is supported by JSPS KAKENHI Grant Number 17H01110 and by a grant
 from NAOJ.
TI has been supported by MEXT as ``Priority Issue on Post-K computer''
 (Elucidation of the Fundamental Laws and Evolution of the Universe),
 JICFuS and JSPS KAKENHI Grant Number 17H04828.
HY is supported by JSPS KAKENHI Grant Number 17H04827. 
MO is supported by World Premier International Research Center
 Initiative (WPI Initiative), MEXT, Japan, and JSPS KAKENHI Grant Number 
 15H02064.

Numerical computations were partially carried out on the K computer at
the RIKEN Advanced Institute for Computational Science (Proposal
numbers hp150226, hp160212, hp170231), and ``Aterui'' supercomputer at
Center for Computational Astrophysics, CfCA, of National Astronomical
Observatory of Japan.  

The NB816 filter was supported by Ehime University (PI: Y.\ Taniguchi).
The NB921 filter was supported by JSPS KAKENHI Grant Number 23244025) (PI: M. Ouchi).

The Hyper Suprime-Cam (HSC) collaboration includes the astronomical communities of Japan and Taiwan, and Princeton University.  The HSC instrumentation and software were developed by the National Astronomical Observatory of Japan (NAOJ), the Kavli Institute for the Physics and Mathematics of the Universe (Kavli IPMU), the University of Tokyo, the High Energy Accelerator Research Organization (KEK), the Academia Sinica Institute for Astronomy and Astrophysics in Taiwan (ASIAA), and Princeton University.  Funding was contributed by the FIRST program from Japanese Cabinet Office, the Ministry of Education, Culture, Sports, Science and Technology (MEXT), the Japan Society for the Promotion of Science (JSPS),  Japan Science and Technology Agency  (JST),  the Toray Science  Foundation, NAOJ, Kavli IPMU, KEK, ASIAA, and Princeton University.

\end{ack}

\appendix

\section{A simple model of Ly$\alpha$ transfer in a halo}

We discuss a possible explanation of equation~(9) in \S3.3 
with a simple model of Ly$\alpha$ transfer in a halo through the ISM and CGM.
In future, the validity of this simple model will be examined 
with Ly$\alpha$ transfer simulations \citep{Abe2018} 
in model galaxies produced by Hasegawa et al.'s RHD simulations.

Suppose a galaxy emitting Ly$\alpha$ and the line-of-sight coordinate to a distant observer. 
The ISM and CGM of the galaxy along the line-of-sight is assumed to be composed of multiple layers of H~{\sc i} gas and the inter-layer medium.
An H~{\sc i} layer is completely opaque against Ly$\alpha$ but has holes where opacity against Ly$\alpha$ is negligible.
The inter-layer medium is also assumed to be optically thin for Ly$\alpha$, i.e. sufficiently low density and/or highly ionized.
The distribution of the H~{\sc i} layers is assumed random. 
We do not distinguish the ISM and CGM for simplicity and divide the ISM/CGM into $n$ portions along the line-of-sight. 
If the probability to have an H~{\sc i} layer in a portion is $p$ (constant), the probability to have $k$ layers in total along the line-of-sight 
is described by a binomial distribution: 
$P(k)={}_n C_k p^k (1-p)^{n-k}$. 
Taking the limit of $n\to\infty$ (i.e. the thickness of the portions $\to0$), 
we obtain $p\to0$ and $P(k)\to e^{-\lambda} \lambda / k!$, 
where $\lambda=np$ is the expected average number of the H~{\sc i} layers along the line-of-sight. 
Therefore, the realized number of the layers along the line-of-sight follows a Poisson probability with the parameter $\lambda$.

Suppose that the H~{\sc i} layers have a covering fraction of $f_{\rm c}$ 
(i.e. the areal fraction of the holes is $1-f_{\rm c}$).
The corresponding effective optical depth for Ly$\alpha$ of a single layer 
$\tau_{\rm layer}=-\ln(1-f_{\rm c})$. The typical $f_{\rm c}$ is uncertain 
but we may consider $f_{\rm c}\sim0.5$ because Ly$\alpha$ escape fractions 
of $\sim50$\% are observed as the highest value (e.g., \cite{Hayes2014})   
and may be regarded as a single H~{\sc i} layer case. In this case, 
we obtain $\tau_{\rm layer}\sim1$ and 
$\langle\tau_\alpha\rangle=\lambda\tau_{\rm layer}\approx\lambda$ 
with the average number of the H~{\sc i} layers along the line-of-sight, 
$\lambda$.
When $\lambda\gg1$, the Poisson distribution for the number of the layers, 
$k$, can be approximated to a Gaussian distribution with the mean of $\lambda$ 
and the dispersion of $\lambda$. Therefore, the realized optical depth for 
Ly$\alpha$ photons, $\tau_\alpha=k\tau_{\rm layer}\approx k$, 
also follows a Gaussian probability distribution 
with both of the mean and dispersion to be 
$\lambda\approx\langle\tau_\alpha\rangle$ 
as equation~(9).

\section{Effect of the narrowband transmission shape}

\begin{figure}
 \begin{center}
  \includegraphics[width=7cm]{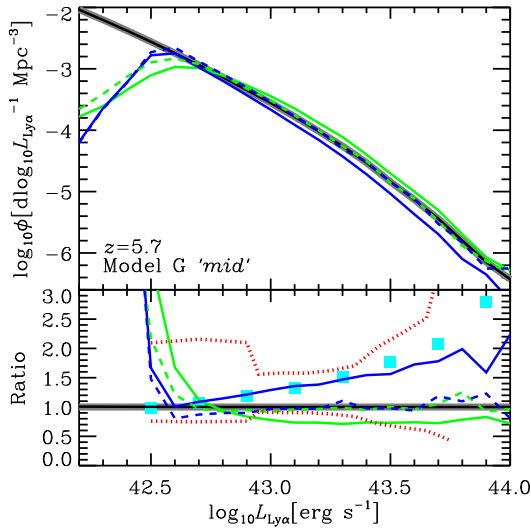}
 \end{center}
 \caption{{\it (Top)}
 The Ly$\alpha$ LFs at $z=5.7$ of the best-model, Model G, with 
 different LAE selections in virtual observations. The thick gray lines 
 are the Ly$\alpha$ LFs of all halos in the simulation box, while the 
 black lines are those of halos with the rest-frame Ly$\alpha$ equivalent 
 width larger than 10~\AA. They are almost identical. The solid lines 
 are the LFs of color-magnitude selected halos($NB816>5\sigma$ and 
 $i-NB816>1.2$): the IGM-transmitted true Ly$\alpha$ luminosity (green)
 and the NB-estimated Ly$\alpha$ luminosity (blue). The dashed lines are
 also the LFs of color-magnitude selected halos but for an ideal box-car
 NB816 shape. The color-code is the same as those of the solid lines. 
 {\it (Bottom)} The ratios of the LFs relative to those of all halos. 
 The line shape and color-code are the same as the top panel, except for
 the dotted lines showing the observational uncertainties taken from
 Konno et al.~(2017). The square marks are the estimates for
 Suprime-Cam/NB921 taken from Matthee et al.~(2015) as a reference.}
 \label{CF_err}
\end{figure}

\citet{Matthee2015} and \citet{Santos2016} have suggested that 
there is an effect of the narrowband (NB) transmission shape on 
Ly$\alpha$ LFs. The actual NB shape is not top-hat box-car but 
rather triangle-like. The assumption of a box-car shape may cause a 
systematic error in the LF derivation. Although the HSC NBs have 
more top-hat shapes than those of Suprime-Cam \citep{Ouchi2017}, 
the NB shape effect may remain. We examine this point through 
virtual observations of our simulations by assuming the actual NB 
transmission and an ideal box-car transmission with the same 
central wavelength and the width. In the following, we show the 
results at $z=5.7$ with the best-model, Model~G, but other models 
at other redshifts give the same results qualitatively.

Observationally, the Ly$\alpha$ luminosity is estimated from the 
NB magnitude if there is no spectroscopy. Assuming an NB whose
transmission is symmetric with respect to the central wavelength, 
a Ly$\alpha$ line located at the central wavelength, a complete
Gunn-Peterson trough below Ly$\alpha$, and a flat continuum (in $F_\nu$)
estimated from a broadband (BB) which does not include the Ly$\alpha$
line, the NB-estimated line flux is expressed as
\begin{equation}
 F_{\rm Ly\alpha}^{\rm NB} 
  = \frac{c \Delta\lambda_{\rm NB}}{\lambda_{\rm NB}^2}
  \left(F_{\nu,\rm NB}-\frac{1}{2}F_{\nu,\rm BB}\right)\,,
\end{equation}
where $F_{\nu,\rm NB}$ and $F_{\nu,\rm BB}$ are the NB and BB 
flux densities, respectively, $\Delta\lambda_{\rm NB}$ and 
$\lambda_{\rm NB}$ are the width and the central wavelength of 
the NB, and $c$ is the light speed.  The NB-estimated Ly$\alpha$
luminosity is expected to be the same as the real one if the above
assumptions are valid. However, the actual line is not always at the
center of the NB but can be at a wavelength where the transmission is
significantly lower than the peak at the center. As a result, the
NB-estimated luminosity is equal to or lower than the real one, causing
underestimation.

There is another effect in the survey volume estimation. We often 
estimate the survey volume simply from the NB width. However, 
this may not be correct because the survey volume depends on the 
luminosity; a low luminosity line can be detected only around the 
transmission peak wavelength and its survey volume is smaller, 
and vice versa.

Fig.~\ref{CF_err} summarizes our results. First, the thick gray line 
is the true LF based on all halos in the simulation box. Note 
that Ly$\alpha$ luminosity in this section is always apparent (or
observable) one. The black line which is almost identical to the
black line is the LF based on halos whose Ly$\alpha$
equivalent width (EW) in the rest-frame is larger than 10~\AA\ 
(i.e. halos selected as LAEs). Therefore, the non-LAEs' contribution 
to the Ly$\alpha$ LF is negligible. The green lines are the LFs based on the
halos selected by the same color-magnitude selection as the observations
\citep{Konno2017} and their actual Ly$\alpha$ luminosity, whereas the
blue lines are the LFs for the same color-magnitude selected LAEs but
their NB-estimated Ly$\alpha$ luminosity. The magnitude limit is set at
$5\sigma$ in the NB magnitude corresponding to 
$L_{\rm Ly\alpha}=10^{42.5}$~erg~s$^{-1}$, 
and we have added photometric errors assuming a Gaussian distribution 
in the $F_{\nu,\rm NB}$ space. The actual NB transmission curves are 
assumed for the solid lines but the ideal top-hat box-car curves are 
assumed for the dashed lines. As expected, the LFs with the ideal 
NB transmission are very consistent with the true one (the gray line), 
except for so-called incompleteness around the detection limit. 

The survey volume effect can be found in the green solid lines. 
Underestimation in the volume at brighter luminosity causes 
overestimation in the LF up to 30\% and a faster drop in the LF 
than the green dashed line around the detection limit indicates 
overestimation in the survey volume and underestimation in the LF.
The underestimation in NB-estimated luminosity can be found in the 
blue solid lines. This effect steepens the LF shape and the 
deviation from the true one becomes larger at brighter luminosity.
Since this effect is compensated partly by the survey volume 
effect, the difference from the true LF remains modest: a factor 
of 1.5 at $10^{43.5}$ erg s$^{-1}$. For $>10^{43.8}$ erg s$^{-1}$, 
it is difficult to derive the difference owing to a limited number of
halos in the luminous end in our simulation. Interestingly, the ratio 
relative to the true LF is similar to (and smaller than) those estimated
by \citet{Matthee2015} at $<10^{43.5}$ ($>10^{43.5}$) erg s$^{-1}$.

In conclusion, there is the NB transmission effect on the Ly$\alpha$ LF
as proposed by \citet{Matthee2015} and \citet{Santos2016}. Although the
correction factor is as small as $<1.5$ at $<10^{43.5}$ erg s$^{-1}$ and
the current observational uncertainties \citep{Konno2017} are larger
than the NB shape effect, this causes a systematic effect and should be
corrected. On the other hand, \citet{Konno2017} (and also
\cite{Shimasaku2006,Ouchi2008,Ouchi2010}) performed an end-to-end Monte
Carlo simulation to derive their Ly$\alpha$ LFs. In their simulations,
they had taken into account the NB shape effect. Therefore, their
Ly$\alpha$ LFs are considered to be those corrected for the NB shape
effect and the best estimations of the true ones.


\end{document}